\documentclass[aps,prd,preprint,draft,showpacs,eqsecnum,nofootinbib]{revtex4}
\usepackage{graphicx,epsf,amssymb}
\newcommand{\be}{\begin{equation}}
\newcommand{\ee}{\end{equation}}

\newcommand{\bea}{\begin{eqnarray}}
\newcommand{\eea}{\end{eqnarray}}
\newcommand{\Tr}{{\rm Tr}}

\def\del{\partial}
\def\sst{\scriptscriptstyle}
\def\tr{\mathop{\rm tr}\nolimits}
\def\Tr{\mathop{\rm Tr}\nolimits}

\def\coeff#1#2{\relax{\textstyle {#1 \over #2}}\displaystyle}

\newif\ifdraft
\drafttrue
\newif\ifpreprint
\preprinttrue

\def\Sect#1{Section~{\ref{#1}}}
\def\sect#1{section~{\ref{#1}}}
\def\app#1{appendix~{\ref{#1}}}

\def\fig#1{fig.~{\ref{#1}}}
\def\figs#1#2{figs.~{\ref{#1}} and {\ref{#2}}}

\def\Tr{\, {\rm Tr}}

\def\hf{{{1\over2}}}

\def\ns{n_{\mskip-2mu s}}
\def\nf{n_{\mskip-2mu f}}

\def\spa#1.#2{\left\langle#1\,#2\right\rangle}
\def\spb#1.#2{\left[#1\,#2\right]}
\def\sand#1.#2.#3{%
\left\langle\smash{#1}{\vphantom1}^{-}\right|{#2}%
\left|\smash{#3}{\vphantom1}^{-}\right\rangle}
\def\sandp#1.#2.#3{%
\left\langle\smash{#1}{\vphantom1}^{-}\right|{#2}%
\left|\smash{#3}{\vphantom1}^{+}\right\rangle}
\def\sandpp#1.#2.#3{%
\left\langle\smash{#1}{\vphantom1}^{+}\right|{#2}%
\left|\smash{#3}{\vphantom1}^{+}\right\rangle}
\def\sandpm#1.#2.#3{%
\left\langle\smash{#1}{\vphantom1}^{+}\right|{#2}%
\left|\smash{#3}{\vphantom1}^{-}\right\rangle}
\def\sandmp#1.#2.#3{%
\left\langle\smash{#1}{\vphantom1}^{-}\right|{#2}%
\left|\smash{#3}{\vphantom1}^{+}\right\rangle}
\def\sandmm#1.#2.#3{%
\left\langle\smash{#1}{\vphantom1}^{-}\right|{#2}%
\left|\smash{#3}{\vphantom1}^{-}\right\rangle}
\def\spab#1.#2.#3{\sandmm#1.#2.#3}
\def\spba#1.#2.#3{\sandpp#1.#2.#3}
\def\spaa#1.#2.#3.#4{\sandmp#1.{#2#3}.#4}
\def\spbb#1.#2.#3.#4{\sandpm#1.{#2#3}.#4}

\newbox\charbox
\newbox\slabox
\def\s#1{{      
         \setbox\charbox=\hbox{$#1$}
         \setbox\slabox=\hbox{$/$}
         \dimen\charbox=\ht\slabox
         \advance\dimen\charbox by -\dp\slabox
         \advance\dimen\charbox by -\ht\charbox
         \advance\dimen\charbox by \dp\charbox
         \divide\dimen\charbox by 2
         \raise-\dimen\charbox\hbox to \wd\charbox{\hss/\hss}
         \llap{$#1$}
}}

\def\eqn#1{eq.~(\ref{#1})}
\def\Eqn#1{Equation~(\ref{#1})}
\def\eqns#1#2{eqs.~(\ref{#1}) and~(\ref{#2})}
\def\Eqns#1#2{Eqs.~(\ref{#1}) and~(\ref{#2})}
\def\qb{{\bar q}}

\def\ib{{\bar\imath}}
\def\e{\epsilon}
\def\eps{\epsilon}

\def\Gr{{\rm Gr}}

\def\sign{{\mathop{\rm sign}\nolimits}}
\def\lr{\leftrightarrow}

\def\Li{\mathop{\rm Li}\nolimits}
\def\Ls{\mathop{\rm Ls}\nolimits}
\def\Split{\mathop{\rm Split}\nolimits}
\def\tree{{(0)}}
\def\oneloop{{(1)}}
\def\lloop{{(l)}}

\def\cg{c_\Gamma}
\def\Ord{{\cal O}}

\def\Kh{{\hat K}}

\def \as {\relax\ifmmode\alpha_s\else{$\alpha_s${ }}\fi}

\def\si{\sigma}
\def\sandp#1.#2.#3{%
\left\langle\smash{#1}{\vphantom1}^{+}\right|{#2}%
\left|\smash{#3}{\vphantom1}^{+}\right\rangle}
\def\ksl{\s{k}}
\def\Ksl{\s{K}}

\def\Soft{{\cal S}}

\def\Re{\mathop{\rm Re}}
\def\Res{\mathop{\rm Res}}
\def\tlambda{{\tilde\lambda}}

\def\f{{\! f}}
\def\ihf{{\textstyle{i\over2}}}

\newbox\ourfigbox
\def\SizedFigureWithCaption#1#2#3{%
\setbox\ourfigbox
   \hbox{\hss\epsfxsize #1 \epsfbox{#2}\hss}
\hbox to \wd\ourfigbox{\vbox{\noindent\copy\ourfigbox\break
\vskip -6mm      \hbox to \wd\ourfigbox{\hss#3\hss}}}
}

\def\llongrightarrow{%
\relbar\mskip-0.5mu\joinrel\mskip-0.5mu\relbar
      \mskip-0.5mu\joinrel\longrightarrow}
\def\inlimit^#1{\buildrel#1\over\llongrightarrow}
\def\dash{\hbox{-\kern-.02em}}

\def\spa#1.#2{\left\langle#1\,#2\right\rangle}
\def\spb#1.#2{\left[#1\,#2\right]}
\def\spash#1.#2{\vphantom{\hat K}\spa{\smash{#1}}.{\smash{#2}}}
\def\spbsh#1.#2{\vphantom{\hat K}\spb{\smash{#1}}.{\smash{#2}}}
\def\lor#1.#2{\left(#1\,#2\right)}
\def\sand#1.#2.#3{%
\left\langle\smash{#1}{\vphantom1}^{-}\right|{#2}%
\left|\smash{#3}{\vphantom1}^{-}\right\rangle}
\def\sandpp#1.#2.#3{%
\left\langle\smash{#1}{\vphantom1}^{+}\right|{#2}%
\left|\smash{#3}{\vphantom1}^{+}\right\rangle}
\def\sandpm#1.#2.#3{%
\left\langle\smash{#1}{\vphantom1}^{+}\right|{#2}%
\left|\smash{#3}{\vphantom1}^{-}\right\rangle}
\def\sandmp#1.#2.#3{%
\left\langle\smash{#1}{\vphantom1}^{-}\right|{#2}%
\left|\smash{#3}{\vphantom1}^{+}\right\rangle}

\begin{document}
\hfuzz 10 pt


\ifpreprint
\noindent
SLAC--PUB--12054
\hfill hep-ph/0608180
\fi

\title{Recursive Construction of Higgs\,$+$\,Multiparton Loop 
Amplitudes: \\
The Last of the $\phi$-nite Loop Amplitudes%
\footnote{Research supported by the US Department of
Energy under contract DE--AC02--76SF00515,
and by the Italian MIUR under contract 2004021808\_009.}}

\author{Carola F. Berger$^1$, Vittorio Del Duca$^2$, and Lance J. Dixon$^1$}

\affiliation{
$^{1}$ Stanford Linear Accelerator Center \\
               Stanford University \\
              Stanford, CA 94309, USA \\
$^{2}$ Istituto Nazionale di Fisica Nucleare \\
              Sez.~di Torino \\
via P. Giuria, I--10125 Torino, Italy }

\date{August, 2006}

\begin{abstract}
We consider a scalar field, such as the Higgs boson $H$, coupled
to gluons via the effective operator $H \tr G_{\mu\nu} G^{\mu\nu}$ induced
by a heavy-quark loop.  We treat $H$ as the real part of a
complex field $\phi$ which couples to the self-dual part of the gluon
field-strength, via the operator
$\phi \tr G_{{\sst SD}\,\mu\nu} G_{\sst SD}^{\mu\nu}$,
whereas the conjugate field $\phi^\dagger$ couples to the
anti-self-dual part.  There are three infinite sequences of
amplitudes coupling $\phi$ to quarks and gluons that vanish at
tree level, and hence are finite at one loop, in the QCD coupling.
Using on-shell recursion relations, we find compact expressions for
these three sequences of amplitudes and discuss their analytic
properties.
\end{abstract}

\pacs{12.38.Bx, 14.80.Bn, 11.15.Bt, 11.55.Bq \hspace 
{1cm}}
\maketitle


\renewcommand{\thefootnote}{\arabic{footnote}}
\setcounter{footnote}{0}


\section{Introduction}
\label{IntroSection}

In the next year, the Large Hadron Collider (LHC) will begin
operation at CERN, ushering in a new regime of directly
probing physics at the shortest distance scales.
The LHC will search for physics beyond the Standard Model,
as well as for the Higgs boson. The Higgs mechanism, 
which describes the breaking of electroweak symmetry
in the Standard Model and its supersymmetric extensions,
is simultaneously the keystone of the Standard Model (SM) 
and its least-well-tested ingredient.

The dominant process for Higgs-boson production, over the 
entire range of Higgs masses relevant for the LHC, 
is the gluon fusion process, $g g\to H$, which is mediated 
by a heavy-quark loop~\cite{Georgi1977gs}. 
The leading contribution comes from the top quark. 
The contributions from other quarks are suppressed by at least a factor
of $\Ord(m_b^2/m_t^2)$, where $m_t, m_b$ are the mass of the top and
of the bottom quark, respectively.  Because the Higgs boson is
produced via a heavy-quark loop, the calculation of the production
rate is quite involved, even at leading order in $\as$. 
The inclusive production rate for $g g\to H + X$ has been computed at
next-to-leading order (NLO) in $\as$~\cite{NLOHiggs}, 
including the full quark-mass dependence~\cite{NLOHiggsExact}, 
which required an evaluation at two-loop accuracy. 
The NLO QCD corrections increase the
production rate by close to 100\%. However, in the {\it
large-$m_t$ limit}, namely if the Higgs mass $m_H$ is smaller than
the threshold for the creation of a top-quark pair, $m_H < 2\,
m_t$, the coupling of the Higgs to the gluons via a top-quark loop
can be replaced by an effective coupling~\cite{NLOHiggs,HggOperator,HggOperator2}.
This approximation simplifies calculations tremendously, because
it effectively reduces the number of loops in a given diagram by one. 

It has been shown that one can approximate the full NLO QCD
corrections quite accurately by computing the NLO QCD correction factor,
$K^{\rm NLO} \equiv \sigma^{\rm NLO}/\sigma^{\rm LO}$, in the
large-$m_t$ limit, and multiplying it by the exact leading-order 
calculation.  This approximation is good to within 10\% in the
entire Higgs-mass range at the LHC, \emph{i.e.} up to
1~TeV~\cite{Kramer1996iq}. 
The reason it works so well is that 
the QCD corrections to $g g\to H$ are
dominated by soft-gluon effects, which do not resolve the
top-quark loop that mediates the coupling of the Higgs boson to the
gluons. The next-to-next-to-leading order (NNLO) corrections to
the production rate for $g g\to H$ have been evaluated in the
large-$m_t$ limit~\cite{NNLOHiggs} and display a modest increase,
less than 20\%, with respect to the NLO evaluation.
The dominant part of the NNLO corrections comes from
virtual, soft, and collinear gluon radiation~\cite{Catani2001ic}, in
agreement with the observations at NLO. In addition, the threshold
resummation of soft-gluon effects~\cite{Kramer1996iq,Catani2003zt}
enhances the NNLO result by less than 10\%, showing that the
calculation has largely stabilized by NNLO.  The soft and collinear terms have
recently been evaluated to one more order (N$^3$LO)~\cite{MochVogt}, 
with partial results to N$^4$LO~\cite{Ravindran},
which further reduces the uncertainty on the inclusive production
cross section.

Backgrounds from SM physics are usually quite
large and  hamper the Higgs-boson search. A process that
promises to have a more amenable background is Higgs production in
association with a high transverse-energy ($E_T$) jet, $pp\to
H\,+$~jet. In addition, this process offers the advantage of being
more flexible in the choice of acceptance cuts to curb the
background.  The $pp\to H\,+$ jet process is known exactly
at leading
order~\cite{Ellis1988xu}, while the NLO
calculation~\cite{deFlorian1999zd} has been performed in the
large-$m_t$ limit. For Higgs $+$ jet production, the large-$m_t$
limit is valid as long as $m_H < 2\, m_t$ and the transverse
energy $E_T$ is smaller than the top-quark mass, $E_T<
m_t$~\cite{Baur1989cm}. As long as these conditions hold,
the jet-Higgs invariant mass can be taken larger than
$m_t$~\cite{H2j}. At $m_H = 120$~GeV, the NLO corrections to the
$pp\to H\,+$ jet process increase the leading-order prediction by
about 60\%, and are thus of the same order as the NLO corrections
to fully inclusive $pp\to H$ production considered above. At
present, the NNLO corrections to $pp\to H\,+$ jet are not known.

An even more interesting process is Higgs production in association
with two jets. In fact, a key component of the program to measure
the Higgs-boson couplings at the LHC is the vector-boson fusion 
(VBF) process, $qq\to qqH$ via $t$-channel $W $ or $Z$
exchange.  This process is characterized by two forward quark 
jets~\cite{Zeppenfeld2000td}.  The NLO corrections to Higgs 
production via VBF fusion in association with
two jets are known to be small~\cite{NLOWBFFOZ}. Thus, 
small theoretical uncertainties are predicted for this 
production process.
The production of $H\,+2$ jets via gluon fusion
is a part of the inclusive Higgs production signal.
However, it constitutes a background when trying to isolate 
the $HWW$ and $HZZ$ couplings responsible for the VBF process. 
A precise description of this background is needed in
order to separate the two major sources of $H\,+$~2~jets events:
one needs to find characteristic distributions that distinguish VBF
from gluon fusion.

The production of $H+2$~jets via gluon fusion is known at leading order
in the large-$m_t$ limit~\cite{DawsonKauffman}
and exactly~\cite{H2j2}. As in the case of $H+1$~jet production,
the large-$m_t$ limit is valid as long as $m_H < 2 \,m_t$ and the
jet transverse energies are smaller than the top-quark mass,
$E_{T,1},E_{T,2} < m_t$, even if the dijet mass, as well as either of
the jet-Higgs masses, is larger than the top-quark mass~\cite{H2j}.
At present, the NLO calculation for the production of $H\,+$~2 jets
via gluon fusion has not yet been completed,\footnote{%
Preliminary NLO results have been reported recently~\cite{Zanderighi06}.} 
not even in the large-$m_t$ limit, although the
necessary amplitudes, namely the one-loop amplitudes for
a Higgs boson plus four partons~\cite{EGZHiggs} and the tree amplitudes for
a Higgs boson plus five partons~\cite{DFM,DGK,BGK}, have been 
computed.
Considering the aforementioned results for the NLO corrections to 
fully inclusive Higgs production and to Higgs production
in association with one jet, and the fact that the largest NLO 
corrections are usually found in gluon-initiated processes, 
there is no reason to expect that the NLO corrections to $H+\,2$~jets 
production via gluon fusion are small.
Thus, such a NLO calculation is highly desirable.

Another distinguishing feature of VBF is that to leading order,
and with a good approximation also to NLO, no color is exchanged
in the $t$ channel~\cite{Dokshitzer1987nc}. The different
gluon-radiation pattern expected for Higgs production via VBF
versus its major backgrounds, namely $t\bar{t}$ production and QCD
$WW + 2$~jet production, is at the core of the central-jet veto
proposal, both for heavy~\cite{Barger1994zq} and
light~\cite{Kauer2000hi} Higgs searches. A veto of any additional
jet activity in the central-rapidity region is expected to
suppress the backgrounds more than the signal, because the QCD
backgrounds are characterized by quark or gluon exchange in the
$t$ channel. The exchange of colored partons should lead to more
central gluon radiation. In the case of Higgs production via gluon
fusion, with two jets separated by a large rapidity interval, the
scattering process is dominated by gluon exchange in the $t$
channel. Thus, as for the QCD backgrounds, the bremsstrahlung
radiation is expected to occur everywhere in rapidity, and Higgs
production via gluon fusion can be likewise checked by requiring a
central-jet veto. For hard radiation, the effectiveness of such a
veto may be analyzed through $H+\,3$~jets production, which for
gluon fusion is known in the large-$m_t$ limit at leading
order~\cite{DFM}.  An analogous study that also includes soft 
radiation has just been completed, by interfacing matrix-element 
calculations of $H+\,2$ and 3 jets production with 
parton-shower effects~\cite{DelDuca2006hk}.

The discussion above motivates us to look for methods to compute
Higgs production in association with many jets at least to NLO
or higher accuracy.
Tree amplitudes for a Higgs boson produced via gluon fusion together
with many partons
have been computed analytically in the large-$m_t$ limit~\cite{DGK,BGK}
and are available numerically through the matrix-element Monte Carlo
generators ALPGEN~\cite{ALPGEN} and MADEVENT~\cite{MADGRAPH}.
However, at one loop only the QCD amplitudes for a Higgs boson plus  
up to four partons are known~\cite{EGZHiggs}. 
The three-parton helicity amplitudes were computed analytically
in the large-$m_t$ limit~\cite{Schmidt}. 
The four-parton results are semi-numerical, except for two cases:  
(1) analytic results have been presented for the case of two 
quark-antiquark pairs~\cite{EGZHiggs}, and 
(2) the case of four gluons all with the same helicity is now
known analytically as well~\cite{BadgerGloverH4g}.

Recently a lot of progress has been made in the computation
of tree-level gauge-theory amplitudes~\cite{DGK,BGK,CSW,Currents,RSVNewTree,%
BCFRecursion,BCFW,LuoWen,BFRSV,BadgerMassive,BadgerVector},
following Witten's
proposal~\cite{WittenTopologicalString} of a weak-weak coupling duality
between ${\cal N} = 4$ supersymmetric gauge theory and the topological
open-string $B$ model in twistor space
(for a review, see ref.~\cite{CSReview}). Recursion relations for
computing tree amplitudes have been written~\cite{BCFRecursion,BCFW}
which employ only on-shell amplitudes at complex values of the external
momenta.
Likewise, progress has been made in the computation of one-loop
amplitudes~\cite{BST,BCF7,NeqFourSevenPoint,BCFII,NeqFourNMHV,%
OtherGaugeCalcs,BBCF,BFM,OnShellRecursionI,LastFinite,Bootstrapping,%
FordeKosower,BBDFK1,BBDFK2}, particularly those with many external legs,
for which there are many different kinematic variables.  If careful
attention is not paid to the analytic structure, very large
analytic expressions may result.  One approach to this problem
is to compute the loop amplitudes numerically or 
semi-numerically~\cite{NumericalLoops,EGZHiggs,Zanderighi06}.
Another approach is to exploit the analytic properties of the
amplitudes in order to facilitate their computation.
The two main analytic properties of loop amplitudes are
branch cuts and factorization poles.

In pure QCD with massless quarks, there are special helicity 
configurations for which the tree amplitudes vanish, 
and the corresponding one-loop amplitudes are finite.
Such helicity amplitudes are of no immediate use in phenomenology, 
because they contribute first at NNLO.  However,
they have interesting analytic properties.
The specific helicity amplitudes in this category are the 
pure-glue one-loop amplitudes for which all the gluons have the
same helicity, or else all but one do, namely,
${\cal A}_n^{\oneloop}(1^\pm,2^+,3^+,\ldots,n^+)$;
and in addition the amplitudes with one pair of massless external quarks
and $(n-2)$ positive-helicity gluons,
${\cal A}_n^{\oneloop}(1_{\bar{q}}^-,2_q^+,3^+,\ldots,n^+)$.
These amplitudes contain no branch cuts; they are purely rational 
functions of the kinematic variables~\cite{AllPlus,Mahlon}.
Compact forms for the amplitudes have been found by constructing on-shell 
recursion relations, along the lines of the tree-level 
relations~\cite{BCFRecursion,BCFW}, and then solving them in closed 
form~\cite{OnShellRecursionI,LastFinite}.

All the other one-loop helicity amplitudes in QCD contain branch cuts.
The terms with branch cuts can be determined from unitarity in four
dimensions; then recursion relations can be established for the
remaining rational terms~\cite{Bootstrapping,BBDFK1}.  This approach
offers an efficient way in principle of building up a generic 
$n$-point one-loop amplitude from one-loop amplitudes with a 
smaller number of points.  It has been applied in practice to 
determine the rational parts of the $n$-gluon amplitudes with two
negative-helicity gluons in arbitrary locations in the 
color ordering~\cite{FordeKosower,BBDFK2}, and also
for cases with three or four color-adjacent negative-helicity 
gluons~\cite{BBDFK1}.
(For $n=6$, the rational terms for three color-nonadjacent
negative-helicity gluons, as well as two negative-helicity ones, 
have been worked out~\cite{XYZ6pt} using Feynman-diagrammatic 
methods~\cite{XYZmethod}.)

A closely-related framework can be constructed for amplitudes
for a scalar field, such as the Higgs boson $H$, interacting with 
an arbitrary number of quarks and gluons. In the large-$m_t$ limit, 
the Higgs field couples to gluons via the effective operator 
$H \tr G_{\mu\nu} G^{\mu \nu}$ induced by the heavy-quark loop.
To make the situation parallel to the pure-QCD case,
we follow refs.~\cite{DGK,BGK} and treat $H$ as the real part of 
a complex field $\phi$.  Here $\phi$ couples to the self-dual 
part of the gluon field strength via the operator 
$\phi \tr G_{{\sst SD}\,\mu\nu} G_{\sst SD}^{\mu\nu}$,
whereas the conjugate field $\phi^\dagger$ couples to the
anti-self-dual part.   
Then, exactly as in the case of pure QCD 
({\it i.e.} when $\phi$ is absent),
the amplitudes ${\cal A}_n(\phi, 1^\pm,2^+,3^+,\ldots,n^+)$ and
${\cal A}_n(\phi, 1_{\bar{q}}^-,2_q^+,3^+,\ldots,n^+)$ vanish at tree
level, and are free of branch cuts at one loop.
(They are also free of infrared and ultraviolet divergences.)
Thus the one-loop amplitudes,
${\cal A}_n^{\oneloop}(\phi, 1^\pm,2^+,3^+,\ldots,n^+)$ and
${\cal A}_n^{\oneloop}(\phi, 1_{\bar{q}}^-,2_q^+,3^+,\ldots,n^+)$,
can be written as rational functions of the kinematic variables.
The $\phi^\dagger$ amplitudes are obtained from the $\phi$ amplitudes
by parity, which also reverses all parton helicities.
Amplitudes containing a scalar Higgs field $H$ are given by 
the sum of the $\phi$ and $\phi^\dagger$ amplitudes;
whereas amplitudes containing a pseudoscalar field $A$ instead 
are found from the difference of the two amplitudes. 
In these sums and differences, the finite $\phi$ amplitudes have to be 
combined with the image under parity of divergent $\phi$ amplitudes
with mostly negative gluon helicities.

The aim of this paper is to construct and solve on-shell recursion 
relations for the finite one-loop amplitudes with a single scalar field $\phi$. 
To assemble complete analytic helicity amplitudes for $H$ or $A$ fields,
other, divergent helicity configurations are also required.
These are the $\phi^\dagger$ amplitudes with the same
helicity configurations as those computed here, or equivalently,
the $\phi$ amplitudes with all parton helicities reversed.
The full set of helicity amplitudes for $H$ or $A$ fields
requires as well $\phi$ amplitudes with two or more negative parton
helicities {\it and} two or more positive helicities.
We leave such computations to future work --- except that we shall
quote the recent results of ref.~\cite{BadgerGloverH4g} for
four identical-helicity gluons.  We stress, however,
that the amplitudes we compute represent one piece 
in the $\phi$-$\phi^\dagger$ decomposition of $H$ or $A$
amplitudes for which the corresponding tree amplitudes are
nonvanishing.   Hence they contribute to NLO Higgs cross sections.
(This is in contrast to the finite amplitudes in pure QCD,
without a $\phi$ field, which first contribute at NNLO.)
A second reason why the finite $\phi$ amplitudes are useful is that
in the recursive approach we pursue, amplitudes with few negative 
helicities are needed to supply factorization information for 
amplitudes with more legs and more negative helicities.

The construction of on-shell recursion relations relies on information
about the factorization properties of amplitudes, as momentum invariants
built from two or more external momenta become null.  
In the ``collinear'' case of two external momenta, this behavior, 
involving complex momenta, is quite subtle at the loop level.
Loop amplitudes can contain double poles in collinear 
invariants~\cite{OnShellRecursionI},
and ``unreal'' poles for which the collinear limit in real 
momenta is not singular, yet nonetheless a single pole 
arises for complex momenta~\cite{LastFinite}.
In our recursive approach to the finite quark-gluon amplitudes,
${\cal A}_n^{\oneloop}(\phi, 1_{\bar{q}}^-,2_q^+,3^+,\ldots,n^+)$,
we encounter both these types of poles.
A knowledge of the residues at these poles is essential
for constructing correct recursion relations.
It might be possible to rigorously analyze such contributions
using ``space-cone'' gauge techniques~\cite{CSandVY}.
In this paper we adopt a much more pragmatic approach;
indeed, we argue that the unreal poles take precisely the same form 
as in the pure-QCD case~\cite{LastFinite}, for which several
types of consistency checks were available.

Using this approach, we present recursion relations for the finite 
$\phi$-quark amplitudes with an arbitrary number of external gluons.
We find compact solutions to the relations, valid for all $n$.
To confirm these relations we perform nontrivial consistency
checks of the factorization properties of the solutions.
One of the factorizations, as the momenta of the quark and
anti-quark become collinear, is onto the one-loop
$\phi$-plus-$n$-gluon amplitudes with a single negative gluon helicity
and the rest positive, and determines this sequence of amplitudes.

The $\phi$-amplitudes with two or more negative helicities can be obtained
by extensions of the methods of
refs.~\cite{Neq4Oneloop,UnitarityMachinery,BCFII,%
Bootstrapping,BBDFK1,BBDFK2}.
Unitarity (or the related loop-level application of
maximally-helicity-violating (MHV) rules~\cite{BST}) 
can be used to determine the cut-containing functions
from known tree-level
amplitudes~\cite{Neq4Oneloop,UnitarityMachinery,BCFII}.
The cut-containing parts of Higgs amplitudes with an arbitrary
number of identical-helicity gluons have been found in this
way~\cite{BadgerGloverH4g}.
The remaining rational functions can then be computed via a
factorization bootstrap approach, in the form of on-shell recursion
relations analogous to the approach of ref.~\cite{Bootstrapping,BBDFK1,BBDFK2}
in the pure-QCD case. In this publication, however, we mainly concentrate
on the amplitudes with one or no negative helicities,
which do not contain any cuts.

The paper is organized as follows: 
After introducing some convenient notation in \sect{NotationSection}, 
we review the on-shell recursion
relations in \sect{RecursionReviewSection} and the previously-known
one-loop finite amplitudes in pure QCD in \sect 
{QCDAmplitudesReviewSection}.
In \sect{PhiAmplitudesReviewSection}
we summarize known results for amplitudes with a $\phi$ field and
present a recursion relation for the
MHV tree-level $\phi$-plus-$n$-gluon amplitudes,
${\cal A}_n^{\tree}(\phi, 1^+,\ldots,i^-,\ldots,j^-,\ldots,n^+)$.
In \sect{SoftHiggsSection} we discuss some interesting features
of the soft-Higgs limit of the $\phi$ amplitudes, as the momentum of
$\phi$ vanishes.  In the tree-level case, the soft-Higgs limit
always produces a fixed, helicity-dependent multiple of the corresponding
QCD amplitude.  At one loop, however, this is not true in general,
although it is true for the finite $\phi$ amplitudes~\cite{DGK}.

In \sect{AllPlusAmplitudesSection} we present and solve
the recursion relation for the one-loop $\phi$-plus-$n$-gluon amplitude,
with all the gluons of like helicity. In \sect{AmplitudesSection} we  
do the same for the one-loop $\phi$-quark-gluon amplitude
${\cal A}_n^{\oneloop}(\phi, 1_{\bar{q}}^-,2_q^+,3^+,\ldots,n^+)$.
Using factorization, we obtain the $\phi$-plus-$n$-gluon 
amplitudes with a single negative-helicity gluon,
${\cal A}_n^{\oneloop}(\phi, 1^-,2^+,3^+,\ldots,n^+)$. 
\Sect{TwoThreeFourSection} collects explicit results for all 
helicity configurations for the 
$\phi\,+\,$2 and 3 parton amplitudes, as well as partial
results for $\phi\,+\,$4 partons, using also results from
refs.~\cite{Schmidt,BadgerGloverH4g}.
In \sect{ConclusionSection} we draw our conclusions. 
In \app{NormalizationAppendix} we discuss the normalization of the one-loop
amplitudes.
In \app{TreeProofAppendix} we prove the validity of the on-shell 
recursion relations for $\phi$-amplitudes at tree level.


\section{Notation}
\label{NotationSection}

In this section
we first discuss the decomposition of amplitudes containing a single Higgs
boson and multiple partons, 
into amplitudes with a $\phi$ and a $\phi^\dagger$ field,  
respectively.  Then we review the color-decomposition of these 
amplitudes at tree level and at one loop.  Finally, we introduce 
some convenient notation for the manipulation
of these amplitudes in the spinor-helicity formalism.

\subsection{$\phi$-$\phi^\dagger$ decomposition of Higgs amplitudes}
\label{DecompositionSection}

In ref.~\cite{DGK}, the MHV rules for pure-gluon amplitudes~\cite{CSW}
were extended to include the coupling to the SM Higgs boson 
in the large-$m_t$ limit. These rules were then
further extended to compute amplitudes with quarks~\cite{BGK}.
Although we pursue in this paper a recursive rather than MHV approach,
the construction of refs.~\cite{DGK,BGK} is still useful,
so we briefly summarize it here.

The coupling of the SM Higgs boson to gluons is
through a fermion loop~\cite{Georgi1977gs,HggOperator,Rizzo}, 
with the dominant contribution coming
from the top quark, because the Higgs coupling to quarks is
proportional to the respective quark masses. For large
top mass, $m_t \rightarrow \infty$, the top quark can be
integrated out, yielding the following effective
interaction~\cite{HggOperator,HggOperator2}
\be
\mathcal{L}_H^{\mbox{\tiny int}} =
\frac{C}{2} H \, \tr G_{\mu\nu} G^{\mu\nu} \,,
\label{Lint1}
\ee
where the strength of the interaction is given, to leading order
in the strong coupling, by $C = \as/(6 \pi v)$, with $v = 246$ GeV.
(The trace is normalized so that 
$\tr G_{\mu\nu} G^{\mu\nu} = G^a_{\mu\nu} G^{a\,\mu\nu}$.)

In ref.~\cite{DGK}, it was found that the MHV or twistor-space  
structure is simplest upon dividing the effective interaction 
Lagrangian into two parts, a holomorphic (self-dual) 
and an antiholomorphic (anti-self-dual) part.
Then the MHV rules for QCD~\cite{CSW} can be extended
straightforwardly to include these new interaction types.
The Higgs boson is considered to be the real part of a complex field,
$\phi = \frac{1}{2} (H + i A)$, so that
\bea
{\cal L}^{\rm int}_{H,A} &=&
{C\over2} \Bigl[ H \tr G_{\mu\nu} G^{\mu\nu}
              + i A \tr G_{\mu\nu}\, {}^*G^{\mu\nu} \Bigr]
\label{effinta}\\
&=&
C \Bigl[ \phi \tr G_{{\sst SD}\,\mu\nu} G_{\sst SD}^{\mu\nu}
+ \phi^\dagger \tr G_{{\sst ASD}\,\mu\nu} G_{\sst ASD}^{\mu\nu} \Bigr]
\ .
\label{effintb}
\eea
Here the field strength has been divided into a self-dual (SD) and an
anti-self-dual (ASD) field strength,
\be
G_{\sst SD}^{\mu\nu} = \hf(G^{\mu\nu}+{}^*G^{\mu\nu}) \ , \quad
G_{\sst ASD}^{\mu\nu} = \hf(G^{\mu\nu}-{}^*G^{\mu\nu}) \ , \quad
{}^*G^{\mu\nu} \equiv \ihf \epsilon^{\mu\nu\rho\sigma} G_{\rho\sigma}  
\ .
\ee
The scalar $H$ and pseudoscalar $A$ are reconstructed from the
complex fields $\phi$ and $\phi^\dagger$ according to,
\be
H = \phi + \phi^\dagger\,,
\qquad\quad
A = {1\over i} ( \phi - \phi^\dagger ) \,.
\label{Heqn}
\ee
{}From \eqn{Heqn}, the amplitude for a single scalar Higgs boson
plus multiple quarks and gluons can be recovered,
at any loop order $l$,
as the sum of the amplitudes with $\phi$ and $\phi^\dagger$,
\begin{equation}
{\cal A}_n^\lloop(H,\ldots) =
{\cal A}_n^\lloop(\phi,\ldots)
+ {\cal A}_n^\lloop(\phi^\dagger,\ldots) \,,
\label{Hreconstruct}
\end{equation}
where ``$\ldots$'' indicates an arbitrary configuration of partons.

As a byproduct, it is trivial to obtain the amplitudes for a
pseudoscalar $A$ plus partons as well, in the limit where the
coupling is described by the effective Lagrangian~(\ref{effinta}),
\begin{equation}
{\cal A}_n^\lloop(A,\ldots) =
{1\over i} \left[ {\cal A}_n^\lloop(\phi,\ldots)
-  {\cal A}_n^\lloop(\phi^\dagger,\ldots) \right] \,.
\label{Areconstruct}
\end{equation}
When using \eqn{Areconstruct} one should keep in mind that the value
of the normalization factor $C$ in the pseudoscalar case differs
from that in the scalar case.  For example, if the pseudoscalar
state arises from a two-Higgs doublet model, and the only surviving
contribution is from the top quark, with $m_t$ taken to be large,
then the leading-order value is $C = \as \cot\beta/(4\pi v)$,
where $\tan\beta = v_2/v_1$ is the ratio of Higgs vacuum
expectation values.   In contrast to the scalar case, this
coupling does not get renormalized by QCD corrections~\cite{ABth,NAABth},
although at order $\as^2$ the pseudoscalar $A$ begins to
couple to the divergence of the light-quark axial current~\cite{CKSB}.
As we shall see in \sect{AxialCurrentDivSubsection}, the amplitudes
produced by this operator vanish in the limit that the masses of the
light quarks go to zero. So there is no contribution to the
cross section at NLO, that is, at order $C\, \as \sim \as^2$.

The amplitudes with $\phi$ are related to those with $\phi^\dagger$
by parity,
\begin{equation}
{\cal A}_n^\lloop(\phi^\dagger,1^{h_1},2^{h_2},\ldots,n^{h_n})
=
\Bigl[ {\cal A}_n^\lloop(\phi,1^{-h_1},2^{-h_2},\ldots,n^{-h_n})
\Bigr] \Bigr|_{ \spa{i}.{j} \lr \spb{j}.{i} } \,.
\label{ParityExch}
\end{equation}
That is, to go from an amplitude with $\phi$
to an amplitude with $\phi^\dagger$ one needs to reverse the helicities
of all gluons, and replace $\spa{i}.{j}$ with $\spb{j}.{i}$.
It is therefore sufficient to compute only the $\phi$ amplitudes,
and get the $\phi^\dagger$ amplitudes by parity.
Note that reconstruction of the scalar $H$ and pseudoscalar $A$
helicity amplitudes~(\ref{Hreconstruct}) and (\ref{Areconstruct})
from $\phi$ amplitudes, with the aid of~\eqn{ParityExch}, 
requires {\it pairs} of $\phi$ amplitudes with reversed parton helicities. 

\subsection{Tree-level color decompositions}
\label{ColorSubsection}

We are interested in calculating tree-level and one-loop amplitudes
for a single color-neutral scalar field, $\phi$,
plus either $n$ gluons or two quarks and $(n-2)$-gluons.
Because $\phi$ is color neutral, the color organization of these
amplitudes is identical to that of the corresponding amplitudes
in pure QCD. We now review the standard, trace-based color
decompositions~\cite 
{TreeColor,BGSixMPX,TreeReview,BKColor,TwoQuarkThreeGluon}
for QCD, but add the argument $\phi$ to all the amplitudes.
By ``QCD'', we actually mean a slight generalization:
SU$(N_c)$ gauge theory with $n_\f$ massless quarks
(fermions in the fundamental $N_c\,+\,\overline{N}_c$ representation)
and $n_s$ massless squarks (complex scalars in the
$N_c\,+\,\overline{N}_c$ representation).

In general, the coefficients of the various color-trace structures,
called {\it partial\/} amplitudes,
are built out of several {\it primitive\/}
amplitudes~\cite{TwoQuarkThreeGluon}.  Primitive amplitudes are
color-ordered building blocks, which are functions only of the
kinematic variables.  We will write recursion relations for
the different primitive amplitudes.

For tree amplitudes containing the field $\phi$ and $n$ external gluons,
the color decomposition is~\cite{TreeColor,BGSixMPX,TreeReview,DGK},
\begin{eqnarray}
&& {\cal A}_n^{\tree}(\phi,1^{h_1},2^{h_2},\ldots,n^{h_n})
\nonumber\\
&&\hskip1cm =
C \, g^{n-2} \sum_{\sigma\in S_n/Z_n}
   \Tr(T^{a_{\sigma(1)}} T^{a_{\sigma(2)}}\ldots T^{a_{\sigma(n)}})\
    A_n^\tree(\phi,\sigma(1^{h_1},2^{h_2},\ldots,n^{h_n}))\,,~~~~
\label{GluonTreeColorDecomposition}
\end{eqnarray}
where $S_n$ is the full permutation group on $n$ elements,
$Z_n$ is the cyclic subgroup preserving the trace, and
$j^{h_j}$ denotes the $j$-th (outgoing) momentum $k_j$ and helicity  
$h_j$.
The $T^a$ are fundamental representation SU$(N_c)$ color matrices
normalized so that $\Tr(T^a T^b) = \delta^{ab}$.
We only need to compute the partial amplitudes
$A_n^\tree(\phi,1^{h_1},2^{h_2},\ldots,n^{h_n})$
having the standard cyclic ordering; the
remaining quantities entering \eqn{GluonTreeColorDecomposition}
are obtained by applying the permutations $\sigma$ to the momentum
labels.
(Whenever the permutation $\si$ acts on a list of indices,
it is understood to be applied to each index separately:
$\si(3,\ldots,n) \equiv \si(3),\ldots,\si(n)$, {\it etc.})
The normalization factor $C$ is to be set to $\as/(6 \pi v)$
for the SM Higgs boson in the large-$m_t$ limit,
and to $1$ in the pure-QCD case.

For tree amplitudes containing a $\phi$ field,
a quark pair in the fundamental representation,
and $(n-2)$ external gluons, the color decomposition
is~\cite{TreeReview},
\begin{equation}
{\cal A}_n^{\tree}(\phi,1_{\bar{q}},2_q,3,\ldots,n)
\ =\  C \, g^{n-2} \sum_{\sigma\in S_{n-2}}
    (T^{a_{\sigma(3)}}\ldots T^{a_{\sigma(n)}})_{i_2}^{~\ib_1}\
     A_n^\tree(\phi,1_{\bar{q}},2_q;\sigma(3,\ldots,n))\,.
\label{TreeColorDecomposition}
\end{equation}
Here we have suppressed the helicity labels, and $S_{n-2}$ is the
permutation group associated with the $n-2$ gluons.
Because the color indices have been removed,
there is no need to distinguish a
quark leg $q$ from an anti-quark leg $\bar{q}$
in the partial amplitudes;
charge conjugation relates the two choices.
Helicity conservation implies that the helicities of the fermionic
legs 1 and 2 are opposite.  We take the helicity of leg 1 to be negative,
and that of leg 2 to be positive.  The other case is obtained by
using a reflection symmetry, 
\be
A_n^\tree(\phi,1_\f^+,2_f^-,3,4,\ldots,n)
 = (-1)^n A_n^\tree(\phi,2_f^-,1_\f^+,n,n-1,\ldots,3)\,,
\label{TreeReflection}
\ee
and then relabelling the external legs.

\subsection{One-loop color decompositions}
\label{OneloopColorSubsection}

At one loop, the color decomposition for a $\phi$ field plus
$n$ external gluons contains double traces as well as single
traces~\cite{BKColor},
\begin{equation}
{\cal A}_n^{\oneloop}(\phi,1,2,\ldots,n) = C \, g^n \, \cg
\sum_{c=1}^{\lfloor{n/2}\rfloor+1}
       \sum_{\sigma \in S_n/S_{n;c}}
      \Gr_{n;c}( \sigma ) \, A_{n;c}(\phi,\sigma(1,2,\ldots,n)) \,,
\label{AdjointColorDecomposition}
\end{equation}
where ${\lfloor{x}\rfloor}$ is the largest integer less than or equal
to $x$.  The leading-color structure,
\begin{equation}
\Gr_{n;1}(1) = N_c\ \Tr (T^{a_1}\cdots T^{a_n} ) \,,
\end{equation}
is $N_c$ times the tree color structure.  The subleading-color
structures are given by
\begin{equation}
\Gr_{n;c}(1) = \Tr( T^{a_1}\cdots T^{a_{c-1}} )\,
\Tr (T^{a_c}\cdots T^{a_n}) \,.
\end{equation}
In \eqn{AdjointColorDecomposition},
$S_{n;c}$ is the subgroup of $S_n$ that
leaves $\Gr_{n;c}$ invariant.
We have extracted a loop factor,
relative to ref.~\cite{BKColor}, of
\begin{equation}
\cg \equiv  {1\over(4\pi)^{2-\eps}}
   {\Gamma(1+\eps)\Gamma^2(1-\eps)\over\Gamma(1-2\eps)} =
   {1\over(4\pi)^2} + \Ord(\e) \,.
\label{cgdefn}
\end{equation}
For the finite helicity amplitudes, which form the main subject of
this paper, $\cg$ may be set to $1/(4\pi)^2$.
The $A_{n;1}$ are the ``leading-color'' partial amplitudes,
while the $A_{n;c}$ for $c>1$ are subleading color,
because for large $N_c$, $A_{n;1}$ alone gives the
leading contribution to the color-summed correction to the cross  
section,
obtained by interfering ${\cal A}_{n}^{\tree}$ with
${\cal A}_{n}^{\oneloop}$.
The $\Ord(\as)$ corrections to the operator coefficient
$C$~\cite{MatchingRef} (see \app{NormalizationAppendix})
can also be included in~\eqn{AdjointColorDecomposition}.

For general helicity configurations, $A_{n;1}$ has the following
dependence on the numbers of fermions and scalars, $n_\f$ and $n_s$,
\begin{equation}
A_{n;1} = A_n^{[g]} + {n_\f \over N_c} A_n^{[f]}
+ {n_s \over N_c} A_n^{[s]} \,,
\label{OneloopnfnsDecomp}
\end{equation}
where $A_n^{[g,f,s]}$ gives the contribution of a (gluon, fermion,  
scalar)
in the loop~\cite{BKColor}.  For the finite helicity configurations,
supersymmetry Ward identities~\cite{SWI}
--- valid for the pure-QCD case only --- imply that
$A_n^{[g]} = - A_n^{[f]} = A_n^{[s]}$, or in other words that,
\begin{equation}
A_{n;1}(1^\pm,2^+,\ldots,n^+) =
\biggl( 1 - {n_\f \over N_c} + {n_s \over N_c} \biggr)
   A_n^{[s]}(1^\pm,2^+,\ldots,n^+)\,.
\label{OneloopFiniteDecomp}
\end{equation}
For the finite helicity amplitudes containing a single $\phi$ field
and $n$ gluons, we shall see in \sect{AllPlusAmplitudesSection} that
the fermion and scalar loop still have equal and opposite contributions,
$A_n^{[f]} = - A_n^{[s]}$.  However, the gluonic loop differs,
so the equation analogous to \eqn{OneloopFiniteDecomp} is,
\begin{eqnarray}
A_{n;1}(\phi,1^\pm,2^+,\ldots,n^+)
&=&  A_n^{[g-s]}(\phi,1^\pm,2^+,\ldots,n^+)
\nonumber\\
&&\hskip0cm
+ \biggl( 1 - {n_\f \over N_c} + {n_s \over N_c} \biggr)
   A_n^{[s]}(\phi,1^\pm,2^+,\ldots,n^+)\,,
\label{OneloopPhiFiniteDecomp}
\end{eqnarray}
where
\begin{equation}
A_n^{[g-s]} \equiv A_n^{[g]} - A_n^{[s]} \,.
\label{Angminussdef}
\end{equation}
We emphasize that the relation
$A_n^{[f]}(\phi,\ldots) = - A_n^{[s]}(\phi,\ldots)$,
{\it i.e.} the decomposition used in \eqn{OneloopPhiFiniteDecomp}, 
is only valid for the finite helicity amplitudes.

The fermion and scalar loop contributions, proportional
to $n_\f$ and $n_s$ respectively, only enter the leading-color partial
amplitudes $A_{n;1}$.  The subleading-color contributions
to the one-loop $n$-gluon amplitudes, $A_{n;c}$ for $c>2$,
come just from purely-gluonic graphs.  This result holds
in pure QCD, and for amplitudes containing a $\phi$ field.
The subleading-color terms are given by sums over permutations 
of the gluonic contributions to the leading-color terms~\cite{Neq4Oneloop}.
The formula is,
\begin{equation}
A_{n;c}(\phi,1,2,\ldots,c-1;c,c+1,\ldots,n)\ =\
(-1)^{c-1} \sum_{\sigma\in COP\{\alpha\}\{\beta\}}
A_{n}^{[g]}(\phi,\sigma(1,2,\ldots,n)) \,,
\label{sublanswer}
\end{equation}
where $\alpha_i \in \{\alpha\} \equiv \{c-1,c-2,\ldots,2,1\}$,
$\beta_i \in \{\beta\} \equiv \{c,c+1,\ldots,n-1,n\}$,
and $COP\{\alpha\}\{\beta\}$ is the set of all
permutations of $\{1,2,\ldots,n\}$ with $n$ held fixed
that preserve the cyclic ordering of the
$\alpha_i$ within $\{\alpha\}$,
and of the $\beta_i$ within $\{\beta\}$,
while allowing for all possible relative orderings
of the $\alpha_i$ with respect to the $\beta_i$.
\Eqn{sublanswer} was established in the pure-QCD case,
but the arguments rely just on the property that the
color structures of the one-loop graphs involve only structure
constants $f^{abc}$.  This property still holds when the vertices
from the interaction
$\phi \tr G_{{\sst SD}\,\mu\nu} G_{\sst SD}^{\mu\nu}$
are included. 
For another version of the argument, which uses
$f^{abc}$-based color structures instead of trace-based ones, and in
addition avoids the introduction of the subleading-color pieces $A_ 
{n;c}$,
see ref.~\cite{DDDM}.

The color decomposition at one loop, for amplitudes containing
a $\phi$ field, a pair of fundamental
representation quarks, and $(n-2)$ gluons,
is~\cite{TwoQuarkThreeGluon},
\begin{equation}
{\cal A}_{n}^{\oneloop}(\phi,1_{\bar{q}},2_q,3,\ldots,n)
\ =\  C \, g^n \, \cg
   \sum_{j=1}^{n-1} \sum_{\sigma\in S_{n-2}/S_{n;j}}
    \!\! \Gr_{n;j}^{(\bar{q}q)}(\sigma(3,\ldots,n))\
   A_{n;j}(\phi,1_{\bar{q}},2_q;\sigma(3,\ldots,n))\ ,
\label{OneLoopColorDecomposition}
\end{equation}
where we have again extracted an extra factor of $\cg$,
and the color structures $\Gr_{n;j}^{(\bar{q}q)}$ are defined by,
\begin{eqnarray}
\Gr_{n;1}^{(\bar{q}q)}(3,\ldots,n)
   \ &=& N_c\ (T^{a_3}\cdots T^{a_n})_{i_2}^{~\ib_1}\,,\nonumber\\
\Gr_{n;2}^{(\bar{q}q)}(3;4,\ldots,n)
   \ &=& 0\ ,\nonumber\\
\Gr_{n;j}^{(\bar{q}q)}(3,\ldots,j+1;j+2,\ldots,n)
\ &=& \Tr(T^{a_3}\cdots T^{a_{j+1}})\ \
    (T^{a_{j+2}}\cdots T^{a_n})_{i_2}^{~\ib_1}\,,
   \quad j=3,\ldots,n-2, \nonumber\\
\Gr_{n;n-1}^{(\bar{q}q)}(3,\ldots,n)\ &=& \Tr(T^{a_3}\cdots T^{a_n})\ \
      \delta_{i_2}^{~\ib_1} \,. \label{ColorStructures}
\end{eqnarray}
As in the $n$-gluon case, the $A_{n;1}$ are leading-color partial  
amplitudes,
and the $A_{n;j}$ for $j>1$ are subleading color.

%
\begin{figure}[t]
\centerline{\epsfxsize 3.0 truein \epsfbox{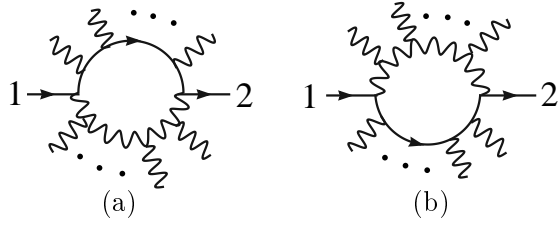}}
\caption{(a) An $L$ type primitive amplitude, in which the fermion line 
turns left on entering the loop (following the arrow).
(b) In an $R$ type primitive amplitude, the fermion line turns right.}
\label{LeftRightAFigure}
\end{figure}

%
\begin{figure}[t]
\centerline{\epsfxsize 3.0 truein \epsfbox{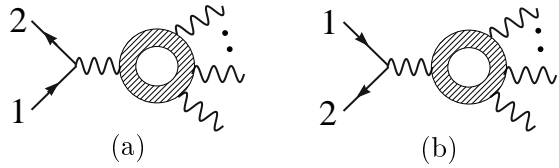}}
\caption{(a) Graphs in which the external fermion line passes to 
the left of the loop are assigned to $L$ type.  
(b) Graphs in which it passes to the right are called $R$ type.
Gluons, fermions or scalars run in the loop.  The same decomposition
is used when gluons are emitted from the external fermion line.}
\label{LeftRightBFigure}
\end{figure}

Following ref.~\cite{TwoQuarkThreeGluon}, 
where more details can be found, 
we introduce the primitive amplitudes,
\begin{equation}
\matrix{
    & A_n^{L,[J]}(\phi,1_\f,3,4,\ldots,j,2_\f,j+1,\ldots,n),  \hfill\cr
    & A_n^{R,[J]}(\phi,1_\f,3,4,\ldots,j,2_\f,j+1,\ldots,n),  \hfill\cr}
    \hskip 1 cm  J=1,\coeff{1}{2},0,
\label{PrimitiveAmplitudes}
\end{equation}
where $J={1\over2}$ and $J=0$ denote the contributions with a closed
fermion loop and closed complex scalar loop, respectively,
and ``$L$'' and ``$R$'' denote the two orientations of the fermion line
in the loop.  Generic diagrams contributing to the $L$ and $R$ terms
are shown in \figs{LeftRightAFigure}{LeftRightBFigure}.
Because the primitive amplitudes can be used to build 
amplitudes with any color representation for the fermions, we label 
them by $f$ to denote a generic fermion in any color representation.
Diagrams without closed fermion or scalar loops are
assigned to $J=1$; they may or may not contain a closed gluon loop, as
the two types of diagrams mix under gauge transformations.
For notational simplicity, we suppress the superscript ``$[1]$'',
$A_n^{L} \equiv A_n^{L,[1]}\,,\,\, A_n^{R} \equiv A_n^{R,[1]}$.
The primitive amplitudes~(\ref{PrimitiveAmplitudes}) are not all
independent.  The set of diagrams where the incoming leg $1$ turns left
is related, up to a sign, to a corresponding set where it turns
right. Thus, the two sets are related by a reflection which
reverses the cyclic ordering,
\begin{equation}
   A_n^{R,[J]}(\phi,1_\f,3,4,\ldots,2_\f,\ldots,n-1,n) =
   (-1)^n  A_n^{L,[J]}(\phi,1_\f,n,n-1,\ldots,2_\f,\ldots,4,3).
\label{FlipSymmetry}
\end{equation}

The leading-color contribution to \eqn{OneLoopColorDecomposition},
$A_{n;1}$, is given in terms of primitive amplitudes by,
\begin{eqnarray}
A_{n;1}(\phi,1_{\bar{q}},2_q;3,\ldots,n) &= &
    A_n^L(\phi,1_\f,2_\f,3,\ldots,n)
   - {1\over N_c^2} A_n^R(\phi,1_\f,2_\f,3,\ldots,n) \nonumber\\
&+& {\nf\over N_c} A_n^{L,[1/2]}(\phi,1_\f,2_\f,3,\ldots,n)
   + {\ns\over N_c} A_n^{L,[0]}(\phi,1_\f,2_\f,3,\ldots,n).~~~~~~~
\label{Anoneformula}
\end{eqnarray}
For QCD the number of scalars vanishes, $\ns = 0$, while $\nf$ is the  
number of light quark flavors.  The subleading-color
partial amplitudes $A_{n;j>1}$ appearing in 
\eqn{OneLoopColorDecomposition}
may be expressed as a permutation sum over primitive amplitudes,
\begin{eqnarray}
&& \hskip -3 cm A_{n;j}(\phi,1_\qb,2_q; 3,\ldots,j+1;j+2,j+3,\ldots,n)
\nonumber\\ \hskip 2 cm   & = &
(-1)^{j-1} \sum_{\sigma\in COP\{\alpha\}\{\beta\}}
     \Biggl[ A_n^{L,[1]} (\phi,\sigma(1_\f,2_\f,3,\ldots,n) )  
\nonumber \\
&&   - {\nf\over N_c} A_n^{R,[1/2]} (\phi,\sigma(1_\f,2_\f,3,\ldots,n) )
    - {\ns\over N_c} A_n^{R,[0]} (\phi,\sigma(1_\f,2_\f,3,\ldots,n) )
      \Biggr]\,,~~~~~~
\label{subltotal}
\end{eqnarray}
where $\alpha_i \in \{\alpha\} \equiv \{j+1,j,\ldots,4,3\}$,
$\beta_i \in \{\beta\} \equiv \{1,2,j+2,j+3,\ldots,n-1,n\}$,
and  $COP\{\alpha\}\{\beta\}$ is the set of all
permutations of $\{1,2,\ldots,n\}$ defined after \eqn{sublanswer},
except that here leg $1$ is held fixed. However, as
in the $n$-gluon case above, 
using $f^{abc}$-based color structures~\cite{DDDM}
instead of trace-based ones, it is
possible to avoid the introduction of the subleading-color
amplitudes $A_{n;j>1}$.

In the case studied below, where all external gluons carry
the same helicity, the fermion loop and scalar loop are the same up  
to a sign,
\begin{eqnarray}
A_n^{L,[0]}(\phi,1_\f^-,2^+,\ldots, j_\f^+, \ldots,n^+)
& = &- A_n^{L,[1/2]}(\phi,1_\f^-,2^+,\ldots, j_\f^+, \ldots,n^+)
   \nonumber \\
& \equiv & A_n^{s}(\phi,1_\f^-,2^+,\ldots, j_\f^+, \ldots,n^+) \,.
\label{AsDef}
\end{eqnarray}
As mentioned above, in contrast to pure QCD, this relation does 
not follow from supersymmetry; instead we will show 
it recursively below.
Then by computing the closed scalar-loop primitive amplitude
we obtain also the closed fermion-loop primitive amplitude.

We shall find it convenient to compute the combinations
\begin{equation}
A_n^s \hskip 1cm \hbox{and} \hskip 1cm  A_n^{L-s} \equiv A_n^L- 
A_n^s\,,
\label{LminusSDef}
\end{equation}
instead of $A_n^s$ and $A_n^L$.
Also in contrast to pure QCD, for which an identity 
relates $A_n^{L-s}(1_\f^-,2^+,\ldots,j_\f^+,\ldots,n^+)$
with the amplitude with the cyclic ordering reversed,
$A_n^{L-s}(1_\f^-,n^+,\ldots,j_\f^+,\ldots,2^+)$~\cite{LastFinite},
here we will have to compute all $A_n^{L-s}$ for $j$ from 2 up to $n$.
The generic recursion relation will only hold for $j\leq n-2$.
However, we will be able to compute 
the cases of $A_n^{L-s}$ for $j=n$ or $j=n-1$
by taking collinear limits of appropriate amplitudes with larger $n$.
The scalar-loop contribution $A_n^s$ vanishes if $j=n$ or $j=n-1$ ---
these configurations have only graphs with tadpoles and bubbles
on external lines, which vanish in dimensional regularization.

In summary, for amplitudes with a single quark pair
and identical-helicity gluon legs, there are two independent classes
of primitive amplitudes to compute,
\bea
A_n^{L-s}(\phi,1_\f^-,2^+,\ldots, j_\f^+,(j+1)^+, \ldots,n^+)
\hskip 0.5cm &\hbox{and}& \hskip 0.5cm
A_n^{s}(\phi,1_\f^-,2^+,\ldots, j_\f^+,(j+1)^+, \ldots,n^+) \,.
\nonumber
\eea
%

\subsection{Spinor-helicity notation}
\label{SpinorNotationSubsection}

The primitive amplitudes are functions of the massive momentum
$k_\phi$ of the $\phi$ particle, and the massless momenta $k_i$ of the
$n$ partons.  All momenta entering the amplitudes are taken to
be outgoing, so the kinematical constraints are,
\begin{eqnarray}
k_\phi + \sum_{i=1}^n k_i &=& 0,
\label{MomConsphi}\\
k_\phi^2 &=& m_H^2,
\qquad
k_i^2 = 0.
\label{OnShellConstraint}
\end{eqnarray}
The amplitudes are best described using
spinor inner-products~\cite{SpinorHelicity,TreeReview}
composed of the partonic momenta,
\begin{equation}
\spa{j}.{l} = \langle j^- | l^+ \rangle = \bar{u}_-(k_j) u_+(k_l)\,,
\hskip 2 cm
\spb{j}.{l} = \langle j^+ | l^- \rangle = \bar{u}_+(k_j) u_-(k_l)\, ,
\label{spinorproddef}
\end{equation}
where $u_\pm(k)$ is a massless Weyl spinor with momentum $k$ and positive
or negative chirality. We use the convention
$\spb{j}.{l} = \sign(k_j^0 k_l^0)\spa{l}.{j}^*$
standard in the QCD literature, so that,
\begin{equation}
\spa{i}.{j} \spb{j}.{i} = 2 k_i \cdot k_j \equiv s_{ij}\,.
\end{equation}

\def\vmu{{\vphantom{\mu}}}
We denote the sums of cyclicly-consecutive external momenta by
\begin{equation}
K^\mu_{i\cdots j} \equiv
    k_i^\mu + k_{i+1}^\mu + \cdots + k_{j-1}^\mu + k_j^\mu,
\label{KDef}
\end{equation}
where all indices are mod $n$ for an $n$-parton amplitude.
The invariant mass of this vector is
$s_{i\cdots j} \equiv K_{i\cdots j}^2$.
Special cases include the two- and three-particle invariant masses,
which are denoted by
\begin{equation}
s_{ij} \equiv K_{i,j}^2
\equiv (k_i+k_j)^2 = 2k_i\cdot k_j,
\qquad \quad
s_{ijk} \equiv (k_i+k_j+k_k)^2.
\label{TwoThreeMassInvariants}
\end{equation}
In color-ordered amplitudes only invariants with cyclicly-consecutive
arguments need appear, {\it e.g.}{} $s_{i,i+1}$ and $s_{i,i+1,i+2}$.
It is convenient to introduce the same combinations as~\eqn{KDef}
but with the $\phi$ momentum added as well,
\begin{equation}
K^\mu_{\phi,i\cdots j} \equiv
    k_\phi^\mu + k_i^\mu + k_{i+1}^\mu + \cdots + k_{j-1}^\mu + k_j^\mu,
\label{KphiDef}
\end{equation}
with invariant-mass $s_{\phi,i\cdots j} \equiv K_{\phi,i\cdots j}^2$.
Longer spinor strings will also appear, such as
\begin{eqnarray}
\spab{k}.{\Ksl_{i\cdots j}}.{l} &=& \sum_{m=i}^j \spa{k}.{m}\spb{m}.{l}
\,,
\label{spabstrings} \\
\sandmp{k}.{\Ksl_{i_1\cdots j_1} \Ksl_{i_2\cdots j_2}}.{l}
&=& \sum_{m_1=i_1}^{j_1} \sum_{m_2=i_2}^{j_2}
      \spa{k}.{m_1} \spb{m_1}.{m_2} \spa{m_2}.{l}
   \,.
\label{spaastrings}
\end{eqnarray}
For small values of $n$, such strings can be written out explicitly as
\begin{eqnarray}
\spab{k}.{(a+b)}.{l} &=& \spa{k}.{a} \spb{a}.{l} + \spa{k}.{b} \spb 
{b}.{l} \,,
           \nonumber   \\
\spba{k}.{(a+b)}.{l} &=& \spb{k}.{a} \spa{a}.{l} + \spb{k}.{b} \spa 
{b}.{l} \,,
           \nonumber   \\
\spaa{k}.{(a+b)}.{(c+d)}.{l} &=&
      \spa{k}.{a} \spba{a}.{(c+d)}.{l}
    + \spa{k}.{b} \spba{b}.{(c+d)}.{l} \,.
\end{eqnarray}
%


\section{Review of recursion relations and factorization}
\label{RecursionReviewSection}

\subsection{On-shell recursion relations}

Here we briefly review the on-shell recursion
relations found and proved in refs.~\cite{BCFRecursion,BCFW}.
For further details we refer to these papers.
The on-shell recursion relations rely on general properties
of complex functions as well as factorization properties of
scattering amplitudes.
The proof~\cite{BCFW} of the relations relies on a
parameter-dependent ``$[k,l\rangle$''
shift of two of the external massless spinors, $k$ and $l$,
in an $n$-point process,
\be
[k,l\rangle: 
\quad \quad \tlambda_k \rightarrow \tlambda_k - z\tlambda_l \,, 
\qquad \quad \lambda_l \rightarrow \lambda_l + z\lambda_k \,,
\label{SpinorShift}
\ee
where $z$ is a complex number.  The corresponding momenta
(labeled by $p_i$ instead of $k_i$ in this section) are shifted
as well,
\begin{eqnarray}
&p_k^\mu &\rightarrow\ p_k^\mu(z) = p_k^\mu -
       {z\over2}{\sand{k}.{\gamma^\mu}.{l}},\nonumber\\
&p_l^\mu &\rightarrow\ p_l^\mu(z) = p_l^\mu +
       {z\over2}{\sand{k}.{\gamma^\mu}.{l}} \,,
\label{MomentumShift}
\end{eqnarray}
so that they remain massless, $p_k^2(z) = 0 = p_l^2(z)$,
and overall momentum conservation is maintained.

An on-shell amplitude containing the momenta $p_k$ and $p_l$
then becomes parameter-dependent as well,
\begin{equation}
A(z) = A(p_1,\ldots,p_k(z),p_{k+1},\ldots,p_l(z),p_{l+1},\ldots,p_n).
\end{equation}
When $A$ is a tree amplitude or finite one-loop
amplitude, $A(z)$ is a rational function of $z$.
The physical amplitude is given by $A(0)$.

If $A(z)\rightarrow 0$ as $z\rightarrow\infty$, as in 
suitable cases in tree-level 
QCD~\cite{BCFRecursion,BCFW,LuoWen,BadgerMassive},
then the contour integral around the circle $C$ at infinity vanishes,
\begin{equation}
{1\over 2\pi i} \oint_C {dz\over z}\,A(z)  = 0\,;
\label{ContourInt}
\end{equation}
that is, there is no ``surface term''.
In appendix~\ref{TreeProofAppendix} we show that a tree amplitude
$A(z)$ containing an extra $\phi$ field also vanishes as 
$z \rightarrow \infty$, for the same choices of shift that lead
to a vanishing $A(z)$ in pure QCD.

Evaluating the integral~(\ref{ContourInt}) as a sum
of residues, we can then solve for $A(0)$ to obtain,
\begin{equation}
A(0) = -\sum_{{\rm poles}\ \alpha} \Res_{z=z_\alpha}  {A(z)\over z}\,.
\label{NoSurface}
\end{equation}
If $A(z)$ only has simple poles, then each residue is given by 
factorizing the shifted amplitude on the appropriate pole in 
momentum invariants~\cite{BCFW}, so that at tree level,
\begin{equation}
A(0) = \sum_{r,s,h}
    A^h_L(z = z_{rs}) { i \over K_{r\cdots s}^2 } A^{-h}_R(z = z_ 
{rs})  \,,
\label{BCFWRepresentation}
\end{equation}
where $h=\pm1$ labels the helicity of the intermediate state.
There is generically a double sum over momentum poles, 
labeled by leg indices $r,s$, with legs $k$ and $l$ always 
appearing on opposite sides of the pole.  By definition,
the leg $l$ is contained in the set $\{r,\ldots,s\}$
and the leg $k$ is not. The squared momentum 
associated with the pole, $K_{r\cdots s}^2$, is evaluated 
in the unshifted kinematics;
whereas the on-shell amplitudes $A_L$ and $A_R$ are evaluated in
kinematics that have been shifted by \eqn{SpinorShift} with
$z=z_{rs}$, where
\begin{equation}
z_{rs} = - {K_{r\cdots s}^2 \over \sand{k}.{\Ksl_{r\ldots s}}.{l} } \,.
\end{equation}
To extend the approach to one loop~\cite{OnShellRecursionI},
the sum (\ref{BCFWRepresentation})
should also be taken over the two ways of assigning the loop to $A_L$
and $A_R$.  This formula assumes that there are no additional poles
present in the amplitude other than the standard poles for real
momenta.  At tree level it is possible to demonstrate the absence of
additional poles, but at loop level it is not true.

Because of the general structure of multiparticle
factorization~\cite{BernChalmers}, only standard single poles in $z$
arise from multiparticle channels, even at one loop.  However, 
double poles in $z$ do arise at one loop due to collinear
factorization~\cite{OnShellRecursionI,BBDFK1}.  The splitting
amplitudes with helicity configuration $(+{}+{}+)$ and $(-{}-{}-)$ (in
an all-outgoing helicity convention) can
lead to double poles in $z$, because their dependence on the spinor
products takes the form $\spb{a}.{b}/\spa{a}.{b}^2$ for $(+{}+{}+)$,
or its complex conjugate $\spa{a}.{b}/\spb{a}.{b}^2$ for
$(-{}-{}-)$~\cite{Neq4Oneloop}.  As discussed in
ref.~\cite{OnShellRecursionI}, this behavior alters the form of the
recursion relation in an essential manner.  In general, underneath
the double pole sits an object of the form,
\begin{equation}
{\spb{a}.b\over\spa{a}.b}\,,
\label{UnrealPole}
\end{equation}
which we call an ``unreal pole'' because there is no pole present when
real momenta are used; it only appears, as a single pole, when we
continue to complex momenta.  As we shall discuss in
\sect{AmplitudesSection}, the finite $\phi\bar{q}qgg\ldots g$ 
amplitudes exhibit similar phenomena, except that in this case, 
just as in the pure-QCD case of finite $\bar{q}qgg\ldots g$
amplitudes~\cite{LastFinite}, one encounters
unreal poles that do not sit underneath a double pole.

\subsection{One-loop factorization properties}
\label{OneLoopFact}

In order to build on-shell recursion relations, we need the
factorization properties of one-loop amplitudes for complex momenta.
It is useful to first review the factorization properties for real
momenta, which we know from general
arguments~\cite{TreeReview,BernChalmers,OneloopSplit}.

As the real momenta of two color-adjacent external partons 
$a$ and $b$ become collinear, the one-loop amplitude factorizes as,
\begin{eqnarray}
A_{n}^{\oneloop} \inlimit^{a \parallel b}\
\sum_{h=\pm}  && \null \hskip -.35 cm \biggl(
\Split^\tree_{-h}   (a^{h_a},b^{h_b};z)\,
          A_{n-1}^{\oneloop}(\ldots,(a+b)^h,\ldots) \nonumber\\
&&\hphantom{\biggl()}
+ \Split^{\oneloop}_{-h}(a^{h_a},b^{h_b};z)\,
          A_{n-1}^\tree(\ldots,(a+b)^h,\ldots) \biggr)\,.
\end{eqnarray}
The tree and loop splitting amplitudes $\Split^\tree$ and
$\Split^{\oneloop}$ are given in ref.~\cite{Neq4Oneloop}.  
In general, there are three types of limits: a pair of gluons becoming
collinear; a gluon becoming collinear with a quark (or anti-quark); and a
quark and anti-quark becoming collinear (if they happen to be adjacent).
When all external gluons have positive helicity, as here, the limits 
simplify.  For $\phi$-amplitudes, no singularity results when a parton
becomes collinear with the $\phi$ state, because the $\phi$ is massive.

When two gluons become collinear, the finite $\phi\bar{q}qgg\ldots g$
amplitudes $A_n^L$ and $A_n^s$ behave respectively as,
\begin{eqnarray}
A_{n}^{L}(\phi,\ldots,a^+,b^+,\ldots) \inlimit^{a \parallel b}\
&&
\Split^\tree_{-}   (a^{+},b^{+};z)\,
          A_{n-1}^{L}(\phi,\ldots,(a+b)^+,\ldots) \nonumber\\
&& + \Split^{\oneloop:g}_{+}(a^{+},b^{+};z)\,
          A_{n-1}^\tree(\phi,\ldots,(a+b)^-,\ldots) \,,
\label{TwoGluonLimitL}\\
A_{n}^{s}(\phi,\ldots,a^+,b^+,\ldots) \inlimit^{a \parallel b}\
&&
\Split^\tree_{-}   (a^{+},b^{+};z)\,
          A_{n-1}^{s}(\phi,\ldots,(a+b)^+,\ldots) \nonumber\\
&& + \Split^{\oneloop:s}_{+}(a^{+},b^{+};z)\,
          A_{n-1}^\tree(\phi,\ldots,(a+b)^-,\ldots) \,.
\label{TwoGluonLimits}
\end{eqnarray}
Here $\Split^{\oneloop:g}$ ($\Split^{\oneloop:s}$) 
is the one-loop splitting amplitude with a gluon (scalar)
circulating in the loop.  The remaining two
terms, with opposite intermediate-gluon helicity, vanish because
$\Split^\tree_{+}(a^{+},b^{+};z)$ and
$A_{n-1}^\tree(\phi,\pm,+,+,\ldots,+)$ are zero.  
Because the two one-loop splitting amplitudes are equal~\cite{Neq4Oneloop},
\begin{equation}
\Split^{\oneloop:g}_{+}(a^{+},b^{+};z) =
\Split^{\oneloop:s}_{+}(a^{+},b^{+};z),
\end{equation}
we are motivated to take the difference between $A_n^L$ and $A_n^s$
to form $A_{n}^{L-s}$.  This combination has the simpler collinear 
behavior,
\begin{equation}
A_{n}^{L-s}(\phi,\ldots,a^+,b^+,\ldots) \inlimit^{a \parallel b}\
\Split^\tree_{-}   (a^{+},b^{+};z)\,
          A_{n-1}^{L-s}(\phi,\ldots,(a+b)^+,\ldots) \,.
\label{DeltaLimitI}
\end{equation}

When a gluon becomes collinear with any fermion, 
both $A_n^L$ and $A_n^s$ behave as (using fermion helicity-conservation
and \eqn{phiffvanishtree}),
\begin{equation}
A_{n}^{L,s}(\phi,a_\f^\pm,b^+,\ldots) \inlimit^{a \parallel b}
\Split^\tree_{(f)\mp}(a_\f^{\pm},b^{+};z)\,
          A_{n-1}^{L,s}(\phi,(a+b)_\f^\pm,\ldots) \,.
\end{equation}
Hence the difference $A_{n}^{L-s}$ has the limiting behavior,
\begin{equation}
A_{n}^{L-s}(\phi,a_\f^\pm,b^+,\ldots) \inlimit^{a \parallel b}
\Split^\tree_{(f)\mp}(a_\f^{\pm},b^{+};z)\,
          A_{n-1}^{L-s}(\phi,(a+b)_\f^\pm,\ldots) \,.
\label{DeltaLimitII}
\end{equation}
When a quark and anti-quark become collinear (for $j_\f=2$), 
the amplitude factorizes onto the finite one-loop $\phi gg\ldots g$
amplitudes,
\begin{eqnarray}
A_{n}^{L,s}(\phi,a_\f^-,b_\f^+,\ldots) \inlimit^{a \parallel b}\
&&
\Split^\tree_{-}(a_\f^-,b_\f^+;z)\,
       A_{n-1}^{[g,s]}(\phi,(a+b)^+,\ldots) \nonumber\\
&& \null  + \Split^\tree_{+}(a_\f^-,b_\f^+;z)\,
          A_{n-1}^{[g,s]}(\phi,(a+b)^-,\ldots)\,.
\label{QuarkAntiQuarkLimit}
\end{eqnarray}
As mentioned in \sect{OneloopColorSubsection}, in contrast with
the pure-QCD case~\cite{LastFinite}, the gluon and scalar loop 
contributions to $A_{n-1;1}(\phi,\pm,+,+,\ldots,+)$
are not the same.  For the difference $A_n^{L-s}$ we find 
from \eqn{QuarkAntiQuarkLimit} the collinear behavior,
\bea
A_{n}^{L-s}(\phi,a_\f^-,b_\f^+,\ldots) \inlimit^{a \parallel b}\
&&
\Split^\tree_{-}(a_\f^-,b_\f^+;z)\,
       A_{n-1}^{[g-s]}(\phi,(a+b)^+,\ldots) \nonumber\\
&& \null  + \Split^\tree_{+}(a_\f^-,b_\f^+;z)\,
          A_{n-1}^{[g-s]}(\phi,(a+b)^-,\ldots)\,.
\label{DeltaLimitIII}
\eea

Next consider multiparticle factorization.  In this case, the vanishing of
the tree amplitudes $A_n^\tree(1_\f^-,2^+,\ldots,j_\f^+,\ldots,n^+)$ and
$A_n^\tree(\phi,1_\f^-,2^+,\ldots,j_\f^+,\ldots,n^+)$ means that the
only singularities are on gluon poles.  There are two
possible ways to attach the $\phi$ field,
\begin{eqnarray}
&& \null \hskip -4 mm
A_n^{L,s}(\phi,1_\f^-,2^+,\ldots,j_\f^+,\ldots,m^+,\ldots,n^+)
\inlimit^{K_{1\cdots m}^2\rightarrow 0}
\label{phiMultiParticle}\\ &&\hskip 5mm
A_{m+1}^\tree(\phi,1_\f^-,2^+,\ldots,j_\f^+,\ldots,m^+,-K_{1\ldots m}^-)
{i \over K_{1\ldots m}^2 }
A_{n-m+1}^{\oneloop}(K_{1\ldots m}^+,(m+1)^+,\ldots,n^+) \,\nonumber
\\ &&\hskip 5mm
+\ A_{m+1}^\tree(1_\f^-,2^+,\ldots,j_\f^+,\ldots,m^+,-K_{1\ldots m}^-)
{i \over K_{1\ldots m}^2 }
A_{n-m+1}^{[g,s]}(\phi,K_{1\ldots m}^+,(m+1)^+,\ldots,n^+) \,,
\nonumber
\end{eqnarray}
with $m\ge3$ and $m\ge j_\f$.  For $A_n^{L-s}$,
the first term in~\eqn{phiMultiParticle} does not contribute,
but the second does.

In addition to the factorization properties for real momenta
just described, the appearance of unreal poles of the 
form~(\ref{UnrealPole}) sometimes has to be taken into account.
While such behavior is not understood in general, in the present
case we can use the similarity of our amplitudes to the corresponding
finite pure-QCD amplitudes $\bar{q}qgg\ldots g$~\cite{LastFinite}.
The contribution of unreal-pole terms to the recursion relations
for those amplitudes could be stringently cross checked against various
QCD and QED amplitudes.  The inclusion of a massive $\phi$ field is quite
mild from the point of view of factorization, so it should
not alter the form of the unreal-pole terms in the recursion relations.
In \sect{AmplitudesSection} we will describe these terms in more detail.

\section{Review of known finite amplitudes}
\label{ReviewAmplitudesSection}

In this section, we collect previously-known results for tree
and one-loop finite amplitudes in pure QCD, and for tree amplitudes
containing a single $\phi$ field.  These results feed into the
recursive formul\ae{} for the finite one-loop amplitudes containing
a single $\phi$ field, to be discussed in~\sect{AmplitudesSection}.

\subsection{Pure-QCD amplitudes}
\label{QCDAmplitudesReviewSection}

The $n$-gluon tree amplitudes with zero or one negative-helicity gluon
all vanish,
\begin{equation}
A_n^\tree(1^\pm, 2^+,3^+, \ldots, n^+) = 0 \,,
\label{vanishtree}
\end{equation}
where the omitted labels (``$\ldots$'') always refer to
positive-helicity gluons.
Also vanishing are the tree amplitudes for a pair of massless quarks
and $(n-2)$ gluons all of the same helicity,
\begin{equation}
A_n^\tree(1_\f^-, 2^+, \ldots, (j-1)^+, j_\f^+, (j+1)^+,
\ldots, n^+) = 0 \,,
\label{ffvanishtree}
\end{equation}
where the subscript $f$ denotes a generic fermion, and
the positive-helicity fermion ($j_\f^+$) can be located at an
arbitrary position with respect to the negative-helicity one ($1_\f^-$).
(Note that only the case $j=2$ is required in the pure-QCD version
of the tree-level formula~(\ref{TreeColorDecomposition}).)

The first nonvanishing tree amplitudes are the
MHV amplitudes~\cite{ParkeTaylor,TreeReview},
\begin{equation}
A_n^\tree(1^+, 2^+, \ldots, m_1^-, \ldots, m_2^-, \ldots, n^+) =
i { \spa{m_1}.{m_2}^4 \over \spa1.2 \spa2.3 \cdots \spa{n}.1} \,,
\label{mhvtree}
\end{equation}
and
\begin{equation}
A_n^\tree(1_\f^-, 2^+, \ldots, j_\f^+, \ldots,  m^-, \ldots, n^+) =
i { \spa{1}.{m}^3 \spa{j}.{m} \over \spa1.2 \spa2.3 \cdots \spa{n}.1}  
\,.
\label{ffmhvtree}
\end{equation}

We also need the one-loop pure-gluon amplitudes with
zero or one negative helicity.
The all-positive-helicity case, for $n\ge4$, is given
by~\cite{AllPlus,Mahlon}
\begin{equation}
A_n^{[g]}(1^+, 2^+, \ldots, n^+) =
A_n^{[s]}(1^+, 2^+, \ldots, n^+) =
{i \over 3}
{H_n\over \spa1.2 \spa2.3\cdots \spa{(n-1)}.n\spa{n}.1} \,,
\label{OneLoopAllPlusAmplitude}
\end{equation}
where
\begin{equation}
H_n = -\sum_{1\leq i_1<i_2<i_3<i_4\leq n}
   \Tr_{-}\Bigl[\ksl_{i_1}\ksl_{i_2}\ksl_{i_3}\ksl_{i_4}\Bigr] \,,
\label{Hndef}
\end{equation}
and
\begin{eqnarray}
\Tr_-\Bigl[\ksl_{i_1}\ksl_{i_2}\ksl_{i_3}\ksl_{i_4}\Bigr] &=& {1\over 2}
\Tr[(1-\gamma_5)\ksl_{i_1}\ksl_{i_2}\ksl_{i_3}\ksl_{i_4}] \nonumber\\
&=& \spa{i_1}.{i_2} \spb{i_2}.{i_3} \spa{i_3}.{i_4} \spb{i_4}.{i_1} \,.
\end{eqnarray}
It is convenient to rewrite \eqn{OneLoopAllPlusAmplitude} using
an identity proved in ref.~\cite{LastFinite},
\begin{equation}
H_n = - {1\over2} \Tr_{-} [ {\cal F}(2,n) ]^2
\,,
\label{HtoFF}
\end{equation}
where we define the more general quantity ${\cal F}(l,p)$,
\begin{equation}
{\cal F}(l,p) = \sum_{i=l}^{p-1} \sum_{m=i+1}^{p} \ksl_i \ksl_m \,,
\label{Flpdef}
\end{equation}
involving sums over ordered pairs ($i<m$) of momenta between $l$ and  
$p$.
In \eqn{HtoFF} for the $n$-point all-plus amplitude, all such pairs  
appear
except those involving leg 1.  Using the cyclic symmetry of this
amplitude, any one of the $n$ legs could be omitted, say leg $l$,
if the ordering in the corresponding sums for ${\cal F}(l+1,l-1)$
is defined modulo $n$ in \eqn{Flpdef}.

For $n=3$ we need to take one leg off shell.  We define a vertex, 
with legs $1$ and $2$ the on-shell external legs~\cite{OnShellRecursionI},
\begin{eqnarray}
A_3^\oneloop(1^+, 2^+, 3^+) & = &
   {V_3^\oneloop(1^+, 2^+, 3^+)   \over K_{1,2}^2} \,,
\label{OneLoopThreeAmplitude}
\end{eqnarray}
where
\begin{equation}
V_3^\oneloop(1^+, 2^+, 3^+)  \equiv
- {i \over 3} \spb1.2 \spb2.3 \spb3.1 \,.
\label{OneLoopThreeVertex}
\end{equation}

The remaining one-loop finite helicity amplitudes in pure QCD
are those for $n$ gluons, with one of negative helicity;
and those for a quark-antiquark pair plus $(n-2)$ gluons of positive
helicity.  They are not actually needed for the recursion relations
for the single-$\phi$ amplitudes considered here.  However,
their structure, in the form given in ref.~\cite{LastFinite},
is so similar to that of the single-$\phi$ amplitudes, that we
reproduce them here.  The one-minus $n$-gluon amplitude is given by,
\begin{equation}
A_{n;1}(1^-,2^+,3^+,\ldots,n^+)= {i\over3}
   {T_1 + T_2 \over \spa1.2\spa2.3\cdots \spa{n}.{1} } \,,
\label{oneminus}
\end{equation}
where
\begin{eqnarray}
T_1 &=& \sum_{l=2}^{n-1}
   { \spa{1}.{l} \spa{1}.{(l+1)}
     \sandmp1.{\Ksl_{l,l+1} \Ksl_{(l+1)\cdots n}}.1
    \over \spa{l}.{(l+1)} } \,,
\label{oneminusT1} \\
T_2 &=& \sum_{l=3}^{n-2} \sum_{p=l+1}^{n-1}
{ \spa{(l-1)}.{l}
    \over \sandmp1.{\Ksl_{(p+1)\cdots n} \Ksl_{l\cdots p}}.{(l-1)}
          \sandmp1.{\Ksl_{(p+1)\cdots n} \Ksl_{l\cdots p}}.{l} }
\nonumber \\
&& \hskip15mm\times
{ \spa{p}.{(p+1)}
    \over \sandmp1.{\Ksl_{2\cdots (l-1)} \Ksl_{l\cdots p}}.{p}
          \sandmp1.{\Ksl_{2\cdots (l-1)} \Ksl_{l\cdots p}}.{(p+1)} }
\nonumber \\
&& \hskip15mm\times
    {\sandmp1.{\Ksl_{l\cdots p} \Ksl_{(p+1)\cdots n}}.1}^3
\nonumber \\
&& \hskip15mm\times
    { \sandmp1.{\Ksl_{2\cdots (l-1)} [ {\cal F}(l,p) ]^2
    \Ksl_{(p+1)\cdots n}}.1
     \over s_{l\cdots p} }
\,.
\label{oneminusT2}
\end{eqnarray}

The finite one-loop QCD amplitudes with quarks are~\cite{LastFinite},
\begin{equation}
A_n^{L-s}(1_\f^-,2^+,\ldots,j_\f^+,\ldots,n^+) = {i \over 2}
{ Q_0 \over \spa1.2\spa2.3\cdots\spa{n}.1} \,,
\label{QCDLs}
\end{equation}
where
\begin{equation}
Q_0 = \spa1.j \sum_{l=3}^{n-1} \sandmp1.{\Ksl_{2\cdots l}\ksl_l}.1
\,,
\label{LsQ0}
\end{equation}
and
\begin{equation}
A_n^{s}(1_\f^-,2^+,\ldots,j_\f^+,\ldots,n^+) = {i\over3}
   {S_1 + S_2 \over \spa1.2\spa2.3\cdots \spa{n}.{1} } \,,
\label{QCDs}
\end{equation}
where
\begin{eqnarray}
S_1 &=& \sum_{l=j+1}^{n-1}
   { \spa{j}.{l} \spa{1}.{(l+1)}
     \sandmp1.{\Ksl_{l,l+1} \Ksl_{(l+1)\cdots n}}.1
    \over \spa{l}.{(l+1)} } \,,
\label{QCDsS1} \\
S_2 &=& \sum_{l=j+1}^{n-2} \sum_{p=l+1}^{n-1}
{ \spa{(l-1)}.{l}
    \over \sandmp1.{\Ksl_{(p+1)\cdots n} \Ksl_{l\cdots p}}.{(l-1)}
          \sandmp1.{\Ksl_{(p+1)\cdots n} \Ksl_{l\cdots p}}.{l} }
\nonumber \\
&& \hskip15mm\times
{ \spa{p}.{(p+1)}
    \over \sandmp1.{\Ksl_{2\cdots (l-1)} \Ksl_{l\cdots p}}.{p}
          \sandmp1.{\Ksl_{2\cdots (l-1)} \Ksl_{l\cdots p}}.{(p+1)} }
\nonumber \\
&& \hskip15mm\times
    {\sandmp1.{\Ksl_{l\cdots p} \Ksl_{(p+1)\cdots n}}.1}^2
        \sandmp{j}.{\Ksl_{l\cdots p} \Ksl_{(p+1)\cdots n}}.1
\nonumber \\
&& \hskip15mm\times
    { \sandmp1.{\Ksl_{2\cdots (l-1)} [ {\cal F}(l,p) ]^2
    \Ksl_{(p+1)\cdots n}}.1
     \over s_{l\cdots p} }
\,.
\label{QCDsS2}
\end{eqnarray}
These expressions coincide for $n=4,5$ with previously-computed 
amplitudes~\cite{KunsztEtAl,GGGGG,TwoQuarkThreeGluon}.
\Eqn{oneminus} agrees numerically with the result of ref.~\cite{Mahlon}
for $n\le18$.
Furthermore, in ref.~\cite{LastFinite} the amplitudes
(\ref{QCDLs}) and (\ref{QCDs}) were related
to QED and mixed QED/QCD amplitudes by converting
gluons into photons using appropriate permutation
sums~\cite{TwoQuarkThreeGluon}. The resulting amplitudes
can be compared to the one-loop QED and mixed photon-gluon
amplitudes computed by Mahlon~\cite{Mahlon,MahlonQED},
which provides a strong cross check of \eqns{QCDLs}{QCDs},
and of the recursion relations used to generate them.

\subsection{Amplitudes with a $\phi$ field}
\label{PhiAmplitudesReviewSection}

Amplitudes containing a single $\phi$ field have an ``MHV structure''
very close to that of pure QCD~\cite{DGK,BGK}.  First of all,
those helicity configurations with almost all positive-helicity gluons,
which vanish in tree-level QCD, \eqns{vanishtree}{ffvanishtree},
continue to vanish when a single $\phi$ is added,
\begin{eqnarray}
A_n^\tree(\phi,1^\pm, 2^+,3^+, \ldots, n^+)
&=& 0 \,,
\label{phivanishtree}\\
A_n^\tree(\phi,1_\f^-, 2^+, \ldots, (j-1)^+, j_\f^+, (j+1)^+, \ldots,  
n^+)
&=& 0 \,.
\label{phiffvanishtree}
\end{eqnarray}
Secondly, the amplitudes with a single $\phi$ field and
two negative-helicity gluons, or a negative-helicity gluon plus
a $\bar{q}q$ pair, and the rest positive-helicity gluons,
are identical in form to the pure-QCD amplitudes,
\eqns{mhvtree}{ffmhvtree}~\cite{DGK,BGK},
\begin{equation}
A_n^\tree(\phi,1^+, 2^+, \ldots, m_1^-, \ldots, m_2^-, \ldots, n^+)=
i { \spa{m_1}.{m_2}^4 \over \spa1.2 \spa2.3 \cdots \spa{n}.1} \,,
\label{phimhvtree}
\end{equation}
and
\begin{equation}
A_n^\tree(\phi,1_\f^-, 2^+, \ldots, j_\f^+, \ldots,  m^-, \ldots, n^+)=
i { \spa{1}.{m}^3 \spa{j}.{m} \over \spa1.2 \spa2.3 \cdots \spa{n}.1}  
\,.
\label{phiffmhvtree}
\end{equation}
(Compared with the conventions in refs.~\cite{DGK,BGK}, the partial
amplitudes here have an extra overall factor of $i$.)

\Eqn{phimhvtree} was asserted in ref.~\cite{DGK} based on comparison
with explicit results for small values of $n$~\cite{DawsonKauffman,KD,DFM}, 
and consistency with collinear
and multiparticle factorization.  It can easily be proved recursively,
however, by adapting the proof given in ref.~\cite{BCFRecursion} for
the corresponding pure-QCD MHV amplitude.
The pure-QCD recursion relation~(\ref{BCFWRepresentation})
is still valid for the amplitudes with a $\phi$ attached, because
these amplitudes continue to vanish as $z \rightarrow \infty$,
as we show in appendix \ref{TreeProofAppendix}.

\begin{figure}[t]
\centerline{\epsfxsize 4.5 truein \epsfbox{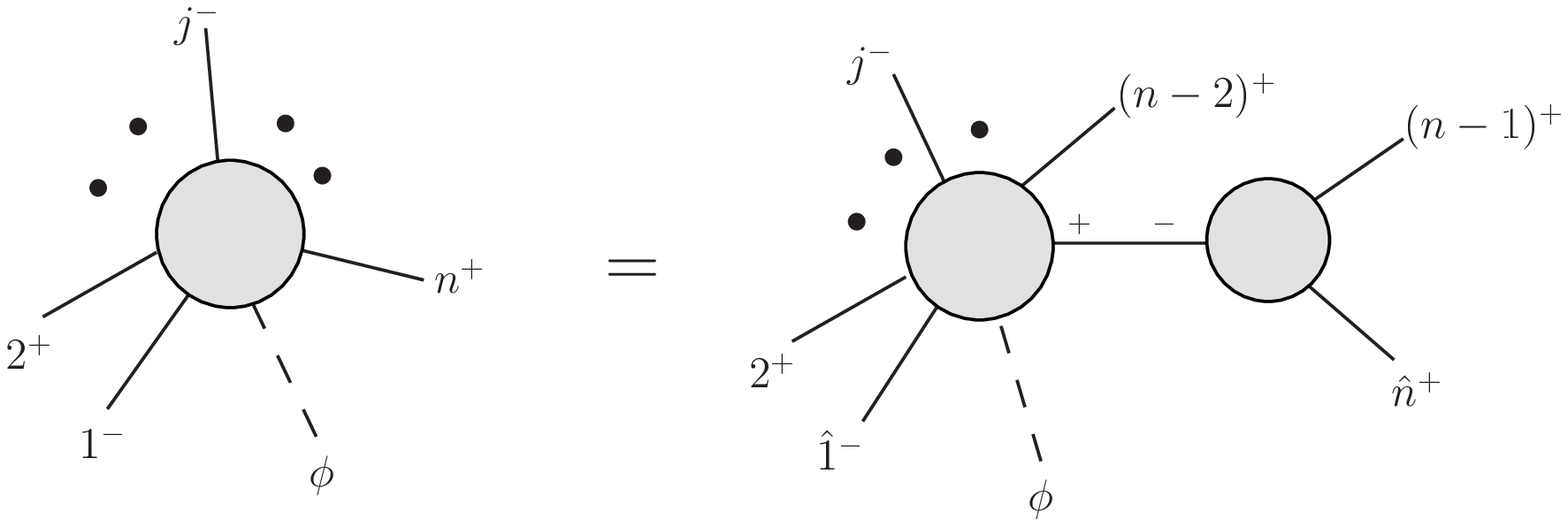}}
\caption{Diagram corresponding to the recursion
relation~(\ref{mhvrec}) for
$A_n^\tree(\phi,1^-,2^+,\dots,j^-,\dots,n^+)$.}
\label{MHVRecursionFigure}
\end{figure}

For convenience, we take the negative-helicity gluons to be legs 
$1$ and $j$.  We use a $[1,n\rangle$ complex momentum shift; 
that is, we set $k=1$ and $l=n$ in \eqn{SpinorShift},
\be
[1,n\rangle: 
\quad \quad \tlambda_1 \rightarrow \tlambda_1 - z\tlambda_n \,, 
\qquad \quad \lambda_n \rightarrow \lambda_n + z\lambda_1 \,.
\label{n1shift}
\ee
Then the on-shell recursion relation~\cite{BCFRecursion,BCFW} has only
one nonvanishing term, depicted in \fig{MHVRecursionFigure},
\bea
&&A_n^\tree(\phi,1^-,2^+,\dots,j^-,\dots,n^+)
\nonumber\\
&&\hskip1.0cm 
=\ A_{n-1}^\tree(\phi,\hat{1}^-,2^+,\dots,j^-,\dots,(n-2)^+,
\hat{K}_{n-1,n}^+)
\, \frac{i}{s_{n-1,n}} \,
A_3^\tree((n-1)^+,\hat{n}^+,-\hat{K}_{n-1,n}^-) \,.
\nonumber\\
\label{mhvrec}
\eea
The form of this relation is precisely the same as for the MHV $n$-gluon
amplitudes in pure QCD:  The only place the shift has any effect is in
$\hat{K}_{n-1,n}$, and it has no dependence on the (implicit) momentum
of $\phi$.  Therefore, the solution of
the recursion relation is exactly the same as in pure QCD, for $n>3$.
We demonstrate the solution~(\ref{phimhvtree}) recursively
by substituting it into the right-hand side of~\eqn{mhvrec},
and recovering the left-hand side,
\bea
&&
{ {\spash{\hat{1}}.{j}}^4 \over
   \spash{\hat{1}}.{2}
   \spa{2}.{3}
   \cdots \spash{(n-2)}.{\hat{K}_{n-1,n}} \spash{\hat{K}_{n-1,n}}.{1} }
  \, \frac{i}{s_{n-1,n}} \,
{ {\spbsh{(n-1)}.{\hat{n}}}^3
   \over \spbsh{\hat{n}}.{(-\hat{K}_{n-1,n})} 
         \spbsh{(-\hat{K}_{n-1,n})}.{(n-1)} }
\nonumber\\
& = & i \, { {\spa{1}.{j}}^4 \,{\spb{(n-1)}.{n}}^2 \over \spa{1}.{2} \cdots
\spa{(n-3)}.{(n-2)} \spa{(n-1)}.{n}
\sand{1}.{\Ksl_{n-1,n}}.{(n-1)} \sand{(n-2)}.{\Ksl_{n-1,n}}.{n} }  
\nonumber \\
& = &  i { \spa{1}.{j}^4 \over \spa1.2 \spa2.3 \cdots \spa{n}.1} \,.
{~}\label{mhvrecsol}
\eea
There is one important exception at the lowest order, $n=3$.
Here the recursion for the $\phi$ case
has one extra term, replacing the term in \eqn{mhvrec},
which vanishes for $n=3$,
\be
A_3^\tree(\phi,1^-,2^-,3^+)\ =\ 
A_2^\tree(\phi,\hat{1}^-,\hat{K}_{2,3}^-) \, \frac{i}{s_{23}} \,
A_3^\tree(2^-,\hat{3}^+,-\hat{K}_{2,3}^+) \, .
\ee
This term is nonvanishing, and the three-parton kinematics are
nonsingular, because of the extra momentum injected by $\phi$.
The proof of the fermionic MHV formula, \eqn{phiffmhvtree},
is completely analogous.

%

Another infinite series of tree-level $\phi$ amplitudes
is the ``maximally googly'' set~\cite{BGK},
\bea
A_n^\tree(\phi,1^-,2^-,3^-,\dots,n^-)
&=& i { (-1)^n (k_\phi^2)^2 \over \spb1.2\spb2.3 \cdots \spb{n}.{1} }
\,, \label{phiantimhvtree}\\
A_n^\tree(\phi^\dagger,1^+,2^+,3^+,\dots,n^+)
&=& i { (k_\phi^2)^2 \over \spa1.2\spa2.3 \cdots \spa{n}.{1} }
\,. \label{phidaggerantimhvtree}
\eea
%

\subsection{Coupling of $A$ to the axial current divergence, 
and the $\phi g^+g^+$ amplitude}
\label{AxialCurrentDivSubsection}

As mentioned in \sect{DecompositionSection},
at order $\as^2$ in the effective theory, the pseudoscalar $A$ begins to
couple to the divergence of the light-quark axial current~\cite{CKSB}.
The full effective Lagrangian is given by
\be
\mathcal{L}_A^{\mbox{\tiny int}} =
\frac{C}{2} A  \, \tr G_{\mu\nu} \mbox{ }^* G^{\mu\nu}
+ \frac{C'}{2} A \,\partial_\mu \sum\limits_{i = 1}^{n_l} \bar{q}_i
\gamma^\mu \gamma_5 q_i  \,,
\label{LintA}
\ee
where the sum is over all light quark flavors $n_l$, and
$C$ and $C'$ denote the appropriate coefficients,
which begin at order $\as$ for $C$ and order $\as^2$ for $C'$.
It was found in ref.~\cite{CKSB} by explicit calculation
that the second term in \eqn{LintA}
does not contribute at NLO, in accordance with
a non-Abelian version~\cite{NAABth}
of the Adler-Bardeen theorem~\cite{ABth}.
This can also be seen via recursion
relations as follows: The amplitude
$A_2^{\tree}(A_{J_5},1_f^-,2_f^+)$
vanishes by helicity conservation for massless fermions
$f$.  We have denoted the contribution from the second term 
in \eqn{LintA} by the subscript $J_5$.  By $A_2^{\tree}$ we mean 
a tree amplitude, which contributes at NLO due to the factor of 
$\as^2$ in $C'$.  The effective $A\bar{q}q$-vertex
is proportional to the momentum that flows through it
due to the derivative in \eqn{LintA}. 
When attaching a gluon to this vertex, one can choose
its polarization vector so that each of the two diagrams
contributing to $A_3^{\tree}(A_{J_5},1_f^-,2_f^+,3^\pm)$
vanishes.  (The amplitude is of course gauge invariant; in other
gauges the two diagrams cancel against each other.) 
Thus $A_3^{\tree}(A_{J_5},1_f^-,2_f^+,3^\pm) = 0$.
By recursion, the amplitudes with an arbitrary number of
gluons also vanish --- there are no lower-point amplitudes
to contribute to the right-hand side of a recursion relation. 
Thus we do not need to consider the contribution of the light-quark
axial current at NLO, and the amplitudes
involving a pseudoscalar $A$ can be constructed via
\eqn{Areconstruct}.

In \app{NormalizationAppendix} we compute the simplest,
finite one-loop amplitude with a single $\phi$ field,
$A_{2;1}(\phi,1^+,2^+)$.
The result can be written in terms of the ``googly''
two-gluon tree amplitude,
\be
A_{2;1}(\phi,1^+,2^+) = - 2 A_2^\tree(\phi^\dagger,1^+,2^+)\,.
\label{phi12Norm}
\ee
This amplitude enters recursively into all the subsequently
constructed amplitudes.


\section{The soft Higgs limit}
\label{SoftHiggsSection}

It is useful to consider the kinematic limit of the single-$\phi$
amplitudes as $k_\phi \to 0$.  Of course this soft-Higgs limit
can only be reached if $m_H=0$.  The limiting behavior of
single-$\phi$ tree amplitudes was studied in ref.~\cite{DGK}.
Here we review that analysis and extend it to the loop level.
First consider amplitudes with a single real scalar field $H$,
as the Higgs momentum goes to zero, so that $H$ has almost no spatial
variation. Then the $H \tr G_{\mu\nu}G^{\mu\nu}$ interaction behaves
just like a constant multiple of the pure-glue Lagrangian
$\tr G_{\mu\nu}G^{\mu\nu}$.
So one might expect the Higgs-plus-$n$-gluon amplitudes to become 
proportional to the pure-gauge-theory amplitudes, according to
the relation,
\be
{\cal A}_n(H,\{k_i,\lambda_i,a_i\}) \longrightarrow
({\rm const.}) \times {\del \over \del g} {\cal A}_n(\{k_i, 
\lambda_i,a_i\})
\qquad\quad \hbox{as $k_H \to 0$,}
\label{softHiggs1}
\ee
for any helicity configuration.  Taking into account the number
of factors of the gauge coupling in the $n$-parton tree amplitudes, 
namely $(n-2)$, \eqn{softHiggs1} leads to
the tree-level soft Higgs limit,
\be
A_n^{\tree}(H,1,2,3,\ldots,n) \longrightarrow
(n-2) A_n^{\tree}(1,2,3,\ldots,n)
\qquad\quad \hbox{as $k_H \to 0$,}
\label{softHiggs2tree}
\ee
which holds color structure by color structure.
Similarly, at $l$ loops, our naive expectation for the soft limit is,
\be
A_n^{\lloop}(H,1,2,3,\ldots,n) \longrightarrow
(n+2\,l-2) A_n^{\lloop}(1,2,3,\ldots,n)
\qquad\quad \hbox{as $k_H \to 0$.}
\label{softHiggs2loop}
\ee
In fact, we will find that \eqn{softHiggs2loop} is not precisely
correct beyond tree level.  A generalization of this equation to the 
case of $\phi$ and $\phi^\dagger$ amplitudes will hold at
loop level, but again it will hold precisely only for the 
finite one-loop amplitudes.  \Eqn{softHiggs2loop} generically 
breaks down at loop level because of an exchange-of-limits
problem, between dimensional regularization of infrared divergences, 
and the momentum injected into the amplitude by the Higgs boson, 
which acts as a kind of effective infrared regulator as well.

Before examining the infrared issue,
let us first propose an extension of \eqn{softHiggs2loop} to the 
case of amplitudes with a single $\phi$, or $\phi^\dagger$.
At tree level, by studying the MHV representation
of the $\phi$ and $\phi^\dagger$ amplitudes, it can be shown~\cite{DGK}
that the factor of $(n-2)$ in \eqn{softHiggs2tree} 
is partitioned into $(n_{-}-1)$ and $(n_{+}-1)$ in the two cases,
\be
A_n^\tree(\phi,1,2,3,\ldots,n) \longrightarrow
(n_- - 1) A_n^\tree(1,2,3,\ldots,n)
\qquad\quad \hbox{as $k_\phi \to 0$,}
\label{softphi}
\ee
and (by parity)
\be
A_n^\tree(\phi^\dagger,1,2,3,\ldots,n) \longrightarrow
(n_+ - 1) A_n^\tree(1,2,3,\ldots,n)
\qquad\quad \hbox{as $k_\phi \to 0$,}
\label{softphidagger}
\ee
where $n_\pm$ are the number of external partons with helicity $\pm1$
for gluons ($\pm{1\over2}$ for fermions).
The natural generalization of \eqns{softphi}{softphidagger}
to $l$ loops would be,
\be
A_n^\lloop(\phi,1,2,3,\ldots,n) \longrightarrow
(n_- + l - 1) A_n^\lloop(1,2,3,\ldots,n)
\qquad\quad \hbox{as $k_\phi \to 0$,}
\label{softphiL}
\ee
and
\be
A_n^\lloop(\phi^\dagger,1,2,3,\ldots,n) \longrightarrow
(n_+ + l - 1) A_n^\lloop(1,2,3,\ldots,n)
\qquad\quad \hbox{as $k_\phi \to 0$.}
\label{softphidaggerL}
\ee
But again, infrared complications can affect these relations.

%
\begin{figure}[t]
\centerline{\epsfxsize 4.5 truein \epsfbox{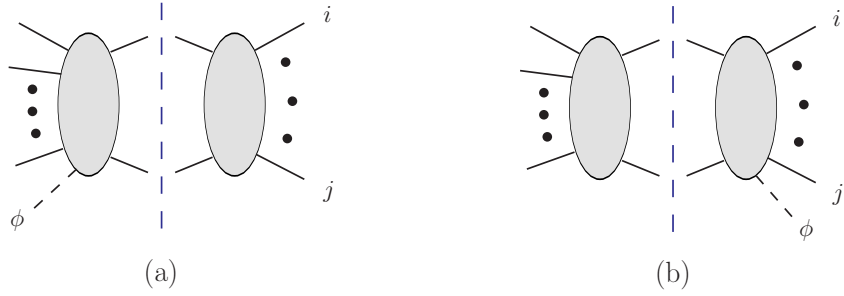}}
\caption{Two cuts of a one-loop $\phi$ amplitude which become
the same cut in the soft limit $k_\phi \to 0$:  
(a) the cut in the $s_{i\cdots j}$ channel, and
(b) the cut in the $s_{\phi,i\cdots j}$ channel.}
\label{SoftCutFigure}
\end{figure}

Consider a particular class of terms in a one-loop Higgs or $\phi$
amplitude, those containing a cut in either the $s_{i\cdots j}$
channel or the $s_{\phi,i\cdots j}$ channel, as illustrated
in~\fig{SoftCutFigure}.  The soft behavior
of these terms can be analyzed using unitarity, plus the 
tree-level soft behavior.  The two cuts in~\fig{SoftCutFigure}
become identical, up to constant factors, as $k_\phi \to 0$.  
Suppose there are $m$ external partons
on the right side of the cuts in \fig{SoftCutFigure},
plus the two crossing the cut.
Then there are $(n-m)$ external partons on the left side of the cut,
plus two more crossing the cut.
(The fields $H$ and $\phi$ do not count as partons.)
For the case of a Higgs field $H$, using \eqn{softHiggs2tree}
in the soft limit $k_H\to0$, 
the cut in \fig{SoftCutFigure}(a) reduces to the pure-QCD cut in the 
$s_{i\cdots j}$ channel, multiplied by a factor of $(n-m)$.
The cut in \fig{SoftCutFigure}(b) reduces to the same pure-QCD cut,
because $s_{\phi,i\cdots j} \to s_{i\cdots j}$, but it is multiplied 
by a factor of $m$.  Summing the two contributions, we find that 
the terms in the loop amplitude containing
$\ln(-s_{i\cdots j})$ and $\ln(-s_{\phi,i\cdots j})$ go smoothly
into the terms containing $\ln(-s_{i\cdots j})$ in the pure-QCD amplitude,
multiplied by a factor of $(n-m) + m = n$.
This argument would seem to prove \eqn{softHiggs2loop} for $l=1$.

A similar argument applies to the $\phi$ and $\phi^\dagger$ amplitudes.
If there are a total of $n_{\pm,L}$ partons with helicity $\pm1$ on the
left-hand side of the cut, and $n_{\pm,R}$ on the right-hand side,
then $n_{\pm,L} + n_{\pm,R} = n_\pm + 2$, because 2 partons of each
helicity are assigned to the cut.  Instead of the multiplicative factor
of $(n-m) + m = n$ found for the case of $H$, for $\phi$ we get from
\eqn{softphi} a factor of $(n_{-,L}-1) + (n_{-,R}-1) = n_{-}$,
which is the factor in~\eqn{softphiL} for $l=1$.

However, both these arguments break down for the case $m=2$, for which only
two partons, $i$ and $i+1$, appear on the right-hand side of the cut.
In this case, the cut in \fig{SoftCutFigure}(a) is generically 
infrared divergent, whereas the cut in \fig{SoftCutFigure}(b) is infrared
finite, for nonzero momentum $k_\phi$.  The generic form of one-loop
divergences for the unrenormalized amplitudes for an 
$H$ or $\phi$ field plus $n$ gluons, for example, 
is~\cite{UniversalIR}
\bea
A_{n;1}(H,1,2,\ldots,n) &=& 
- {1\over \e^2} 
\sum_{i=1}^{n} \left( {\mu^2\over-s_{i,i+1}} \right)^\e
A_n^\tree(H,1,2,\ldots,n)\ +\ \hbox{finite,}
\label{HIR}\\
A_{n;1}(\phi,1,2,\ldots,n) &=& 
- {1\over \e^2} 
\sum_{i=1}^{n} \left( {\mu^2\over-s_{i,i+1}} \right)^\e
A_n^\tree(\phi,1,2,\ldots,n)\ +\ \hbox{finite.}
\label{phiIR}
\eea
The infrared divergences have precisely the same form as for pure QCD,
\be
A_{n;1}(1,2,\ldots,n) = 
\Biggl[ - {1\over \e^2} 
\sum_{i=1}^{n} \left( {\mu^2\over-s_{i,i+1}} \right)^\e
 - {\beta_0\over \e} \Biggr]
A_n^\tree(1,2,\ldots,n)\ +\ \hbox{finite,}
\label{QCDIR}
\ee
where $\beta_0 = (11 N_c -  2 n_\f - n_s)/(3 N_c)$.
(After ultraviolet renormalization the remaining infrared $\beta_0$ terms
match between eqs.~(\ref{HIR})--(\ref{QCDIR}).)
Hence the divergent parts of the (renormalized) one-loop $H$ and $\phi$ 
amplitudes have the soft behavior characteristic of the 
tree-level amplitudes,
\bea
A_{n;1}(H,1,2,\ldots,n)\Big|_{1/\e\ {\rm poles}} 
&\inlimit^{k_H \to 0}&
(n-2) \, A_{n;1}(1,2,\ldots,n)\Big|_{1/\e\ {\rm poles}} \ ,
\label{HdivIR}\\
A_{n;1}(\phi,1,2,\ldots,n)\Big|_{1/\e\ {\rm poles}}
&\inlimit^{k_\phi \to 0}&
(n_{-}-1) \, A_{n;1}(1,2,\ldots,n)\Big|_{1/\e\ {\rm poles}} \ .
\label{phidivIR}
\eea
This behavior does not agree with the naive expectation for the one-loop
behavior, \eqns{softHiggs2loop}{softphiL} for $l=1$.  
However, it is consistent with the behavior of
the divergent cuts for $j=i+1$ ($m=2$) in \fig{SoftCutFigure}.
For example the factor of $(n-2)$ in \eqn{HdivIR}, for the cut of the 
term proportional to $\ln(-s_{i,i+1})/\e \times A_n^\tree$,
comes just from \fig{SoftCutFigure}(a), because the cut of 
\fig{SoftCutFigure}(b) has no pole in $\e$.  
It is the exchange of limits, $\e\to0$
and $k_H\to0$, in \fig{SoftCutFigure}(b), which is at the root
of the non-uniform soft behavior.

This argument does not cover the cases $m=1$ ($\ln(-s_{\phi,i})$ terms)
or $m=0$ ($\ln(-k_\phi^2)$ terms).  These cases have only
the cuts shown in \fig{SoftCutFigure}(b), but the cuts become
quite singular in the $k_\phi \to 0$ limit, so we may expect
non-uniform soft behavior.
The cut-based argument also does not tell us what to expect 
for the rational terms in generic $H$ or $\phi$ amplitudes.  
Indeed, for the one-loop $H\bar{q}q\bar{Q}\bar{Q}$ amplitudes 
presented in ref.~\cite{EGZHiggs}, the soft Higgs limits of 
the generic cut terms (all of which have $m=0$, 1, or 2)
and rational parts at order $\eps^0$ do not appear to be uniform.

On the other hand, for the finite $\phi$ amplitudes we focus 
on in this paper, there should be no infrared issues, 
and hence we should find the simple, uniform behavior,
valid for each color structure,
\be
A_n^\oneloop(\phi,1,2,3,\ldots,n) \inlimit^{k_\phi \to 0}
n_- \, A_n^\oneloop(1,2,3,\ldots,n)
\qquad\quad \hbox{(finite helicity configurations).}
\label{softphi1}
\ee
We shall verify \eqn{softphi1} explicitly below for the finite 
helicity amplitudes.

Assuming that \eqn{softphi1} holds, then
the generic one-loop amplitude with $\phi$ present
must be at least as complicated as the corresponding pure-QCD
amplitude with $\phi$ omitted, because the former amplitude must
reduce to the latter amplitude, multiplied by a factor of $n_-$,
as $k_\phi\to0$.  The one possible exception is the all-plus
amplitude, $A_n^\oneloop(\phi,1^+,2^+,\ldots,n^+)$, because
the multiplicative factor vanishes in this case.  Indeed,
as we shall see in the next section,
these amplitudes are given by a much simpler formula than
the corresponding QCD result~(\ref{OneLoopAllPlusAmplitude}).
\\
{~}

\section{Recursion relation for all-plus case}
\label{AllPlusAmplitudesSection}

Using the results of \sect{RecursionReviewSection}, and the explicit
calculation of the two-gluon one-loop amplitude~(\ref{phi12Norm}),
we can recursively construct the following form of the one-loop
all-plus $\phi$ amplitude,
\bea
A_n^{[f]}(\phi,1^+,2^+,\dots,n^+)
&=& A_n^{[s]}(\phi,1^+,2^+,\dots,n^+) = 0 \,,
\label{allplusnfnsvanish}\\
A_n^{[g]}(\phi,1^+,2^+,\dots,n^+)
&=& - 2 i { (k_\phi^2)^2 \over
\spa{1}.{2} \spa{2}.{3} \cdots \spa{n}.{1}} =
- 2 A_n^\tree(\phi^\dagger,1^+,2^+,\dots,n^+) \,.
\label{allplusphi}
\eea
The proof of \eqn{allplusphi} is analogous to the above proof of
the tree-level MHV amplitude via an on-shell recursion
relation~\cite{BCFRecursion}.

%
\begin{figure}[t]
\centerline{\epsfxsize 4.5
truein \epsfbox{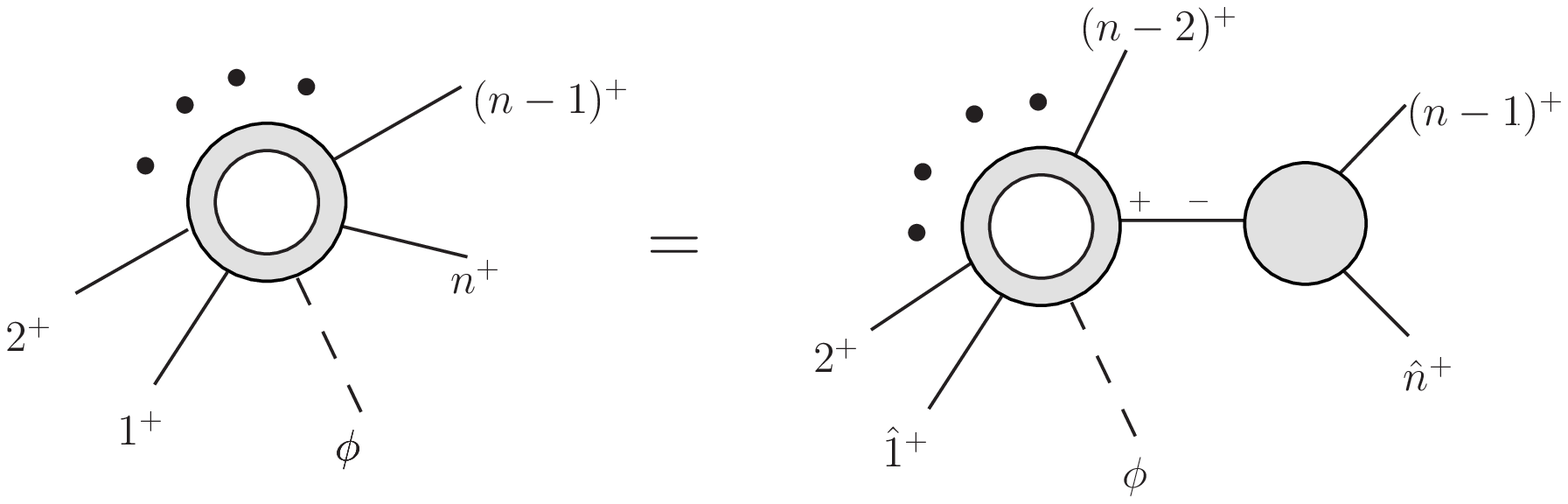}}
\caption{Diagram corresponding to the recursion
relation~(\ref{recallplusphi}) for
$A_n^{(1)}(\phi,1^+,2^+,\ldots,n^+)$.}
\label{AllPlusRecursionFigure}
\end{figure}

Here we use again the $[1,n\rangle$ shift~(\ref{n1shift}).
As in \sect{PhiAmplitudesReviewSection}
there is only one term contributing to the recursion,
as illustrated in \fig{AllPlusRecursionFigure},
\bea
&&A_{n;1}(\phi,1^+,\dots,n^+)
\nonumber\\
&&\hskip0.3cm =
A_{n-1;1}(\phi,\hat{1}^+,2^+,\dots,(n-2)^+,\hat{K}_{n-1,n}^+)
\,\frac{i}{s_{n-1,n}} \,
A_3^\tree((n-1)^+,\hat{n}^+,-\hat{K}_{n-1,n}^-)
\,.~~~~
\label{recallplusphi}
\eea
It is straightforward to show that \eqn{allplusphi}
satisfies \eqn{recallplusphi}, using the same algebra as 
in~\eqn{mhvrecsol}.
The amplitude $A_n^{[g]}(\phi,1^+,2^+,\dots,n^+)$ in \eqn{allplusphi}
vanishes in the soft limit, just as predicted by \eqn{softphi1} 
for $n_{-}=0$.

Note that the lack of $n_\f$ or $n_s$ dependence in 
$A_{2;1}(\phi,1^+,2^+)$
translates recursively into the vanishing of $A_n^{[f]}$ and $A_n^{[s]}$
in \eqn{allplusnfnsvanish}, for all values of $n$. Now consider the
dependence of the one-loop one-minus $\phi$ amplitude on $n_\f$ or $n_s$.
This amplitude factorizes onto two types of one-loop amplitudes,
either the all-plus $\phi$ amplitude, which obeys~\eqn{allplusnfnsvanish},
or the all-plus pure-QCD amplitude. But the pure-QCD amplitude 
obeys $A_n^{[f]} = -A_n^{[s]}$, thanks to 
a supersymmetric Ward identity~\cite{SWI}.
Using this relation and \eqn{allplusnfnsvanish} in a recursive
argument, we must have $A_n^{[f]} = -A_n^{[s]}$ also for the
one-minus $\phi$ amplitude, as mentioned in \sect{NotationSection}, 
and as assumed in the color
decomposition~(\ref{OneloopPhiFiniteDecomp}).

\section{Quark recursion relations}
\label{AmplitudesSection}

As we discussed in \sect{NotationSection}, for the finite
helicity amplitudes --- one quark pair, with all gluons having
positive helicity --- there are two independent primitive amplitudes
that need to be computed,
\begin{eqnarray}
A_n^{L-s}(\phi,j_\f^+)
&\equiv& A_n^{L-s}(\phi,1_\f^-,2^+,\ldots,j_\f^+,\ldots,n^+) \,,
\qquad 2\leq j\leq n,
\label{Lminussshort} \\
A_n^{s}(\phi,j_\f^+)
&\equiv& A_n^{s}(\phi,1_\f^-,2^+,\ldots,j_\f^+,\ldots,n^+) \,,
\qquad 2\leq j \leq n-2\, .
\label{sshort}
\end{eqnarray}
We abbreviate the notation as in ref.~\cite{LastFinite},
retaining only the label of the positive-helicity fermion (and $\phi$).
A computation of these primitive
amplitudes determines all the finite one-loop quark amplitudes
for a single $\phi$ boson coupled to QCD.

\subsection{Recursion relation for $L-s$ contribution}

As mentioned in the introduction and in 
section~\ref{RecursionReviewSection}, finite quark-gluon amplitudes  
in both pure QCD and $\phi$-amplitudes contain ``unreal'' 
poles.  The contributions of the unreal poles can be described
using a ``soft factor'' $\Soft^\tree$ reminiscent of the propagation
of an internal soft gluon~\cite{LastFinite}.

For the recursive construction we need the
``$L\!-\!s$ loop vertex'' introduced in ref.~\cite{LastFinite},
\begin{equation}
V_3^{L-s}(1^+,2^+,3^+) \equiv
    - {i \over 2} \spb1.2 \spb2.3 \spb3.1 \,.
\end{equation}
The soft factor $\Soft^\tree(a,s,b)$ traditionally describes the insertion 
of a soft gluon $s$ between two hard partons $a$ and $b$
in a color-ordered amplitude.  It depends only on the helicity
of the soft gluon and is given by~\cite{TreeReview},
\begin{eqnarray}
\Soft^\tree( a,  s^+, b) & =& {\spa{a}.{b} \over \spa{a}.{s} \spa{s}. 
{b}} \,,\\
\Soft^\tree( a,  s^-, b) & =& -{\spb{a}.{b} \over \spb{a}.{s} \spb{s}. 
{b}} \,.
\label{softdef}
\end{eqnarray}

Choosing the shifted legs in \eqn{SpinorShift} to be
$[k,l\rangle=[1,n\rangle$, \eqn{n1shift},
the recursion relation for the pure-QCD amplitude 
$A_n^{L-s}(j_\f^+) \equiv A_n^{L-s}(1_\f^-,2^+,\ldots,j_\f^+,\ldots,n^+)$
was found to be,
\begin{eqnarray}
&& \null \hskip -10 mm A_n^{L-s}(j_\f^+)
\nonumber\\
&=&
A_{n-1}^{L-s}(\hat 1_\f^-,2^+,\ldots,j_\f^+,\ldots,(n-2)^+,\Kh_{n-1,n}^+)
\, {i\over s_{n-1,n}} \,
  A_3^{\tree}(-\Kh_{n-1,n}^-,(n-1)^+,\hat n^+)
\nonumber\\
&&\hskip-4mm \null + 
A_{n-1}^\tree(\hat 1_\f^-,2^+,\ldots,j_\f^+,\ldots,(n-2)^+,\Kh_{n-1,n}^-)
\, {i\over s_{n-1,n}} \,
  V_3^{L-s}(-\Kh_{n-1,n}^+,(n-1)^+,\hat n^+)
\nonumber\\ &&\hskip 20mm  \null\times
\Soft^\tree(j, \Kh_{n-1,n}^+,\hat 1) \,
\Soft^\tree(\hat n, -\Kh_{n-1,n}^-, n-1) \,,
\label{LsRecursion}
\end{eqnarray}
where the hatted momenta in \eqn{LsRecursion} are defined
by the shift~(\ref{n1shift}), with
\begin{equation}
z = - { K_{n-1,n}^2 \over \spab1.{\Ksl_{n-1,n}}.n } =
- { \spa{(n-1)}.{n} \over \spa{(n-1)}.{1} } \,,
\end{equation}
in each term.  The second term in \eqn{LsRecursion} represents
the contribution of an unreal pole.
The solution to \eqn{LsRecursion} is given by \eqn{QCDLs}.

%
\begin{figure}[t]
\centerline{\epsfxsize 6. truein \epsfbox{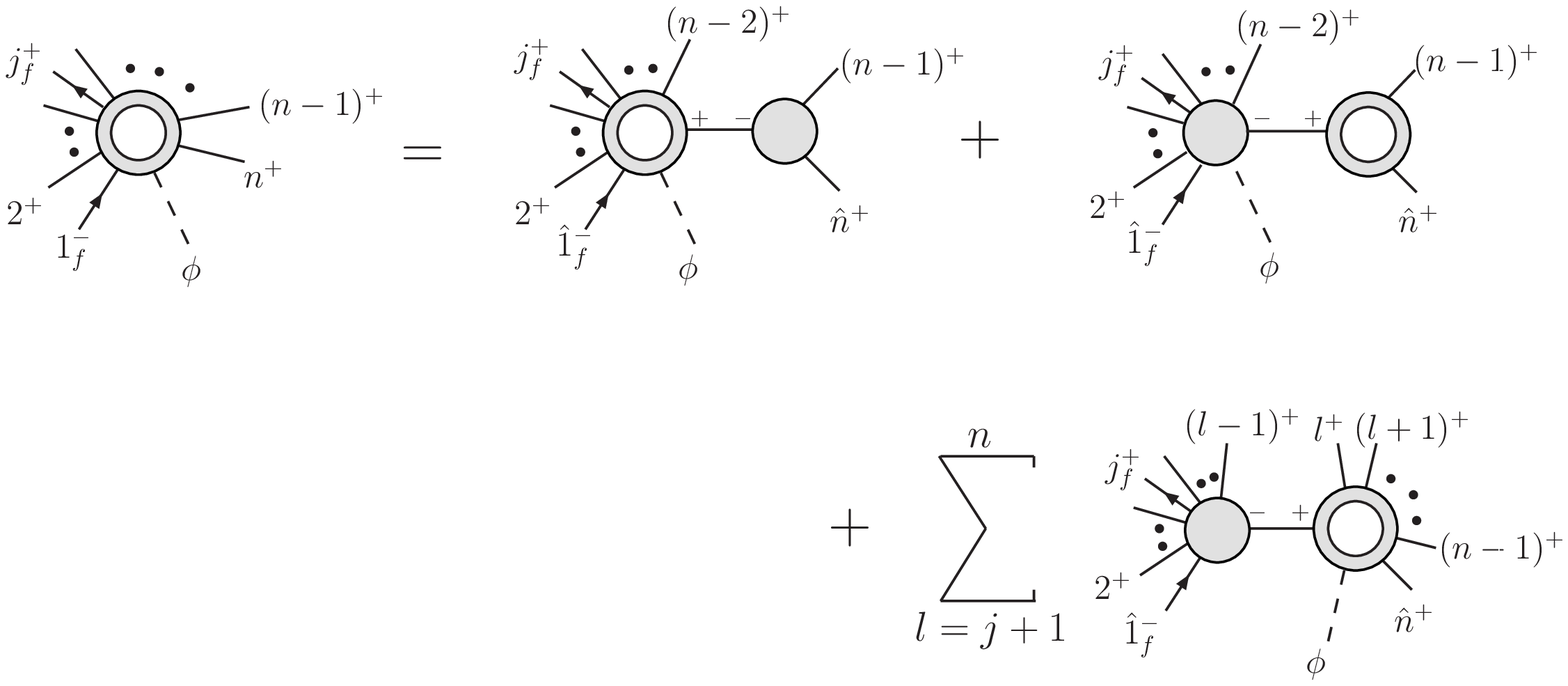}}
\caption{Diagrams corresponding to the terms in the recursion
relation~(\ref{phiLsRecursion}) for $A_n^{L-s}(\phi,j_\f^+)$,
for $j\leq n-2$.  The second term contains an unreal pole.
In the last term, the sum over $l$
runs from $j+1$ to $n$, inclusive.}
\label{LsRecursionFigure}
\end{figure}

For the analogous single-$\phi$ amplitude $A_n^{L-s}(\phi,j_\f^+)$,
using the same $[1,n\rangle$ shift, the $n$-point on-shell 
recursion relation is, for $j\leq n-2$,
\begin{eqnarray}
&& \null \hskip -10 mm A_n^{L-s}(\phi,j_\f^+)
\nonumber\\
&=& 
A_{n-1}^{L-s}(\phi,\hat 1_\f^-,2^+,
               \ldots,j_\f^+,\ldots,(n-2)^+,\Kh_{n-1, n}^+)
\, {i\over s_{n-1,n}} \,  
A_3^{\tree}(-\Kh_{n-1,n}^-,(n-1)^+,\hat n^+)
\nonumber\\
&&\hskip-4mm \null + 
A_{n-1}^\tree(\phi,\hat 1_\f^-,2^+,
               \ldots,j_\f^+,\ldots,(n-2)^+,\Kh_{n-1, n}^-)
\, {i\over s_{n-1,n}} \,
V_3^{L-s}(-\Kh_{n-1,n}^+,(n-1)^+,\hat n^+)
\nonumber\\ &&\hskip 20mm  \null\times
\Soft^\tree(j, \Kh_{n-1,n}^+,\hat 1) \,
\Soft^\tree(\hat n, -\Kh_{n-1,n}^-, n-1)
\nonumber\\
&&\hskip-4mm \null + \sum_{l=j+1}^{n}
A_{l}^\tree(\hat{1}_\f^-,2^+,
             \ldots,j_\f^+,\ldots,(l-1)^+,\Kh_{\phi,l\cdots n}^-)
\, {i\over s_{\phi,l\cdots n}} \,
A_{n-l+2}^{[g]}(\phi,-\Kh_{\phi,l\cdots n}^+,l^+,\ldots,\hat n^+)
\,,
\nonumber\\
~~
\label{phiLsRecursion}
\end{eqnarray}
where in the $s_{\phi,l \cdots n}$ channel we set,
\begin{equation}
z = - {s_{\phi,l \cdots n} \over \sand1.{\Ksl_{\phi,l \cdots n}}.n} =
   {s_{1 \cdots (l-1)} \over \sand1.{\Ksl_{1 \cdots (l-1)}}.n} \,.
\label{zlnChannel}
\end{equation}
The recursion relation is shown in \fig{LsRecursionFigure}.
Again the second term represents the unreal-pole contribution,
which we take to have the same form when $\phi$ is present
as it does in pure QCD.
The additional third term in \eqn{phiLsRecursion}, containing a sum
over $l$, arises because the amplitudes 
$A_m^{[g]}(\phi,+,+,\ldots,+) = A_m^{[g-s]}(\phi,+,+,\ldots,+)$
are nonvanishing when $\phi$ is present.  The corresponding
pure-QCD amplitude combination $A_m^{[g-s]}(+,+,\ldots,+)$ vanishes
according to~\eqn{OneLoopAllPlusAmplitude}, so there is no such
term in the pure-QCD recursion relation~(\ref{LsRecursion}).

The solution to \eqn{phiLsRecursion} is,
\begin{equation}
A_n^{L-s}(\phi,1_\f^-,2^+,\ldots,j_\f^+,\ldots,n^+) = {i \over 2}
{ Q_0^\phi + Q_1^\phi + Q_2^\phi
\over \spa1.2\spa2.3\cdots\spa{n}.1 }
\,,
\label{phiLs}
\end{equation}
where
\begin{eqnarray}
Q_0^\phi &=&
\spa1.j \sum_{l=2}^{n-1} \sandmp1.{\ksl_l \Ksl_{l \cdots n}}.1
\,,
\label{phiLsQ0}\\
Q_1^\phi &=& - 4
\sum_{l=j+1}^{n}
{ \spa{(l-1)}.{l}
\over \sandmp1.{\Ksl_{l\cdots n} \Ksl_{1\cdots n}}.{(l-1)}
        \sandmp1.{\Ksl_{2\cdots (l-1)} \Ksl_{1\cdots n}}.{l} }
\nonumber\\
&&\hskip 10mm \null
   \times {\sandmp1.{\Ksl_{2\cdots (l-1)} \Ksl_{l\cdots n}}.1}^2
    \sandmp{j}.{\Ksl_{1\cdots n} \Ksl_{l\cdots n}}.1
\,,
\label{phiLsQ1}\\
Q_2^\phi &=& - 4
\sum_{l=j+1}^{n-1} \sum_{p=l+1}^{n}
{ \spa{(l-1)}.{l}
\over \sandmp1.{\Ksl_{p\cdots n}
                 (\Ksl_{1\cdots (l-1)} + \Ksl_{p\cdots n})}.{(l-1)}
        \sandmp1.{\Ksl_{p\cdots n}
                 (\Ksl_{1\cdots (l-1)} + \Ksl_{p\cdots n})}.{l} }
\nonumber\\
&&\hskip 6mm \null
   \times
{ \spa{(p-1)}.{p}
\over \sandmp1.{\Ksl_{2\cdots (l-1)}
                 (\Ksl_{1\cdots (l-1)} + \Ksl_{p\cdots n})}.{(p-1)}
        \sandmp1.{\Ksl_{2\cdots (l-1)}
                 (\Ksl_{1\cdots (l-1)} + \Ksl_{p\cdots n})}.{p} }
\nonumber\\
&&\hskip 6mm \null
\times { \sandmp1.{\Ksl_{2\cdots (l-1)} \Ksl_{p\cdots n}}.1 }^3
\sandmp{j}.{(\Ksl_{1\cdots (l-1)} + \Ksl_{p\cdots n})
             \Ksl_{p\cdots n}}.1
\nonumber\\
&&\hskip 6mm \null
\times { (k_\phi^2)^2 \over (K_{1\cdots (l-1)} + K_{p\cdots n})^2 }
\,.
\label{phiLsQ2}
\end{eqnarray}
Making use of momentum conservation, \eqn{MomConsphi}, antisymmetry
of the spinor products, and re-indexing the sum over $p$ in $Q_2^\phi$,
we can rewrite these quantities as
\begin{eqnarray}
Q_0^\phi &=&
\spa1.j \sum_{l=2}^{n-1} \sandmp1.{\Ksl_{\phi,2 \cdots l} \ksl_l}.1
\,,
\label{phiLsQ0ALT}\\
Q_1^\phi &=& - 4
\sum_{l=j+1}^{n}
{ \spa{(l-1)}.{l}
\over \sandmp1.{\Ksl_{l\cdots n} \ksl_\phi}.{(l-1)}
        \sandmp1.{\Ksl_{2\cdots (l-1)} \ksl_\phi}.{l} }
\nonumber\\
&&\hskip 10mm \null
   \times {\sandmp1.{\Ksl_{l\cdots n} \ksl_\phi}.1}^2
    \sandmp{1}.{\Ksl_{l\cdots n} \ksl_\phi}.j
\,,
\label{phiLsQ1ALT}\\
Q_2^\phi &=& - 4
\sum_{l=j+1}^{n-1} \sum_{p=l}^{n-1}
{ \spa{(l-1)}.{l}
\over \sandmp1.{\Ksl_{(p+1)\cdots n} \Ksl_{\phi,l\cdots p}}.{(l-1)}
        \sandmp1.{\Ksl_{(p+1)\cdots n} \Ksl_{\phi,l\cdots p}}.{l} }
\nonumber\\
&&\hskip 6mm \null
   \times
{ \spa{p}.{(p+1)}
\over \sandmp1.{\Ksl_{2\cdots (l-1)} \Ksl_{\phi,l\cdots p}}.{p}
        \sandmp1.{\Ksl_{2\cdots (l-1)} \Ksl_{\phi,l\cdots p}}.{(p+1)} }
\nonumber\\
&&\hskip 6mm \null
\times { \sandmp1.{\Ksl_{(p+1)\cdots n} \Ksl_{\phi,l\cdots p}}.1 }^3
\sandmp{1}.{\Ksl_{(p+1)\cdots n} \Ksl_{\phi,l\cdots p}}.j
\ { (k_\phi^2)^2 \over s_{\phi,l\cdots p} }
\,.
\label{phiLsQ2ALT}
\end{eqnarray}
The latter forms make the soft-$\phi$ limit, $k_\phi \to 0$,
somewhat more manifest.  In this limit, we see that $Q_0^\phi$ reduces
to $Q_0$ as given in \eqn{LsQ0}, and that $Q_1^\phi$ and $Q_2^\phi$ both
vanish.  (Note that there is one more factor of $k_\phi$ in the  
numerator than in the denominator of $Q_1^\phi$.)  Thus we have,
\be
A_n^{L-s}(\phi,j_\f^+) \longrightarrow
A_n^{L-s}(j_\f^+)
\qquad\quad \hbox{as $k_\phi \to 0$,}
\label{softphi1Ls}
\ee
in agreement with \eqn{softphi1} for $n_{-}=1$.

The recursion relation~(\ref{phiLsRecursion}),
and thus its solution~(\ref{phiLs}), is strictly valid only 
for $j \leq n-2$.  However, the cases $j = n-1$
and $j = n$ can be extracted from amplitudes with 
a larger number of legs $n$ and the same value of $j$, 
via appropriate collinear limits. In the limit that
gluons $n-1$ and $n$ become collinear in the amplitude
$A_n^{L-s}(\phi,(n-2)_f^+)$, we obtain the amplitude
$A_{n-1}^{L-s}(\phi,(n-2)_f^+)$ via \eqn{DeltaLimitI}.
Similarly, taking the fermion $n-1$ and gluon $n$
to be collinear in the amplitude
$A_n^{L-s}(\phi,(n-1)_f^+)$ gives the amplitude
$A_{n-1}^{L-s}(\phi,(n-1)_f^+)$ according to \eqn{DeltaLimitII}.
Taking these limits, we find
that \eqn{phiLs} continues to be valid for $j > n-2$. That is,
$Q_2^\phi$ vanishes for $j = n-1$, and
$Q_1^\phi$ and $Q_2^\phi$ vanish for $j = n$,
as can be seen from the limits
of the sums.


\subsection{Verification of solution for $A_n^{L-s}(\phi,j_\f^+)$}
\label{SolVerificationSubSection}

In this subsection we demonstrate that the set of amplitudes
$A_n^{L-s}(\phi,j_\f^+)$ satisfies the recursion
relation~(\ref{phiLsRecursion}).
The analysis parallels section V\,D of ref.~\cite{LastFinite},
which verified the solution 
$A_n^s(j_\f^+) \equiv A_n^s(1_\f^-,2^+,\ldots,j_\f^+,\ldots,n^+)$ 
in \eqn{QCDs} for the
scalar-loop contributions to the pure-QCD finite quark amplitudes.
However, the present analysis is simpler, because the all-plus gluon 
amplitudes~(\ref{allplusphi}) with a single $\phi$ present 
are simpler than those in pure QCD.
The first term on the right-hand side of the recursion
relation~(\ref{phiLsRecursion}), shown in~\fig{LsRecursionFigure}, 
includes the three-point tree amplitude
$A_3^{\tree}(-\Kh_{n-1,n}^-,(n-1)^+,\hat n^+)$.
Let $\hat{Q}_i^\phi$, $i=0,1,2$, stand for
the shifted versions of $Q_i^\phi$ for the appropriate
$(n-1)$-point one-loop quark amplitude,
$A_{n-1}^{L-s}(\phi,\hat 1_\f^-,2^+,
\ldots,j_\f^+,\ldots,(n-2)^+,\Kh_{n-1, n}^+)$.
The first term in \eqn{phiLsRecursion} can be simplified to
\begin{eqnarray}
&&
- {i\over2} { \hat{Q}_0^\phi + \hat{Q}_1^\phi + \hat{Q}_2^\phi
\over \spash{\hat K_{n-1,n}}.{1} \spa1.2 \spa2.3 \cdots
          \spa{(n-3)}.{(n-2)} \spash{(n-2)}.{\hat{K}_{n-1,n}} }
{1\over K_{n-1,n}^2}
\nonumber\\
&&\hskip2cm 
\times  { {\spb{(n-1)}.{n}}^3
   \over \spbsh{n}.{\hat K_{n-1,n}} \spbsh{\hat{K}_{n-1,n}}.{(n-1)} }
\nonumber \\
&=& {i\over2} { \hat{Q}_0^\phi + \hat{Q}_1^\phi + \hat{Q}_2^\phi
    \over \spa1.2 \spa2.3 \cdots \spa{n}.1 }
{ \spa{(n-2)}.{(n-1)} \spa{(n-1)}.{n} \spa{n}.1
   {\spb{(n-1)}.{n}}^3
   \over \spa{(n-1)}.{n} \spb{(n-1)}.{n}
   \spab1.{\Ksl_{n-1,n}}.{{(n-1)}} \spab{(n-2)}.{\Ksl_{n-1,n}}.{n} }
\nonumber\\
&=&  {i\over2} { \hat{Q}_0^\phi + \hat{Q}_1^\phi + \hat{Q}_2^\phi
        \over \spa1.2 \spa2.3 \cdots \spa{n}.1 } \,.
\label{PrefactorBehavior}
\end{eqnarray}
Thus the correct spinor denominator factor is reproduced.

Next we need to determine how the quantities $\hat{Q}_i^\phi$
are affected by the shift~(\ref{n1shift}).
First consider $\hat{Q}_0^\phi$, as given in \eqn{phiLsQ0ALT}.
The $(n-1)$-point expression $\hat{Q}_0^\phi$ is a
single sum over $l$ containing $(n-3)$ terms, which is
one fewer than the number of terms in the $n$-point
expression $Q_0^\phi$ on the left-hand side of the recursion
relation~(\ref{phiLsRecursion}).
All terms except the last in $\hat{Q}_0^\phi$ behave simply
under the shift~(\ref{n1shift}) of $\tlambda_1$ and
$\lambda_n$.  They depend on $\lambda_1$ through $\langle 1^-|$
and $|1^+\rangle$, but are independent of $\tlambda_1$.
The dependence on $\lambda_n$ is solely via the factor
\begin{equation}
\langle 1^- | (\Ksl_{(l+1)\cdots (n-2)} + \hat{\Ksl}_{n-1,n}) \ldots =
\langle 1^- | \Ksl_{(l+1)\cdots n} \ldots \,,
\label{spec1}
\end{equation}
because the shift in $\Ksl_{n-1,n}$ is proportional to $\lambda_1$.
So each such term directly yields the corresponding
term in the $n$-point sum $Q_0^\phi$.

The missing last term in $Q_0^\phi$ (with $l=n-1$)
is provided by the unreal-pole term containing
$V_3^{L-s}(-\Kh_{n-1,n}^+,(n-1)^+,\hat n^+)$
in the recursion relation~(\ref{phiLsRecursion}).
This term can be simplified to,
\begin{eqnarray}
&&
{i\over2}
{ {\spash{1}.{\hat K_{n-1,n}}}^2 \spash{j}.{\hat{K}_{n-1,n}}
   \spbsh{\hat{K}_{n-1,n}}.{(n-1)} \spb{(n-1)}.{n} \spbsh{n}.{\hat K_ 
{n-1,n}}
   \over \spa1.2 \spa2.3 \cdots \spa{(n-3)}.{(n-2)}
   \spash{(n-2)}.{\hat{K}_{n-1,n}} }
\, {1\over s_{n-1,n}}
\nonumber \\
&& \hskip7mm
\times
{ \spa{j}.{1} \spb{n}.{(n-1)}
      \over \spash{j}.{\hat{K}_{n-1,n}} \spash{\hat K_{n-1,n}}.{1}
       \spbsh{n}.{\hat K_{n-1,n}} \spbsh{\hat{K}_{n-1,n}}.{(n-1)}  }
\nonumber\\
&=&
{i\over2} {
\spa{1}.{j} \spab1.{\Ksl_{n-1,n}}.{n} \spb{(n-1)}.{n}
\over \spa1.2\spa2.3\cdots \spa{(n-3)}.{(n-2)}
   \ \spa{(n-1)}.{n} \spab{(n-2)}.{\Ksl_{n-1,n}}.{n} }
\nonumber\\
&=&
-{i\over2} { \spa{1}.{j}
\spa{1}.{n} \spb{n}.{(n-1)} \spa{(n-1)}.{1}
\over \spa1.2\spa2.3\cdots \spa{n}.1 }
\nonumber\\
&=&
{i\over2} { \spa{1}.{j}
   \sandmp1.{\Ksl_{\phi,2\cdots(n-1)} \ksl_{n-1}}.1
\over \spa1.2\spa2.3\cdots \spa{n}.1 } \,,
\label{UnrealTermQ0}
\end{eqnarray}
which clearly agrees with the term with $l=n-1$ in the
expression~(\ref{phiLsQ0ALT}) for $Q_0^\phi$.

Note that the same algebra, with $k_\phi$ set to zero,
demonstrates that the pure-QCD amplitude $A_n^{L-s}(j_f^+)$
in \eqn{QCDLs} solves the corresponding recursion
relation~(\ref{LsRecursion}).  Compared to the single-$\phi$
recursion relation~(\ref{phiLsRecursion}), the pure-QCD
relation~(\ref{LsRecursion}) lacks the set of terms containing
the amplitudes for a $\phi$ particle plus all positive-helicity gluons,
$A_{n-l+2}^{[g]}(\phi,-\Kh_{\phi,l\cdots n}^+,l^+,\ldots,\hat n^+)$.
Thus these latter terms should serve as ``sources'' for the
$Q_1^\phi$ and $Q_2^\phi$ parts of the $A_n^{L-s}(\phi,j_f^+)$
solution.

Let us now turn to $Q_2^\phi$, as given in \eqn{phiLsQ2ALT}.
We split the terms in $Q_2^\phi$ into those
with $p<n-1$ and those with $p=n-1$.  The terms
with $p<n-1$ come from the $(n-1)$-point $\hat{Q}_2^\phi$
contribution in \eqn{PrefactorBehavior}.  To show this,
we observe that, just as for the $\hat{Q}_0^\phi$
terms, $\tlambda_1$ never appears in $\hat{Q}_2^\phi$.
Also, $k_n$ only appears in \eqn{phiLsQ2ALT} for $Q_2^\phi$ via
$\langle 1^- | \Ksl_{(p+1)\cdots n} \ldots$.
Thus we may use \eqn{spec1} once again (with $l$ replaced by $p$),
in order to see that every term with $p<n-1$ in $Q_2^\phi$
is generated directly from the corresponding term in $\hat{Q}_2^\phi$.
Similarly, every term in $Q_1^\phi$ with $l<n$ comes
from the corresponding term in $\hat{Q}_1^\phi$.

The $Q_2^\phi$ terms with $p=n-1$, and the $Q_1^\phi$ term with $l=n$,
originate instead from the terms containing
the amplitudes for a $\phi$ particle plus all positive-helicity gluons,
$A_{n-l+2}^{[g]}(\phi,-\Kh_{\phi,l\cdots n}^+,l^+,\ldots,\hat n^+)$,
in \eqn{phiLsRecursion}.
The $l^{\rm th}$ term in $Q_2^\phi$ comes from the $l^{\rm th}$ term in
the recursion relation, for $l<n$.  For $l=n$, it gives the $l=n$ term
in $Q_1^\phi$.

To write the $l^{\rm th}$ term in the proper form, we use the
following identities, for $l<n$,
\begin{eqnarray}
\spa{n}.{1} \, s_{\phi,l\cdots n} &=&
\sandmp1.{\Ksl_{2\cdots(l-1)} \Ksl_{\phi,l\cdots(n-1)}}.{n} \,,
\label{moreid1}\\
\spab{\hat{n}}.{\Ksl_{\phi,l\cdots n}}.{n} &=&
\spab{n}.{\Ksl_{\phi,l\cdots n}}.{n}
- { K^2_{\phi,l\cdots n} \over \spab1.{\Ksl_{\phi,l\cdots n}}.{n} }
   \spab1.{\Ksl_{\phi,l\cdots n}}.{n}
\nonumber\\
&=& - s_{\phi,l\cdots(n-1)} \,,
\label{moreid2}\\
\spa{(n-1)}.{\hat{n}} &=&
\spa{(n-1)}.{n}
- { K^2_{\phi,l\cdots n} \over \spab1.{\Ksl_{\phi,l\cdots n}}.{n} }
   \spa{(n-1)}.{1}
\nonumber\\
&=& - { \sandmp1.{\Ksl_{2\cdots(l-1)} \Ksl_{\phi,l\cdots(n-1)}}. 
{{(n-1)}}
         \over \spab1.{\Ksl_{\phi,l\cdots n}}.{n} } \,.
\label{moreid3}
\end{eqnarray}
Then the $l^{\rm th}$ term in the recursion
relation~(\ref{phiLsRecursion}) becomes,
\begin{eqnarray}
&&
2i \,
{ {\spash1.{\hat K_{\phi,l\cdots n}}}^2 \spash{j}.{\hat{K}_{\phi,l 
\cdots n}}
   \over \spa1.2 \spa2.3 \cdots \spa{(l-2)}.{(l-1)}
    \spash{(l-1)}.{\hat{K}_{\phi,l\cdots n}} }
{1 \over s_{\phi,l\cdots n}}
\nonumber\\
&& \hskip1cm
\null \times
{ (k_\phi^2)^2
   \over \spash{\hat{K}_{\phi,l\cdots n}}.{l} \spa{l}.{(l+1)} \cdots
    \spa{(n-1)}.{\hat{n}} \spash{\hat{n}}.{\hat{K}_{\phi,l\cdots n}} }
\nonumber\\
&=& - {2i \over \spa1.2 \spa2.3 \cdots \spa{n}.1}
{ \spa{(l-1)}.{l} \spa{(n-1)}.{n} \spa{n}.1
   {\spab1.{\Ksl_{\phi,l\cdots n}}.{n}}^2
    \spab{j}.{\Ksl_{\phi,l\cdots n}}.{n}
   \over \spba{n}.{\Ksl_{\phi,l\cdots n}}.{{(l-1)}}
         \spba{n}.{\Ksl_{\phi,l\cdots n}}.{{l}} }
\nonumber\\
&& \hskip1cm
\null \times
{ \spa{n}.1
    \over \sandmp1.{\Ksl_{2\cdots(l-1)} \Ksl_{\phi,l\cdots(n-1)}}.{n} }
{ \spab1.{\Ksl_{\phi,l\cdots n}}.{n}
\over \sandmp1.{\Ksl_{2\cdots(l-1)} \Ksl_{\phi,l\cdots(n-1)}}. 
{{(n-1)}} }
\ { (k_\phi^2)^2 \over s_{\phi,l\cdots(n-1)} }
\nonumber\\
&=& - {2i \over \spa1.2 \spa2.3 \cdots \spa{n}.1}
{ \spa{(l-1)}.{l}
    \over \sandmp1.{\ksl_n \Ksl_{\phi,l\cdots (n-1)}}.{(l-1)}
          \sandmp1.{\ksl_n \Ksl_{\phi,l\cdots (n-1)}}.{l} }
\nonumber \\
&& \hskip1cm \null \times
{ \spa{(n-1)}.{n}
    \over \sandmp1.{\Ksl_{2\cdots (l-1)} \Ksl_{\phi,l\cdots (n-1)}}. 
{{(n-1)}}
          \sandmp1.{\Ksl_{2\cdots (l-1)} \Ksl_{\phi,l\cdots (n-1)}}. 
{n} }
\nonumber \\
&& \hskip1cm \null \times
    {\sandmp1.{\ksl_n \Ksl_{\phi,l\cdots(n-1)}}.1}^3
     \sandmp{1}.{\ksl_n \Ksl_{\phi,l\cdots(n-1)}}.{j}
\ { (k_\phi^2)^2 \over s_{\phi,l\cdots (n-1)} } \,.
\label{Finallthterm}
\end{eqnarray}
The final form is just the $l^{\rm th}$ term with $p=n-1$ in
formula (\ref{phiLsQ2ALT}) for $Q_2^\phi$.

The case $l=n$ is a little different.
The relevant identities~(\ref{moreid1})--(\ref{moreid3})
are slightly modified, to
\begin{eqnarray}
\spa{n}.{1} \, s_{\phi,n} &=&
\sandmp1.{\Ksl_{2\cdots(n-1)} \ksl_\phi}.{n} \,,
\label{lnmoreid1}\\
\spab{\hat{n}}.{\Ksl_{\phi,n}}.{n} &=&
\spab{n}.{\ksl_\phi}.{n}
- { K^2_{\phi,n} \over \spab1.{\ksl_\phi}.{n} }
   \spab1.{\ksl_\phi}.{n}
\nonumber\\
&=& - k_\phi^2 \,.
\label{lnmoreid2}
\end{eqnarray}
We simplify the $l^{\rm th}$ term in the recursion
relation~(\ref{phiLsRecursion}), for $l=n$, as follows,
\begin{eqnarray}
&&
- 2i \,
{ {\spash1.{\hat K_{\phi,n}}}^2 \spash{j}.{\hat{K}_{\phi,n}}
   \over \spa1.2 \spa2.3 \cdots \spa{(n-2)}.{(n-1)}
    \spash{(n-1)}.{\hat{K}_{\phi,n}} }
\, {1 \over s_{\phi,n}} \, { (k_\phi^2)^2
   \over { \spash{\hat{n}}.{\hat{K}_{\phi,n}} }^2 }
\nonumber\\
&=& - {2i \over \spa1.2 \spa2.3 \cdots \spa{n}.1}
{ \spa{(n-1)}.{n} \spa{n}.{1}
   {\spab1.{\ksl_\phi}.{n}}^2
    \spab{j}.{\ksl_\phi}.{n}
   \over \spba{n}.{\ksl_\phi}.{{(n-1)}} }
\nonumber\\
&&\hskip1cm
\times
     { \spa{n}.{1}
     \over \sandmp1.{\Ksl_{2\cdots(n-1)} \ksl_\phi}.{n} }
   { (k_\phi^2)^2 \over (k_\phi^2)^2 }
\nonumber\\
&=& - {2i \over \spa1.2 \spa2.3 \cdots \spa{n}.1}
     { \spa{(n-1)}.{n}
     { \sandmp1.{\ksl_n \ksl_\phi}.1 }^2
       \sandmp1.{\ksl_n \ksl_\phi}.{j}
   \over \sandmp1.{\ksl_n \ksl_\phi}.{{(n-1)}}
    \sandmp1.{\Ksl_{2\cdots(n-1)} \ksl_\phi}.{n} } \,.
\label{leqnterm}
\end{eqnarray}
The final form is identical to the term with $l=n$ in
formula~(\ref{phiLsQ1ALT}) for $Q_1^\phi$.
We have now accounted for all the terms in
$Q_1^\phi+Q_2^\phi$ that were not present in
$\hat Q_1^\phi + \hat Q_2^\phi$,
thus completing the proof that \eqn{phiLs} obeys
the recursion relation~(\ref{phiLsRecursion}).

\subsection{Recursion relation and solution for $A_n^s$}

The recursion relation for the pure-QCD primitive amplitude
$A_n^s(j_\f^+)$ reads~\cite{LastFinite},
\begin{eqnarray}
&&\null \hskip -5 mm  A_n^s(j_\f^+) = \nonumber \\
&&
A_{n-1}^{s}(\hat 1_\f^-,2^+,\ldots,j_\f^+,\ldots,(n-2)^+,\Kh_{n-1, n}^+)
\, {i\over s_{n-1,n}} \,
 A_3^{\tree}(-\Kh_{n-1,n}^-,(n-1)^+,\hat n^+)
\nonumber\\
&&\null +
A_{n-1}^\tree(\hat 1_\f^-,2^+,\ldots,j_\f^+,\ldots,(n-2)^+,\Kh_{n-1,n}^-) 
\, {i\over s_{n-1,n}^2} \,
 V_3^{\oneloop}(-\Kh_{n-1,n}^+,(n-1)^+,\hat n^+)
\nonumber\\ &&\hskip 5mm \null\times
\Bigl(1 + s_{n-1,n}\, \Soft^\tree(n-2, \Kh_{n-1,n}^+,j) \,
\Soft^\tree(\hat n, -\Kh_{n-1,n}^-, n-1) \Bigr)
\nonumber\\
&&\null + \sum_{l=j+1}^{n-2}
A_{l}^\tree(\hat{1}_\f^-,2^+,\ldots,j_\f^+,\ldots,(l-1)^+,\Kh_{l\cdots n}^-)
\, {i\over s_{l\cdots n}} \, 
A_{n-l+2}^{\oneloop}(-\Kh_{l\ldots n}^+,l^+,\ldots,\hat n^+)
\,.
\nonumber\\
~~~
\label{sRecursion}
\end{eqnarray}
The hatted legs undergo the shift in \eqn{n1shift}, where
in the $s_{l \ldots n}$ channel we set $z$ to the value
\begin{equation}
z = - {s_{l \cdots n} \over \sand1.{\Ksl_{l \cdots n}}.n} \,.
\label{zlnChannelnophi}
\end{equation}
%

%
\begin{figure}[t]
\centerline{\epsfxsize 6. truein \epsfbox{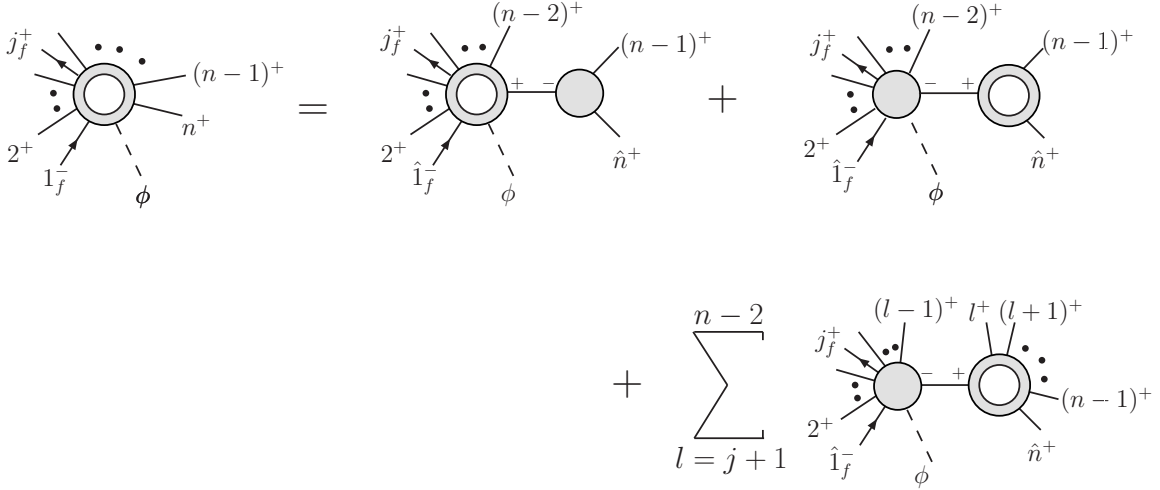}}
\caption{Diagrams corresponding to the terms in the recursion
relation~(\ref{phisRecursion}) for $A_n^s(\phi,j_\f^+)$.
The second diagram contains a double pole,
and an unreal pole underneath it.
In the last diagram, the sum over
$l$ runs from $j+1$ to $n-2$, inclusive.}
\label{sRecursionFigure}
\end{figure}

The recursion relation for the corresponding single-$\phi$ amplitudes,
$A_n^s(\phi,j_\f^+)$, depicted in \fig{sRecursionFigure},
is very similar,
\begin{eqnarray}
&&\null \hskip -5 mm  A_n^s(\phi,j_\f^+) = \nonumber \\
&& 
A_{n-1}^{s}(\phi, \hat 1_\f^-,2^+,\ldots,j_\f^+,\ldots,(n-2)^+,\Kh_{n-1,n}^+)
\, {i\over s_{n-1,n}} \,
 A_3^{\tree}(-\Kh_{n-1,n}^-,(n-1)^+,\hat n^+)
\nonumber\\
&&\null +
A_{n-1}^\tree(\phi,\hat 1_\f^-,2^+,
               \ldots,j_\f^+,\ldots,(n-2)^+,\Kh_{n-1, n}^-)
\, {i\over s_{n-1,n}^2} \,
 V_3^{\oneloop}(-\Kh_{n-1,n}^+,(n-1)^+,\hat n^+)
\nonumber\\ &&\hskip 5mm \null\times
\Bigl(1 + s_{n-1,n}\, \Soft^\tree(n-2, \Kh_{n-1,n}^+,j) \,
\Soft^\tree(\hat n, -\Kh_{n-1,n}^-, n-1) \Bigr)
\nonumber\\
&&\null + \sum_{l=j+1}^{n-2}
A_{l}^\tree(\phi,\hat{1}_\f^-,2^+,
             \ldots,j_\f^+,\ldots,(l-1)^+,\Kh_{l\cdots n}^-)
\, {i\over s_{l\cdots n}} \,
A_{n-l+2}^{\oneloop}(-\Kh_{l\cdots n}^+,l^+,\ldots,\hat n^+)
\,.
\nonumber\\
~~~
\label{phisRecursion}
\end{eqnarray}
Notice that in this case $\phi$
only makes its appearance in the ``source terms'' ---
the second and third terms in \fig{sRecursionFigure} ---
through tree amplitudes.
Recall also that the relevant $\phi$-containing MHV tree
amplitudes~(\ref{phiffmhvtree}) are identical in form to
the ones without $\phi$, \eqn{ffmhvtree}.
Thus we should expect the solution to the recursion relation
to have essentially the same form as in the pure-QCD case.

Indeed, the solution to \eqn{phisRecursion} is
\begin{equation}
A_n^{s}(\phi;j_\f^+) = {i\over3}
   {S_1^\phi + S_2^\phi \over \spa1.2\spa2.3\cdots \spa{n}.{1} } \,,
\label{phis}
\end{equation}
where
\begin{eqnarray}
S_1^\phi &=& \sum_{l=j+1}^{n-1}
   { \spa{j}.{l} \spa{1}.{(l+1)}
     \sandmp1.{\Ksl_{l,l+1} \Ksl_{(l+1)\cdots n}}.1
    \over \spa{l}.{(l+1)} } \,,
\label{phisS1} \\
S_2^\phi &=& - \sum_{l=j+1}^{n-2} \sum_{p=l+1}^{n-1}
{ \spa{(l-1)}.{l}
    \over \sandmp1.{\Ksl_{(p+1)\cdots n} \Ksl_{l\cdots p}}.{(l-1)}
          \sandmp1.{\Ksl_{(p+1)\cdots n} \Ksl_{l\cdots p}}.{l} }
\nonumber \\
&& \hskip15mm\times
{ \spa{p}.{(p+1)}
    \over \sandmp1.{\Ksl_{l\cdots n} \Ksl_{l\cdots p}}.{p}
          \sandmp1.{\Ksl_{l\cdots n} \Ksl_{l\cdots p}}.{(p+1)} }
\nonumber \\
&& \hskip15mm\times
    {\sandmp1.{\Ksl_{l\cdots p} \Ksl_{(p+1)\cdots n}}.1}^2
        \sandmp{j}.{\Ksl_{l\cdots p} \Ksl_{(p+1)\cdots n}}.1
\nonumber \\
&& \hskip15mm\times
    { \sandmp1.{\Ksl_{l\cdots n} [ {\cal F}(l,p) ]^2  \Ksl_{(p+1) 
\cdots n}}.1
     \over s_{l\cdots p} }
\,.
\label{phisS2}
\end{eqnarray}

The $S_1^\phi$ term manifestly has the same form as the pure-QCD
result $S_1$ in \eqn{QCDsS1}, apart from the implicit $\phi$ momentum.
For the $S_2^\phi$ term, we use momentum conservation to substitute
\be
\Ksl_{2\cdots (l-1)} = - \Ksl_{l\cdots n} \,,
\label{MomConsln}
\ee
in the expression for $S_2$ in \eqn{QCDsS2}.
Then the $S_2^\phi$ term is identical in form to $S_2$ as well.

Because of the simple relation between $S_i^\phi$ and $S_i$,
it is trivial to show that
the solution~(\ref{phis}) for $A_n^{s}(\phi,j_\f^+)$,
like the solution~(\ref{phiLs}) for $A_n^{L-s}(\phi,j_\f^+)$,
obeys the expected soft-Higgs limit~(\ref{softphi1}) with $n_{-}=1$,
\be
A_n^{s}(\phi,j_\f^+) \longrightarrow
A_n^{s}(j_\f^+)
\qquad\quad \hbox{as $k_\phi \to 0$.}
\label{softphi1s}
\ee

The simple relation between $S_i^\phi$ and $S_i$
also makes it very easy to demonstrate that the set of amplitudes
$A_n^s(\phi,j_\f^+)$ given in \eqn{phis} satisfy the recursion
relation~(\ref{phisRecursion}).  We just use  the analysis
in section V\,D of ref.~\cite{LastFinite}, which proved the analogous
result for QCD, namely that $A_n^s(j_\f^+)$ as given in \eqn{QCDs}
satisfies the recursion relation~(\ref{sRecursion}).
The only modification that needs to be made to the arguments and
identities is that $\Ksl_{2\cdots (l-1)}$ should be replaced by
$- \Ksl_{l\cdots n}$ everywhere, according to \eqn{MomConsln}.
After doing that, all expressions appearing in the two recursion
relations and solutions, for $A_n^s$ with $\phi$ and without,
have exactly the same form.  The momentum of $\phi$ only appears
implicitly in $A_n(\phi,j_\f^+)$, through momentum conservation.
In addition, after replacing $\Ksl_{2\cdots (l-1)}$, none of
the identities used in ref.~\cite{LastFinite} require momentum
conservation (which could have introduced the $\phi$ momentum).
In the analysis of the contribution of the $l^{\rm th}$ term in
the last diagram of \fig{sRecursionFigure},
for example, only legs $l$ to $n$ are
involved in an essential way.

\subsection{Factorization properties of solution, 
and a $\phi$-plus-$n$-gluon amplitude}
\label{FactorSolutionSubSection}

In appendix A of ref.~\cite{LastFinite} it was shown that
the pure-QCD formula~(\ref{QCDs}) has all
the correct multiparticle poles, factorizing properly onto products
of the quark-containing MHV tree amplitudes~(\ref{ffmhvtree}) and the
one-loop all-plus pure-gluon amplitudes~(\ref{OneLoopAllPlusAmplitude}).
The corresponding demonstration for the $\phi$ amplitude 
$A_n^s(\phi,j_\f^+)$ is completely analogous, so we do not present it here.
The collinear singularities of \eqns{phiLs}{phis} work in exactly the  
same way;
there are no collinear singularities between the $\phi$ particle
and another parton, only between pairs of partons, and the corresponding
$(n-1)$-point amplitudes again have exactly the same form as in the
pure-QCD case.  By taking the anti-quark and quark to be collinear
in the amplitude in which they are adjacent ($j=2$), we can extract
the scalar and $[g-s]$ loop contributions to the one-loop amplitudes for
$\phi g^\pm g^+\ldots g^+$.

A term containing $1/\spb{1}.{2}$, corresponding
to the splitting amplitude $\Split_{-}^\tree(1_f^-,2_f^+;z)$,
indicates a factorization onto the all-plus amplitude, 
$\phi g^+g^+\ldots g^+$.  We observe that \eqn{phis}
for $A_n^s(\phi,2_f^+)$
does not contain a factor of $1/\spb{1}.{2}$, thus confirming
the observation in \sect{AllPlusAmplitudesSection} that
the scalar $\phi$-amplitude with gluons of positive helicity
vanishes, \eqn{allplusnfnsvanish}. Similarly,
in this collinear limit, $A_n^{L-s}(\phi,2_f^+)$ as given 
in~\eqn{phiLs} factorizes onto the
gluonic $\phi$-all-plus amplitudes, \eqn{allplusphi}.
The relevant contribution comes from the term with $l=3$
and $p = n$ of $Q_2^\phi$, as given in~\eqn{phiLsQ2}.

Analogously, we can extract the scalar and $[g-s]$
contributions to the amplitude for a $\phi$ plus
$n$ gluons, one of which has negative helicity,
from \eqns{phis}{phiLs}, respectively.
The result for the scalar contribution is,
\begin{equation}
A_{n}^{[s]}(\phi,1^-,2^+,3^+,\ldots,n^+) = {i\over3}
   {T_1^\phi + T_2^\phi \over \spa1.2\spa2.3\cdots \spa{n}.{1} } \,,
\label{phioneminus}
\end{equation}
where
\begin{eqnarray}
T_1^\phi &=& \sum_{l=2}^{n-1}
   { \spa{1}.{l} \spa{1}.{(l+1)}
     \sandmp1.{\Ksl_{l,l+1} \Ksl_{(l+1)\cdots n}}.1
    \over \spa{l}.{(l+1)} } \,,
\label{phioneminusT1} \\
T_2^\phi &=& - \, \sum_{l=2}^{n-2} \sum_{p=l+1}^{n-1}
{ \spa{(l-1)}.{l}
    \over \sandmp1.{\Ksl_{(p+1)\cdots n} \Ksl_{l\cdots p}}.{(l-1)}
          \sandmp1.{\Ksl_{(p+1)\cdots n} \Ksl_{l\cdots p}}.{l} }
\nonumber \\
&& \hskip15mm\times
{ \spa{p}.{(p+1)}
    \over \sandmp1.{\Ksl_{l\cdots n} \Ksl_{l\cdots p}}.{p}
          \sandmp1.{\Ksl_{l\cdots n} \Ksl_{l\cdots p}}.{(p+1)} }
\nonumber \\
&& \hskip15mm\times
    {\sandmp1.{\Ksl_{l\cdots p} \Ksl_{(p+1)\cdots n}}.1}^3
\nonumber \\
&& \hskip15mm\times
    { \sandmp1.{\Ksl_{l\cdots n} [ {\cal F}(l,p) ]^2  \Ksl_{(p+1) 
\cdots n}}.1
     \over s_{l\cdots p} }
\,.
\label{phioneminusT2}
\end{eqnarray}
Similarly, from \eqn{phiLs} we obtain for the $[g-s]$ contribution,
\begin{equation}
A_n^{[g-s]}(\phi,1^-,2^+,\ldots,n^+) = {i \over 2}
{ R_1^\phi + R_2^\phi
\over \spa1.2\spa2.3\cdots\spa{n}.1 }
\,,
\label{phigsoneminus}
\end{equation}
where
\begin{eqnarray}
R_1^\phi &=& - 4
\sum_{l=3}^{n}
{ \spa{(l-1)}.{l}\,
{\sandmp1.{\Ksl_{l\cdots n} \ksl_\phi}.1}^3
\over \sandmp1.{\Ksl_{l\cdots n} \ksl_\phi}.{(l-1)}
        \sandmp1.{\Ksl_{2\cdots (l-1)} \ksl_\phi}.{l} }
\,,
\label{phigsR1}\\
R_2^\phi &=& - 4
\sum_{l=3}^{n-1} \sum_{p=l}^{n-1}
{ \spa{(l-1)}.{l}
\over \sandmp1.{\Ksl_{(p+1)\cdots n} \Ksl_{\phi,l\cdots p}}.{(l-1)}
        \sandmp1.{\Ksl_{(p+1)\cdots n} \Ksl_{\phi,l\cdots p}}.{l} }
\nonumber\\
&&\hskip 6mm \null
   \times
{ \spa{p}.{(p+1)}
\over \sandmp1.{\Ksl_{2\cdots (l-1)} \Ksl_{\phi,l\cdots p}}.{p}
        \sandmp1.{\Ksl_{2\cdots (l-1)} \Ksl_{\phi,l\cdots p}}.{(p+1)} }
\nonumber\\
&&\hskip 6mm \null
\times { \sandmp1.{\Ksl_{(p+1)\cdots n} \Ksl_{\phi,l\cdots p}}.1 }^4
\ { (k_\phi^2)^2 \over s_{\phi,l\cdots p} }
%
\,.
\label{phigsR2}
\end{eqnarray}
As indicated by \eqn{OneloopPhiFiniteDecomp}, and as explained in
\sect{AllPlusAmplitudesSection}, the amplitude with a fermion in the loop,
$A_{n}^{[f]}(\phi,1^-,2^+,3^+,\ldots,n^+)$, is equal and opposite in sign
to the scalar-loop contribution 
$A_{n}^{[s]}(\phi,1^-,2^+,3^+,\ldots,n^+)$ given in \eqn{phioneminus}.
The soft Higgs limit of \eqns{phioneminus}{phigsoneminus} onto
the corresponding pure-QCD amplitudes~(\ref{oneminus}) can
easily be verified to obey~\eqn{softphi1}, just as in the quark case.

\Eqn{phigsoneminus} completes the construction of all the finite 
one-loop QCD amplitudes containing a $\phi$ field plus multiple partons.

\section{Results for two, three and four partons}
\label{TwoThreeFourSection}

In this section we collect explicit analytical results for up
to four partons, using formul\ae{} from the previous section,
as well as some results for divergent helicity configurations
which have appeared elsewhere, in particular in 
refs.~\cite{Schmidt,BadgerGloverH4g}.

\subsection{Complete results for two and three partons}

In \app{NormalizationAppendix},
eqs.~(\ref{phiPP}), (\ref{phiMP}), and (\ref{phiMM}),
we record the complete set of $\phi gg$ amplitudes,
$\phi g^+g^+$, $\phi g^\mp g^\pm$, and $\phi g^- g^-$.
The $\phi\bar{q}q$ amplitudes vanish (for massless quarks)
by angular momentum conservation, so, rather trivially,
we have the complete set of amplitudes for a single $\phi$
plus 2 partons.

The purpose of this subsection is to give the complete
set of amplitudes for a single $\phi$ plus 3 partons.
We start with the $\phi ggg$ amplitudes.
In the color decomposition~(\ref{AdjointColorDecomposition}),
there is only one color structure, corresponding to the partial
amplitude $A_{3;1}$.
First we record the finite helicity configurations,
using our all-$n$ results.  Then, using these formul\ae{} and
\eqn{Hreconstruct}, we decompose
the one-loop $Hggg$-amplitudes computed in ref.~\cite{Schmidt}
into $\phi ggg$ and $\phi^\dagger ggg$ amplitudes.  Finally we
use parity to recast the results just in terms of $\phi ggg$
amplitudes.  These amplitudes will play a role in the recursive
construction of one-loop Higgs amplitudes with more external partons.

We express the one-loop results in terms of the corresponding
$Hggg$ or $\phi^\dagger ggg$ tree-level amplitudes,
\bea
A_3^\tree(H,1^+,2^+,3^+)
= A_3^\tree(\phi^\dagger,1^+,2^+,3^+) & = & i \frac{ (k_\phi^2)^2}
{\spa{1}.{2} \spa{2}.{3} \spa{3}.{1}} \, ,\\
A_3^\tree(H,1^-,2^+,3^+)
= A_3^\tree(\phi^\dagger,1^-,2^+,3^+)  & = & - i
\frac{\spb{2}.{3}^3}{\spb{1}.{2} \spb{3}.{1}} \, .
\eea
These amplitudes can be read off from eqs.~(\ref{phivanishtree}),
(\ref{phimhvtree}), and (\ref{phiantimhvtree}),
with the help of eqs.~(\ref{Hreconstruct}) and (\ref{ParityExch}).
The $[s]$ and $[g-s]$ components of the finite loop amplitudes are,
using eqs.~(\ref{allplusnfnsvanish}), (\ref{allplusphi}), 
(\ref{phioneminus}), and
(\ref{phigsoneminus}),
\bea
A_3^{[s]}(\phi,1^+,2^+,3^+) & = & 0 \, ,
\label{A3sppp}\\
A_3^{[g-s]}(\phi,1^+,2^+,3^+)
& = & - 2 \,A_3^\tree(\phi^\dagger,1^+,2^+,3^+) \, ,
\label{A3gminussppp}\\
A_3^{[s]}(\phi,1^-,2^+,3^+)
& = & \frac{1}{3} \frac{s_{31} s_{12}}{s_{23}^2}
    A_3^\tree(\phi^\dagger,1^-,2^+,3^+) \, ,
\label{A3smpp}\\
A_3^{[g-s]}(\phi,1^-,2^+,3^+)
& = & - 2 A_3^\tree(\phi^\dagger,1^-,2^+,3^+) \, .
\label{A3gminussmpp}
\eea
Using \eqn{OneloopPhiFiniteDecomp}, the full $\phi ggg$ one-loop
amplitudes are,
\bea
A_{3;1}(\phi,1^+,2^+,3^+) & = & - 2 \, A_3^\tree(\phi^\dagger,1^+,2^+, 
3^+) \, ,
\label{phiallplus3} \\
A_{3;1}(\phi,1^-,2^+,3^+) & = & A_3^\tree(\phi^\dagger,1^-,2^+,3^+)
\left[ - 2 + \frac{1}{3} \left(1 - \frac{n_\f}{N_c} + \frac{n_s}{N_c} 
\right)
\frac{s_{31} s_{12}}{s_{23}^2} \right] \, . \label{phioneminus3}
\eea

The one-loop $Hggg$ amplitudes are~\cite{Schmidt},
in our notation,
\bea
A_{3;1}(H,1^+,2^+,3^+) & = & A_{3}^\tree(\phi^\dagger,1^+,2^+,3^+)
\Biggl[ U_3
+ {1\over3} \biggl( 1 - {n_\f \over N_c} + {n_s \over N_c } \biggr)
{ s_{12}s_{23} + s_{23}s_{31} + s_{31}s_{12} \over (k_\phi^2)^2 }
\Biggr]
\,,
\nonumber\\
\label{Hppp1l}\\
A_{3;1}(H,1^-,2^+,3^+) & = & A_{3}^\tree(\phi^\dagger,1^-,2^+,3^+)
\Biggl[ U_3
+ {1\over3} \biggl( 1 - {n_\f \over N_c} + {n_s \over N_c } \biggr)
{ s_{31}s_{12} \over s_{23}^2 }
\Biggr]
\,,
\label{Hmpp1l}
\eea
where
\bea
U_3 &\equiv&
- {1\over\e^2} \Biggl[ \biggl( {\mu^2 \over -s_{12}} \biggr)^\e
                      + \biggl( {\mu^2 \over -s_{23}} \biggr)^\e
                      + \biggl( {\mu^2 \over -s_{31}} \biggr)^\e \Biggr]
+ { \pi^2 \over 2 }
\nonumber \\
&&\hskip0cm
- \ln\biggl({-s_{12} \over -k_\phi^2} \biggr)
   \ln\biggl({-s_{23} \over -k_\phi^2} \biggr)
- \ln\biggl({-s_{23} \over -k_\phi^2} \biggr)
   \ln\biggl({-s_{31} \over -k_\phi^2} \biggr)
- \ln\biggl({-s_{31} \over -k_\phi^2} \biggr)
   \ln\biggl({-s_{12} \over -k_\phi^2} \biggr)
\nonumber \\
&&\hskip0cm
- 2 \, \Li_2\biggl(1 - {s_{12} \over k_\phi^2} \biggr)
- 2 \, \Li_2\biggl(1 - {s_{23} \over k_\phi^2} \biggr)
- 2 \, \Li_2\biggl(1 - {s_{31} \over k_\phi^2} \biggr)
\nonumber \\
& = &
- {1\over\e^2} \Biggl[ \biggl( {\mu^2 \over -s_{12}} \biggr)^\e
                      + \biggl( {\mu^2 \over -s_{23}} \biggr)^\e
                      + \biggl( {\mu^2 \over -s_{31}} \biggr)^\e \Biggr]
\nonumber \\
&&\hskip0cm
- \Ls_{-1}\biggl( {-s_{12}\over-k_\phi^2}, {-s_{23}\over-k_\phi^2} \biggr)
- \Ls_{-1}\biggl( {-s_{23}\over-k_\phi^2}, {-s_{31}\over-k_\phi^2} \biggr)
- \Ls_{-1}\biggl( {-s_{31}\over-k_\phi^2}, {-s_{12}\over-k_\phi^2} \biggr)
\, ,
\label{U3def}
\eea
with $k_\phi^2 = s_{123}$, and $\mu$ is the scale originating from dimensional
regularization.
The one-mass box function $\Ls_{-1}(r_1,r_2)$ is defined
as~\cite{eeFourPartons},
\be
\Ls_{-1}(r_1,r_2) \equiv
\Li_2(1-r_1) + \Li_2(1-r_2) + \ln r_1 \ln r_2 - { \pi^2 \over 6 }\,.
\label{Lsm1}
\ee
We have extended the results of ref.~\cite{Schmidt} to nonzero 
values of $n_s$ by observing that the $n_\f$ and $n_s$ terms
are proportional to an amplitude for an off-shell gluon to 
split, via a fermion or scalar loop, into two on-shell,
identical-helicity gluons.  (The opposite-helicity case vanishes.)
Such contributions are opposite in sign for fermions and scalars,
as can be seen by contracting the off-shell gluon with an external
$\bar{q}q$ pair, and then using a supersymmetry identity~\cite{SWI}.

Using \eqn{Hreconstruct}, and subtracting \eqn{phiallplus3}
from \eqn{Hppp1l}, and \eqn{phioneminus3} from \eqn{Hmpp1l},
we have,
\bea
A_{3;1}(\phi^\dagger,1^+,2^+,3^+)
& = & A_{3}^\tree(\phi^\dagger,1^+,2^+,3^+)
\nonumber\\
&&\hskip0.3cm
\times \Biggl[ U_3 + 2
+ {1\over3} \biggl( 1 - {n_\f \over N_c} + {n_s \over N_c } \biggr)
{ s_{12}s_{23} + s_{23}s_{31} + s_{31}s_{12} \over (k_\phi^2)^2 }
\Biggr]
\,,~~~~
\label{phidaggerppp1l}\\
A_{3;1}(\phi^\dagger,1^-,2^+,3^+)
& = & A_{3}^\tree(\phi^\dagger,1^-,2^+,3^+) [ U_3 + 2 ]
\,.
\label{phidaggermpp1l}
\eea
The parity conjugates of these results constitute the
remaining helicity amplitudes for $\phi$ plus three gluons,
\bea
A_{3;1}(\phi,1^-,2^-,3^+)
& = & A_{3}^\tree(\phi,1^-,2^-,3^+) [ U_3 + 2 ]
\,,
\label{phimmp1l}\\
A_{3;1}(\phi,1^-,2^-,3^-)
& = & A_{3}^\tree(\phi,1^-,2^-,3^-)
\nonumber\\
&&\hskip0.3cm
\times \Biggl[ U_3 + 2
+ {1\over3} \biggl( 1 - {n_\f \over N_c} + {n_s \over N_c } \biggr)
{ s_{12}s_{23} + s_{23}s_{31} + s_{31}s_{12} \over (k_\phi^2)^2 }
\Biggr]
\,.~~~~
\label{phimmm1l}
\eea
It is trivial to obtain the corresponding amplitudes for 
a pseudoscalar $A$ plus three gluons via \eqn{Areconstruct},
\bea
A_{3;1}(A,1^+,2^+,3^+) & = & i\,A_{3}^\tree(\phi^\dagger,1^+,2^+,3^+)
\nonumber\\
&&\hskip0.3cm  \times
\Biggl[ U_3 + 4
+ {1\over3} \, \biggl( 1 - {n_\f \over N_c} + {n_s \over N_c } \biggr)
{ s_{12}s_{23} + s_{23}s_{31} + s_{31}s_{12} \over (k_\phi^2)^2 }
\Biggr]
\,,~~~~~
\label{Appp1l}\\
A_{3;1}(A,1^-,2^+,3^+) & = & i\, A_{3}^\tree(\phi^\dagger,1^-,2^+,3^+)
\Biggl[ U_3 + 4
- {1\over3} \biggl( 1 - {n_\f \over N_c} + {n_s \over N_c } \biggr)
{ s_{31}s_{12} \over s_{23}^2 }
\Biggr]
\, .
\label{Ampp1l}
\eea

Let us now turn to the $\phi \bar{q} q g$ amplitudes. 
There is only one nonvanishing color structure 
in~\eqn{OneLoopColorDecomposition},
with partial amplitude $A_{3;1}$.  We again
express the one-loop results in terms of the corresponding
$H \bar{q} q g$, or equivalently, $\phi^\dagger \bar{q} q g$
tree amplitude,
\be
A_3^\tree(H,1_f^-,2_f^+,3^+) =
A_3^\tree(\phi^\dagger,1_f^-,2_f^+,3^+) =
- i \frac{\spb{2}.{3}^2}{\spb{1}.{2}} \, .
\ee
This amplitude can be obtained from eqs.~(\ref{phiffvanishtree})
and (\ref{phiffmhvtree}),
by using eqs.~(\ref{Hreconstruct}) and (\ref{ParityExch}). We find
from \eqn{phiLs},
\be
A_3^{L-s}(\phi,1_f^-,2_f^+,3^+)
= 2 \, i \, \frac{\spb{2}.{3}^2}{\spb{1}.{2}}
 + {i\over2} {\spa1.2\spb2.3\over \spa2.3} 
= - 2 A_3^\tree(\phi^\dagger,1_f^-,2_f^+,3^+)
  + {i\over2} {\spa1.2\spb2.3\over \spa2.3} \,.
\ee
The corresponding scalar amplitude $A_3^{s}(\phi;1_f^-,2_f^+,3^+)$ 
vanishes, but the $R$ type amplitude does {\it not},
$A_3^R(\phi,1_f^-,2_f^+,3^+) = - A_3^L(\phi,1_f^-,3^+,2_f^+)
= -(i/2)\spa1.2\spb2.3/\spa2.3$.
\Eqn{Anoneformula} then gives the full $\phi \bar{q}qg$
one-loop amplitude as,
\bea
A_{3;1}(\phi,1_f^-,2_f^+,3^+) &=&
- 2 A_3^\tree(\phi^\dagger,1_f^-,2_f^+,3^+)
 + {i\over2} \left( 1 + {1\over N_c^2} \right)
   {\spa1.2\spb2.3\over \spa2.3} 
\nonumber\\
 &=&
A_3^\tree(\phi^\dagger,1_f^-,2_f^+,3^+) \left[
  - \, 2 
  - \left( 1 + {1\over N_c^2} \right)  {s_{12}\over 2 \, s_{23}}
   \right]  \, .
\label{phiff3}
\eea

{}From the one-loop $H\bar{q}qg$ amplitude~\cite{Schmidt} 
and \eqn{phiff3} we can extract the
corresponding $\phi^\dagger$-contribution as above. 
The $H\bar{q}qg$ amplitude is
\be
A_{3;1}(H,1_f^-,2_f^+,3^+) =
A_3^\tree(\phi^\dagger,1_f^-,2_f^+,3^+)
\Biggl[ V_1 + {1\over N_c^2} V_2
+ {n_\f \over N_c} V_3 
+ {n_s \over N_c} V_4 \Biggr]\ ,
\label{quarkloop}
\ee
with
\bea
V_{1}&\equiv&
- {1\over\e^2}
\biggl[ \biggl(  {\mu^2 \over -s_{23}} \biggr)^\e
                      + \biggl( {\mu^2 \over -s_{31}} \biggr)^\e \biggr]
\,+\,{13\over6\epsilon}\biggl({ \mu^2 \over-s_{12}}\biggr)^\e
\nonumber\\
&&\hskip0cm
- \Ls_{-1}\biggl( {-s_{12}\over-k_\phi^2}, {-s_{23}\over-k_\phi^2} \biggr)
- \Ls_{-1}\biggl( {-s_{31}\over-k_\phi^2}, {-s_{12}\over-k_\phi^2} \biggr)
\,+\,{83\over18}\,-\,{\delta_{R}\over6}\,
-\,{1\over2}{s_{12}\over s_{23}}\ ,
\label{Universal1} \\
V_{2}&\equiv&
\biggl[{1\over\e^2}+{3\over2\e}\biggr]
\biggl({\mu^2\over-s_{12}}\biggr)^\e
\,+\,
\Ls_{-1}\biggl( {-s_{23}\over-k_\phi^2}, {-s_{31}\over-k_\phi^2} \biggr)
\,+\,{7\over2}\,+\,{\delta_{R}\over2}\,
-\,{1\over2}{s_{12}\over s_{23}}\ ,
\label{Universal2} \\
V_{3}&\equiv&
-\,{2\over3\e}
\biggl({\mu^2\over-s_{12}}\biggr)^\e
\,-\,{10\over9} \ ,\label{Universal3}\\
V_{4}&\equiv&
-\,{1\over3\e}
\biggl({\mu^2\over-s_{12}}\biggr)^\e
\,-\,{8\over9} \ .\label{Universal4}
\eea
Here $\delta_R$ is a regularization-scheme dependent parameter, which fixes
the number of helicity states of the gluons running in the loop to 
$(4 - 2 \delta_R \epsilon)$.  For the 't~Hooft-Veltman scheme~\cite{HV}
$\delta_R = 1$, while in the four-dimensional helicity (FDH)
scheme~\cite{BKStringBased,OtherFDH}
$\delta_R = 0$.  (Note that in the FDH scheme, the combination
$V_1-V_2+V_3$ is equal to the function $U_3$ appearing in the 
$Hggg$ amplitude.)  The $n_s$ term can be deduced from the $n_\f$ term
using the fact that both arise from vacuum-polarization contributions.
For this helicity configuration, the fermion and scalar loop 
contributions are {\it not} equal and opposite in sign.

{}From eqs. (\ref{Hreconstruct}), (\ref{phiff3}), and (\ref{quarkloop})
then follows,
\be
A_{3;1}(\phi^\dagger,1_f^-,2_f^+,3^+) =
A_3^\tree(\phi^\dagger,1_f^-,2_f^+,3^+)
\Biggl[ V_1 + {s_{12}\over 2 \, s_{23}} + 2 
+ {1\over N_c^2} \left( V_2 + {s_{12}\over 2 \, s_{23}} \right)
+ {n_\f \over N_c} V_3 
+ {n_s \over N_c}  V_4 \Biggr]\,
\label{phidaggerff3}.
\ee
The $L$, $R$, $f$ and $s$ pieces of $A_{3;1}(\phi,1_f^-,2_f^+,3^-)$
can easily be extracted from the various color components 
of~\eqn{phidaggerff3} after applying parity, \eqn{ParityExch}.
Note that the $s_{12}/(2s_{23})$ terms in \eqn{phidaggerff3} 
cancel similar terms in $V_1$ and $V_2$ in \eqns{Universal1}{Universal2}.

Using \eqn{Areconstruct}, the corresponding amplitude with a pseudoscalar 
$A$ is found to be,
\be
 A_{3;1}(A,1_f^-,2_f^+,3^+) =
i\, A_3^\tree(\phi^\dagger,1_f^-,2_f^+,3^+)
\Biggl[ V_{1} + {s_{12}\over s_{23}} + 4 
+ {1\over N_c^2} \left( V_{2} + {s_{12}\over s_{23}} \right)
+ {n_\f \over N_c} V_3
+ {n_s \over N_c} V_4 \Biggr]\, .
\label{quarkloopA}
\ee

\subsection{Partial results for four partons}

The finite helicity amplitudes for a $\phi$ plus four gluons
are given by,
\bea
A_4^{[s]}(\phi,1^+,2^+,3^+,4^+) & = & 0 \, ,
\label{A4spppp}\\
A_4^{[g-s]}(\phi,1^+,2^+,3^+,4^+)
& = & - 2 \,A_4^\tree(\phi^\dagger,1^+,2^+,3^+,4^+) \, ,
\label{A4gminusspppp}\\
A_4^{[s]}(\phi,1^-,2^+,3^+,4^+)
& = & \frac{i}{3} \Biggl[
   - { \spa1.2 \spb2.3 {\spa1.3}^2
     \over {\spa2.3}^2 \spa3.4 \spa4.1 }
   - { \spa1.4 \spb4.3 {\spa1.3}^2
     \over \spa1.2 \spa2.3 {\spa3.4}^2 }
\nonumber \\
&&\hskip0.5cm
   + { \spab1.{(2+3)}.4 \spab1.{(3+4)}.2
      \over \spa2.3 \spa3.4 \, s_{234} } \Biggr] \, ,
\label{A4smppp}\\
A_4^{[g-s]}(\phi,1^-,2^+,3^+,4^+)
& = & - 2\, i \, \Biggl[
     { {\spab1.{(2+3)}.4}^3
      \over \spa1.2 \spa2.3 \spab3.{(1+2)}.4 \, s_{123} }
\nonumber \\
&&\hskip0.9cm
   + { {\spab1.{(3+4)}.2}^3
      \over \spa3.4 \spa4.1 \spab3.{(1+4)}.2 \, s_{134} }
\nonumber \\
&&\hskip0.9cm
   + { (k_\phi^2)^2 \, {\spb2.4}^4
     \over \spb4.1 \spb1.2 \, \spab3.{(1+2)}.4 \spab3.{(1+4)}.2
     \, s_{124} }
\Biggr]
\,.~~~~~~~
\label{A4gminussmppp}
\eea
We have also computed the terms containing cuts in the following
all-minus amplitude,
\be
A_4^{[g-s]}(\phi,1^-,2^-,3^-,4^-) =
A_{4}^\tree(\phi,1^-,2^-,3^-,4^-) [ U_4 + 2 ]
\, ,
\label{A4gminussmmmm}
\ee
where
\bea
U_4 &\equiv&
- {1\over\e^2} \Biggl[ \biggl( {\mu^2 \over -s_{12}} \biggr)^\e
                      + \biggl( {\mu^2 \over -s_{23}} \biggr)^\e
                      + \biggl( {\mu^2 \over -s_{34}} \biggr)^\e
                      + \biggl( {\mu^2 \over -s_{41}} \biggr)^\e \Biggr]
\nonumber \\
&&\hskip0cm
- \Ls_{-1}\biggl( {-s_{12}\over-s_{123}}, {-s_{23}\over-s_{123}} \biggr)
- \Ls_{-1}\biggl( {-s_{23}\over-s_{234}}, {-s_{34}\over-s_{234}} \biggr)
\nonumber \\
&&\hskip0cm
- \Ls_{-1}\biggl( {-s_{34}\over-s_{341}}, {-s_{41}\over-s_{341}} \biggr)
- \Ls_{-1}\biggl( {-s_{41}\over-s_{412}}, {-s_{12}\over-s_{412}} \biggr)
\nonumber \\
&&\hskip0cm
- \Ls_{-1}^{2{\rm m}e}(s_{123},s_{234};s_{23},k_\phi^2)
- \Ls_{-1}^{2{\rm m}e}(s_{234},s_{341};s_{34},k_\phi^2)
\nonumber \\
&&\hskip0cm
- \Ls_{-1}^{2{\rm m}e}(s_{341},s_{412};s_{41},k_\phi^2)
- \Ls_{-1}^{2{\rm m}e}(s_{412},s_{123};s_{12},k_\phi^2)
\,.
\label{U4def}
\eea
The two-mass box function $\Ls_{-1}^{2{\rm m}e}(s,t;m_1^2,m_3^2)$
is defined as~\cite{eeFourPartons},
\bea
\Ls_{-1}^{2{\rm m}e}(s,t;m_1^2,m_3^2)  & \equiv &
     -\Li_2\left(1-{m_1^2\over s}\right)
     -\Li_2\left(1-{m_1^2\over t}\right)
     -\Li_2\left(1-{m_3^2\over s}\right)
     -\Li_2\left(1-{m_3^2\over t}\right)
\nonumber\\
&& \hskip0.0cm
+ \Li_2\left(1-{m_1^2m_3^2\over st}\right)
     -{1\over2}\ln^2\left({-s\over-t}\right) \, .
\label{Ls2me}
\eea
This result agrees with the all-$n$ cut-constructible
expression $A_n^{(1),CC}$ found previously in 
ref.~\cite{BadgerGloverH4g}, when one substitutes $n=4$.
As in the case of the $Hggg$ amplitudes, the ``$2$'' added to $U_4$
in \eqn{A4gminussmmmm} for $A_4^{[g-s]}(\phi,1^-,2^-,3^-,4^-)$
cancels against a ``$-2$'' in $A_4^{[g-s]}(\phi^\dagger,1^-,2^-,3^-,4^-)$
(the image under parity of $A_4^{[g-s]}(\phi,1^+,2^+,3^+,4^+)$
from \eqns{allplusnfnsvanish}{allplusphi}), 
when the $\phi$ and $\phi^\dagger$ amplitudes are added together
to produce $A_4^{[g-s]}(H,1^-,2^-,3^-,4^-)$, or equivalently
$A_4^{(1),CC}$ of ref.~\cite{BadgerGloverH4g}.

The non-cut-constructible contribution to this amplitude was
computed in ref.~\cite{BadgerGloverH4g}.  The result is
\bea
A_4^{[s]}(\phi,1^-,2^-,3^-,4^-) &=& 
  S^{----}(1,2,3,4) + S^{----}(2,3,4,1)
\nonumber\\
&&\hskip0.0cm
+ \, S^{----}(3,4,1,2) + S^{----}(4,1,2,3) \, ,
\label{A4smmmm}
\eea
where
\bea
S^{----}(1,2,3,4) &=& 
{ i \over 3 } \Biggl[
- \, { s_{13} \, {\spab4.(1+3).2}^2 \over {\spb1.2}^2 {\spb2.3}^2 \, s_{123} }
+ { {\spa3.4}^2 \over {\spb1.2}^2 } 
+ 2 \, { \spa3.4 \spa4.1 \over \spb1.2 \spb2.3 }
\nonumber\\
&&\hskip0.5cm
+ \, {1\over2} \, { s_{12} s_{34} + s_{123} s_{234} - s_{12}^2 
            \over \spb1.2 \spb2.3 \spb3.4 \spb4.1 } \Biggr] \,,
\label{Smmmm}
\eea
and the four terms in \eqn{A4smmmm} are the images of $S^{----}(1,2,3,4)$
under cyclic permutations of the gluon legs.  This result was computed
as a gluon and fermion loop contribution, but it can be extended to 
incorporate a scalar loop contribution using factorization properties.
We checked the collinear and multiparticle 
factorization limits of $A_4^{[g-s]}(\phi,1^-,2^-,3^-,4^-)$
and $A_4^{[s]}(\phi,1^-,2^-,3^-,4^-)$
onto the corresponding $\phi ggg$ and $\phi gg$ amplitudes.

The soft Higgs limit of the scalar loop term 
$A_4^{[s]}(\phi,1^-,2^-,3^-,4^-)$ in \eqn{A4smmmm}
can easily be seen to obey~\eqn{softphi1} with a factor
of $n_{-}=4$. The target pure-QCD amplitude, 
from \eqn{OneLoopAllPlusAmplitude}, is
\be
A_4^{[s]}(1^-,2^-,3^-,4^-) 
= - {i\over3} { \spa1.2 \spa3.4 \over \spb1.2 \spb3.4 } \,.
\label{OneLoopFourMinusAmplitude}
\ee
Equivalently, the Higgs amplitude $A_4^{[s]}(H,1^-,2^-,3^-,4^-)$ 
obeys~\eqn{softHiggs2loop} for $l=1$ and $n=4$, {\it i.e.} with
a factor of $n+2l-2 = 4$~\cite{BadgerGloverH4g}.
The $[g-s]$ term $A_4^{[g-s]}(\phi,1^-,2^-,3^-,4^-)$ 
in \eqn{A4gminussmmmm} vanishes in the soft limit, because the 
tree amplitude $A_{4}^\tree(1^-,2^-,3^-,4^-)$ vanishes,
so that \eqns{HdivIR}{phidivIR} for the soft limits of the
divergent parts hold rather trivially.

Finally, the amplitude for a pseudoscalar $A$ plus four 
negative-helicity gluons is given via \eqn{Areconstruct} as
\bea
A_{4;1}(A,1^-,2^-,3^-,4^-) 
&=& - i \, A_{4;1}(H,1^-,2^-,3^-,4^-)
+ 2 \, i \, A_4(\phi^\dagger,1^-,2^-,3^-,4^-)~~~~ \label{AA4minus1}\\
&=& - i \Biggl[ A_{4}^\tree(\phi,1^-,2^-,3^-,4^-) [ U_4 + 4 ]
\nonumber\\
&&\hskip0.7cm
+ \, \biggl( 1 - {n_\f \over N_c} + {n_s \over N_c } \biggr)
  A_4^{[s]}(\phi,1^-,2^-,3^-,4^-) \Biggr] \,.
\label{AA4minus2}
\eea
%

\section{Conclusions}
\label{ConclusionSection}

The production via gluon fusion of a Higgs boson in association
with multiple jets 
provides an important background to Higgs production via electroweak
vector boson fusion.  Analytic representations of one-loop amplitudes 
containing a Higgs boson and four or more partons, interacting via the
operator $H \tr G_{\mu\nu} G^{\mu\nu}$, may be useful in
quantifying this background. In this paper we have shown that on-shell
recursive methods can be applied at the loop level, to amplitudes
containing a complex $\phi$ field plus multiple partons, where
$\phi = H + i A$, with $H$ a scalar Higgs boson and $A$ a pseudoscalar.
Previous applications of such loop-level relations were to pure QCD.
There are three infinite sequences of finite one-loop amplitudes
containing a single $\phi$ field and multiple QCD partons.
We constructed recursion relations for all three sequences and 
provided the solutions.

In particular, we presented compact formul\ae{} for the finite
one-loop amplitudes with a single complex scalar $\phi$, one quark pair,
and $n-2$ gluons of positive helicity, which 
are built from the primitive amplitudes $A_n^{L-s}(\phi;j_\f)$ 
and $A_n^s(\phi;j_\f)$.  \Eqns{phiLs}{phis} provide the all-$n$ forms  
of these primitive amplitudes with positive-helicity gluons. 
The corresponding tree-level quark-gluon amplitudes vanish, and
hence these amplitudes are both infrared- and ultraviolet-finite.
The amplitudes for a $\phi$ field and $n$ gluons,
all or all but one having positive helicity, were presented
respectively in \eqns{allplusphi}{phioneminus}.  The case
of one negative helicity was obtained by factorization from 
the finite quark-containing series of amplitudes.
Together these amplitudes constitute all of the finite loop amplitudes 
for a single $\phi$ field and multiple partons.
Loop amplitudes containing additional quark pairs
are always infrared and ultraviolet divergent, because the corresponding
tree amplitudes are nonzero.

Even though the corresponding tree-level amplitudes vanish, the finite
one-loop amplitudes we computed enter next-to-leading order cross
sections for Higgs-plus-jet production at hadron colliders,
because the Higgs amplitude is a sum of $\phi$ and $\phi^\dagger$ amplitudes.
However, first the corresponding $\phi^\dagger$ amplitudes have to be
computed.  Or equivalently (by parity), the $\phi$ amplitudes for
configurations with multiple negative-helicity gluons are required.

We collected the complete set of one-loop $\phi$ amplitudes for two 
and three partons, as well as giving partial analytic results for four gluons, 
using also results from refs.~\cite{Schmidt,BadgerGloverH4g}.  
The four-quark Higgs results were presented analytically in cross-section 
form in ref.~\cite{EGZHiggs}.
The $\phi$ amplitudes can be used to obtain
amplitudes for a pseudoscalar $A$, as well as a scalar Higgs.
We discussed the analytic properties of generic $\phi$ and $H$
amplitudes in the soft Higgs limit, as the Higgs momentum vanishes. 
For the finite $\phi$ amplitudes this behavior is quite simple.
The corresponding amplitude in pure QCD is recovered, multiplied
by a factor of the number of negative helicities, $n_{-}$.
However, in the generic case the soft behavior is complicated,
for many components of the amplitudes, by infrared divergences.

The finite $\phi$ amplitudes also appear in factorization limits 
of the remaining, divergent one-loop $\phi$ amplitudes.
Because of this property, their values will serve as input into 
a recursive analytic construction of the latter amplitudes. 
We anticipate no conceptual problems in implementing such a 
unitarity-factorization bootstrap program~\cite{Bootstrapping,BBDFK1,BBDFK2},
order-by-order in the number of negative-helicity gluons.


{\bf Note added}

Soon after this work appeared, the first computation at NLO of Higgs
production via gluon fusion in association with two jets
was completed~\cite{CEZ}.


\section*{Acknowledgments}

We are grateful to Simon Badger, Zvi Bern, Nigel Glover, David Kosower 
and Kasper Risager for stimulating discussions.
We thank Yorgos Sofianatos for helping us uncover an error
in a previous version of this article.
L.D. thanks the INFN, Torino, Roma ``La Sapienza'', and Rome III 
for hospitality, and V.D.D. thanks SLAC for hospitality, during the
course of this work.  The figures were generated using 
Jaxodraw~\cite{Jaxo}, based on Axodraw~\cite{Axo}.


\appendix
\section{Normalization of $A_{2;1}(\phi,1^+,2^+)$}
\label{NormalizationAppendix}

In this appendix we give the result for the
finite one-loop amplitude $A_{2;1}(\phi,1^+,2^+)$
produced by the effective Lagrangian~(\ref{effintb}),
including the normalization factor, which feeds into
all the other finite amplitudes discussed in the paper.
We show that this normalization is consistent with results in
the literature for the difference between the NLO QCD
corrections to cross sections for inclusive production of
a pseudoscalar Higgs boson, versus a scalar one, at hadron colliders.

We have computed the one-loop helicity amplitudes for a
scalar field $H$ and pseudoscalar $A$,
along with two positive-helicity gluons,
$A_{2;1}(H,1^+,2^+)$ and $A_{2;1}(A,1^+,2^+)$, whose normalization
is defined according to \eqn{AdjointColorDecomposition}.
In dimensional regularization, bubbles on external lines vanish.
So there are no contributions from massless quark or squark loops
because there is no direct coupling between the $\phi$ field and
massless fermions; in other words, $A_{2;1}$ has no dependence on 
$n_\f$ or $n_s$.  Accordingly (see \eqn{OneloopnfnsDecomp}),
\bea
A_2^{[f]}(\phi,1^+,2^+) &=& A_2^{[s]}(\phi,1^+,2^+)\ =\ 0 \,,
\label{loopnfnsvanish}\\
A_{2;1}(\phi,1^+,2^+) &=& A_2^{[g]}(\phi,1^+,2^+)\,.
\label{loop2g}
\eea
There are only two nonvanishing
diagrams in the scalar case:  a triangle diagram, and a bubble
diagram involving the four-gluon vertex in pure QCD.
In the pseudoscalar case, the bubble diagram vanishes.
In quoting the results, we do not perform any coupling renormalization,
or renormalization of the effective Lagrangian operators in
\eqn{effinta}.  The results are,
\begin{eqnarray}
A_{2;1}(H,1^+,2^+) &=& \biggl( {\mu^2 \over -s_{12}} \biggr)^\e
\biggl[ - {2 \over \e^2} + \Ord(\e) \biggr]
A_2^\tree(H,1^+,2^+) \,,
\label{scalar2gluons}\\
A_{2;1}(A,1^+,2^+) &=& \biggl( {\mu^2 \over -s_{12}} \biggr)^\e
\biggl[ - {2 \over \e^2} + 4 + \Ord(\e) \biggr]
A_2^\tree(A,1^+,2^+) \,.
\label{pseudo2gluons}
\end{eqnarray}
The normalization of the leading poles in $\e$ agrees
with the general structure of one-loop infrared
divergences~\cite{UniversalIR}.

Because $A_2^\tree(\phi,1^+,2^+)$ vanishes, and using \eqn 
{phivanishtree},
we have
$A_2^\tree(A,1^+,2^+) = i A_2^\tree(H,1^+,2^+)
= i A_2^\tree(\phi^\dagger,1^+,2^+)$.
We can combine the $H$ and $A$ amplitudes~(\ref{scalar2gluons})
and (\ref{pseudo2gluons}) into the finite amplitude containing
$\phi = {1\over2}(H+iA)$,
\begin{eqnarray}
A_{2;1}(\phi,1^+,2^+)
&=&
{1\over2} \Bigl[ A_{2;1}(H,1^+,2^+) + i A_{2;1}(A,1^+,2^+) \Bigr]
\nonumber\\
&=&
{1\over2} \biggl( {\mu^2 \over -s_{12}} \biggr)^\e
\Biggl\{
\biggl[ - {2 \over \e^2} + \Ord(\e) \biggr] A_2^\tree(H,1^+,2^+)
\nonumber\\
&&\hskip2.0cm
+ i \biggl[ - {2 \over \e^2}  + 4 + \Ord(\e) \biggr]
i A_2^\tree(H,1^+,2^+) \Biggr\}
\nonumber\\
&=& - 2 A_2^\tree(H,1^+,2^+) + \Ord(\e)
\nonumber\\
&=& - 2 A_2^\tree(\phi^\dagger,1^+,2^+) + \Ord(\e) \,.
\label{phi12norm}
\end{eqnarray}
Using \eqn{Hreconstruct}, we can obtain $A_{2;1}(\phi^\dagger,1^+,2^+)$
by subtracting \eqn{phi12norm} from \eqn{scalar2gluons}.  We can
then use parity, \eqn{ParityExch}, to get $A_{2;1}(\phi,1^-,2^-)$.
Now $A_{2;1}(\phi,1^\mp,2^\pm)$ vanishes by angular-momentum  
conservation.
So the complete set of $\phi gg$ amplitudes is given, through $\Ord 
(\e)$,
by
\bea
A_{2;1}(\phi,1^+,2^+)  &=& - 2 A_2^\tree(\phi^\dagger,1^+,2^+) \, ,
\label{phiPP}\\
A_{2;1}(\phi,1^\mp,2^\pm)  &=& 0 \, ,
\label{phiMP}\\
A_{2;1}(\phi,1^-,2^-) &=& \biggl( {\mu^2 \over -s_{12}} \biggr)^\e
\biggl[ - {2 \over \e^2} + 2 \biggr]  A_2^\tree(\phi,1^-,2^-) \, .
\label{phiMM}
\eea

The finite one-loop amplitude~(\ref{phiPP}) also enters the
difference in the NLO corrections to the inclusive
hadron-collider production of a pseudoscalar {\it vs.} a scalar
Higgs boson. In the large-$m_t$ limit, the part of the NLO corrections
coming from real emission is identical in the pseudoscalar
and scalar cases~\cite{NLOPseudoSDGZ,NLOPseudoKS}.
In terms of $\phi$ and $\phi^\dagger$ amplitudes, this result follows
from the fact that the individual tree-level helicity amplitudes  
entering the
real-emission subprocesses, $gg\to Hg$, $qg\to Hq$,
$gg\to Ag$, $qg\to Aq$, plus processes related by crossing,
are {\it either} MHV {\it or} anti-MHV --- but not both.
For every possible parton helicity configuration, either
the $\phi$ amplitude $A_3^\tree(\phi,\ldots)$ vanishes according to
\eqn{phivanishtree} or \eqn{phiffvanishtree}, or else the
parity conjugate $\phi^\dagger$ amplitude
$A_3^\tree(\phi^\dagger,\ldots)$ vanishes.
(Only when four partons enter the amplitude, do there begin to be
helicity configurations for which the $\phi$ and $\phi^\dagger$
amplitudes are simultaneously nonvanishing.  Such configurations,
for example $g^-g^-g^+g^+$, are in the ``overlap'' of the
tree-level MHV and anti-MHV towers~\cite{DGK}.)

If either $A_3(\phi,\ldots)$ or $A_3(\phi^\dagger,\ldots)$
always vanishes, then according to \eqns{Hreconstruct}{Areconstruct},
the squared amplitudes for $A$ must equal those for $H$,
\begin{equation}
|A_3^\tree(A,1^\pm,2^\pm,3^\pm)|^2
= |A_3^\tree(H,1^\pm,2^\pm,3^\pm)|^2 \,,
\label{HAident}
\end{equation}
and similarly for the $qg\to Hq$ ($qg\to Aq$) process.
Thus the real-emission contributions (normalized to the
respective Born-level cross sections) are identical.

The finite one-loop amplitude~(\ref{phi12norm}) gives
rise to a difference in the virtual corrections,
\begin{eqnarray}
&&{ d\tilde\sigma^{A,\,{\rm 1-loop}} \over d\sigma^{A,\,{\rm tree}} }
- { d\tilde\sigma^{H,\,{\rm 1-loop}} \over d\sigma^{H,\,{\rm tree}} }
\nonumber\\
&=&
{ g^2 N_c \over (4\pi)^2 } { 1 \over
\sum_{h_1,h_2} \Bigl| A_2^\tree(\phi,1^{h_1},2^{h_2})
              + A_2^\tree(\phi^\dagger,1^{h_1},2^{h_2}) \Bigr|^2 }
\nonumber\\
&&\hskip0cm
\times
2 \Re \sum_{h_1,h_2}
\Biggl\{ \Bigl[ A_{2;1}(\phi,1^{h_1},2^{h_2})
              - A_{2;1}(\phi^\dagger,1^{h_1},2^{h_2}) \Bigr]
           \Bigl[ A_2^\tree(\phi,1^{h_1},2^{h_2})
              - A_2^\tree(\phi^\dagger,1^{h_1},2^{h_2}) \Bigr]^*
\nonumber\\
&&\hskip2cm
       - \Bigl[ A_{2;1}(\phi,1^{h_1},2^{h_2})
              + A_{2;1}(\phi^\dagger,1^{h_1},2^{h_2}) \Bigr]
    \Bigl[ A_2^\tree(\phi,1^{h_1},2^{h_2})
              + A_2^\tree(\phi^\dagger,1^{h_1},2^{h_2}) \Bigr]^*
\Biggr\}
\nonumber\\
&=& {3 \as \over 4\pi}
{-4 \Re \biggl[ A_{2;1}(\phi,1^+,2^+) A_2^\tree(\phi^\dagger,1^+,2^+)^*
               + A_{2;1}(\phi^\dagger,1^-,2^-) A_2^\tree(\phi,1^-,2^-)^*
         \biggr]
\over  A_2^\tree(\phi^\dagger,1^+,2^+) A_2^\tree(\phi^\dagger,1^+,2^+)^*
      + A_2^\tree(\phi,1^-,2^-) A_2^\tree(\phi,1^-,2^-)^* }
\nonumber\\
&=& {3 \as \over 4\pi} (-4) (-2)
\nonumber\\
&=& {6 \as \over \pi} \,.
\label{checknorm}
\end{eqnarray}
However, \eqn{checknorm} does not take into account the
difference in QCD corrections to the matching coefficients for
the operators $\tr G_{\mu\nu} G^{\mu\nu}$ and
$\tr G_{\mu\nu}\, {}^*G^{\mu\nu}$ (which is why we put tildes
over $d\tilde\sigma^{H(A),\,{\rm 1-loop}}$).
The operator $\tr G_{\mu\nu} G^{\mu\nu}$ has coefficient
$C \times [1 + 11\as/(4\pi)]$ at NLO; the operator
$\tr G_{\mu\nu}\, {}^*G^{\mu\nu}$ is unrenormalized.
(The induced coupling of the pseudoscalar to the divergence of the
axial current does not contribute to the cross section at NLO,
as discussed in \sect{AxialCurrentDivSubsection}.)
Including this correction, we find
\begin{eqnarray}
{ d\sigma^{A,\,{\rm 1-loop}} \over d\sigma^{A,\,{\rm tree}} }
- { d\sigma^{H,\,{\rm 1-loop}} \over d\sigma^{H,\,{\rm tree}} }
&=&
{ d\tilde\sigma^{A,\,{\rm 1-loop}} \over d\sigma^{A,\,{\rm tree}} }
- { d\tilde\sigma^{H,\,{\rm 1-loop}} \over d\sigma^{H,\,{\rm tree}} }
- 2 \, {11 \as \over 4 \pi}
\nonumber\\
&=& \biggl( 6 - {11\over2} \biggr) { \as\over\pi }
= {\as\over2\pi} \,.
\label{NLOshift}
\end{eqnarray}
This result agrees with the NLO calculations in
refs.~\cite{NLOPseudoSDGZ,NLOPseudoKS,PseudoField},
which were also reconfirmed in the course of computing the NNLO
corrections~\cite{NNLOPseudo}.

\section{Validity of tree-level on-shell recursion relations with $\phi$}
\label{TreeProofAppendix}

As outlined in \sect{RecursionReviewSection}, the derivation of
the on-shell recursion relations relies on the vanishing of the
amplitude $A(z)$ as $z \rightarrow \infty$. For the pure-QCD case
this result was shown in refs.~\cite{BCFW,BadgerMassive}. Below we
extend the argumentation to amplitudes with a $\phi$ field.

\subsection{Vanishing of $A(z \rightarrow \infty)$ for the $[-,+\rangle$ case}
\label{minuspluscaseSection}

In ref.~\cite{BCFW} it was argued, based on the analysis of standard
Feynman diagrams, that the amplitude $A(z)$ in the pure-QCD case
vanishes as $z \rightarrow \infty$ for the case that the shifted
helicities are $h_k = -,\, h_l = +$, with the corresponding
shifted spinors given in~\eqn{SpinorShift}. The momenta are shifted
according to \eqn{MomentumShift}. Let us recall this argument: Any
Feynman diagram contributing to a pure-gluon amplitude $A(z)$ is
linear in the polarization vectors of external gluons. Because only
gluons $k$ and $l$ are shifted, only their polarization vectors
can depend on $z$. These polarization vectors are given in
spinor-helicity notation by
\be \varepsilon_{\alpha \dot{\alpha}}^{-} 
= - { \lambda_\alpha \tilde{\eta}_{\dot{\alpha}}
\over {\spbsh{ \tilde{\lambda}}.{\tilde{\eta}}} } \ ,
\qquad 
\varepsilon_{\alpha \dot{\alpha}}^{+} 
= \frac{\eta_\alpha \tilde{\lambda}_{\dot{\alpha}}}{\spa{\eta}.{\lambda}} \ ,
\ee
where $\eta$, $\tilde{\eta}$ are fixed but arbitrary reference
spinors. We see that these two polarization vectors behave as
$1/z$ as $z \rightarrow \infty$ for the helicity configurations
$h_k = -,\, h_l = +$, contributing a factor of $z^{-2}$ to the
amplitude. The propagators each contribute a factor of $1/z$,
whereas triple-gluon vertices are proportional to $z$ in the $z
\rightarrow \infty$ limit. Quartic vertices have no momentum
factors.  Therefore the worst possible contribution to $A(z)$ stems
from diagrams with only cubic vertices. A graph with $(r+1)$ cubic
vertices has $r$ propagators. Taking into account the polarization
vectors of the external gluons, the complete behavior as $z\rightarrow
\infty$ is $\sim z^{r+1}/z^{r} \times\,z^{-2}$. Thus the amplitude
vanishes as $1/z$ as $z \rightarrow \infty$ for the helicity
configuration $h_k = -,\, h_l = +$\,.

Now we extend this argument to the case where we attach a complex
scalar $\phi$ to the amplitude, again with the above helicity configuration.
Vertices with 3 and 4 gluons have the same $z \rightarrow \infty$
behavior with and without a $\phi$ attached, thus the above
argument applies directly to those Feynman graphs where the
$z$-dependent momentum flows through only such vertices. However,
the effective Lagrangian~(\ref{effintb}) also contains a vertex 
$V_{\phi gg}$, coupling a $\phi$ field to two gluons.  This vertex 
could potentially spoil the above power counting.

Consider the set of graphs where the large $z$-dependent
momentum flows through $r$ triple-gluon vertices, 
a $V_{\phi gg}$ vertex, and $r$ gluon propagators. 
The $V_{\phi gg}$ vertex is given by
\be 
V_{\phi gg}^{\mu_1 \mu_2} (k_1,k_2)= k_1
\cdot k_2 \, \eta^{\mu_1 \mu_2} - k_2^{\mu_1} k_1^{\mu_2} + i
\varepsilon^{\mu_1\mu_2 \nu_1 \nu_2} k_{1\, \nu_1}k_{2\, \nu_2} 
\,,
\label{Vphigg} 
\ee
where $k_1$ and $k_2$ are the momenta of the two gluon lines.
Naively, one might suspect such graphs to
give nonvanishing contributions as $z \rightarrow \infty$,
because of the $k_2^{\mu_1} k_1^{\mu_2}$-term in $V_{\phi gg}$,
which seems to behave as $z^2$ after the shift~(\ref{MomentumShift}). 
However, the $z^2$ contribution from the term 
$\sim k_2^{\mu_1} k_1^{\mu_2}$ vanishes using current
conservation, after summing over the subset of graphs for which a fixed
set of gluons attaches to each gluon line emanating from 
$V_{\phi gg}$.  That is, the two momenta
$k_1$ and $k_2$ are proportional to each other in the large-$z$
limit, with a difference of order $z^0$, because 
$k_2 = - (k_1 + k_\phi)$, and $k_\phi$ carries no $z$-dependence. 
Thus we can replace $k_2^{\mu_1}(z) k_1^{\mu_2}(z)$ by $-k_1^{\mu_1}(z)
k_1^{\mu_2}(z)$ in the large-$z$ limit. The sum of graphs with a
fixed set of on-shell external gluons attached to the momentum line 
$k_1$ forms a current $J_1^{\mu_1}(z)$.  Because current conservation,
$k_1 \cdot J_1 = 0$, holds also for complex momenta shifted according to
\eqn{MomentumShift}, this sum of graphs vanishes as $z \rightarrow
\infty$. The $z^2$-contribution from the
first term in \eqn{Vphigg}
vanishes on a graph-by-graph basis using \eqn{MomentumShift}, and
the $z^2$-contribution from the last term vanishes on a
graph-by-graph basis as well,
using the antisymmetry of the Levi-Civita symbol. Thus
the vertex behaves again at worst as $z^1$, and the counting
proceeds analogously to the pure-QCD case considered above, with
the difference that a sum over certain subsets of graphs is
required. In summary, tree-level amplitudes $A(z)$ with and
without $\phi$ vanish at least as $1/z$ as $z \rightarrow \infty$
for the helicity configuration $h_k = -,\, h_l = +$\,.

\subsection{Vanishing of $A(z \rightarrow \infty)$ for the 
$[+,+\rangle$ and $[-,-\rangle$ cases}

The cases where the shifted gluons are both positive or both
negative are slightly more complicated and require a more
involved argumentation along the lines of ref.~\cite{BadgerMassive}.
In these cases, certain subsets of Feynman diagrams need to be
summed to show the vanishing of $A(z)$ as $z \rightarrow \infty$
already in the pure-QCD case. The extension to the $\phi$-amplitude
case works as in \sect{minuspluscaseSection}, summing over
additional subsets of graphs.

Here we recall the argument of ref.~\cite{BadgerMassive}, without going
into too much detail, but pointing out the necessary modifications
for the $\phi$-case. We begin with the $[+,+\rangle$ case, and consider
amplitudes which, besides gluons $g_i^+,\, g_j^+$, have also at
least two gluons of negative helicity $g_k^-,g_l^-$. Amplitudes
with fewer than two negative-helicity gluons can be obtained via
soft-gluon limits of such amplitudes.  The argument is inductive in
the number of external gluons.  We proceed in two steps.  First we
perform a generalized shift on the
spinors $\lambda_i, \tilde{\lambda}_k, \tilde{\lambda}_l$,
\be
\lambda_i \rightarrow \lambda_i + z \lambda_k +
z \lambda_l \, , \qquad
\tilde{\lambda}_k \rightarrow 
\tilde{\lambda}_k - z \tilde{\lambda}_i \, , \qquad
\tilde{\lambda}_l \rightarrow \tilde{\lambda}_l - z
\tilde{\lambda}_i \, . \label{tripleshift} 
\ee 
This triple shift preserves the sum of the momenta involved.
Performing the same counting as in
\sect{minuspluscaseSection}, the amplitude vanishes now as
$1/z^2$, or, without summing over the subset attaching to a
$V_{\phi gg}$ vertex, as $1/z$ if the large-$z$ momentum flow hits
such a vertex. In the latter case, however, for the argumentation
below, we need to sum again over this subset as above, to obtain
$1/z^2$ behavior. As usual, we obtain the recursion 
relation~(\ref{BCFWRepresentation}).

Now we perform a \emph{second shift}, this time only involving the
positive-helicity gluons $g_i^+, g_j^+$, 
\be
\lambda_i \rightarrow \lambda_i + y \lambda_j \, , \qquad
\tilde{\lambda}_j \rightarrow \tilde{\lambda}_j - y
\tilde{\lambda}_i \, . \label{secondshift} 
\ee
If, in a given term of \eqn{BCFWRepresentation}, $i$ and $j$ are
both contained in one of the amplitudes $A_L$ or $A_R$, 
then the entire $y$-dependence of the term is contained in 
one of these amplitudes; the other amplitude and the 
propagator are independent of $y$. 
But $A_L$ and $A_R$ have fewer legs than the full amplitude,
and thus by the induction hypothesis the amplitude vanishes 
as $y\rightarrow\infty$.  If $i$ is on one side of the pole
and $j$ is on the other side, in a term of \eqn{BCFWRepresentation}, 
then the propagator, and therefore the value of the shift $z$ at
the pole, will obtain $y$-dependence. 
In the large-$y$ limit, $z \sim y$. By the above
counting, the full amplitude behaves as $1/z^2 \sim 1/y^2$, where
we have counted powers of internal vertices, propagators, and the
polarization vectors of gluons $i,k,l$. Additionally, with the
shift~(\ref{secondshift}) we have to take into account the
polarization vector of gluon $j$, which now gives a factor of $y$
because it has positive helicity. In total, the amplitude now vanishes
as $1/y$ for large $y$. The argument for the $[-,-\rangle$ helicity-case
is analogous.

To obtain this behavior, sums over two different subsets of graphs
are required.  In the pure-QCD case~\cite{BadgerMassive}, 
only the sum over the graphs on the left and on the right 
of the recursion relation~(\ref{BCFWRepresentation}) is necessary. 
For the $\phi$ case an additional sum over the subsets forming 
the current attaching to $V_{\phi gg}$ is required, if the large
shift-dependent momentum flows through such a vertex.

The above argument works if there are two extra gluons of opposite
helicity to the shifted ones. If the amplitude under consideration
does not have such a helicity configuration, then we can argue as
in ref.~\cite{BadgerMassive} that we can obtain the desired
configuration by taking the soft-gluon limit from an auxiliary amplitude
with two extra gluons, as long as the auxiliary gluons are not
adjacent to the marked gluons $i$ and $j$. Moreover, in ref.~\cite{DGK}
it was shown that the configurations $A(\phi,\pm, +,\ldots,+)$
vanish completely at tree level, \eqn{phivanishtree}, without relying on
on-shell recursion relations, and thus such an auxiliary
construction is only necessary for the $[-,-\rangle$ case.

\end{document}